\documentclass[a4paper,11pt]{article}
\usepackage{jheppub} 
\usepackage[utf8]{inputenc}
\usepackage{physics}
\usepackage{slashed}
\usepackage{caption}
\usepackage{xcolor}
\usepackage{comment}
\usepackage{multirow}
\usepackage{graphics}
\usepackage{float}
\usepackage{cancel}
\usepackage{soul}
\usepackage{cases}
\usepackage{array}
\usepackage{mathtools}  
\usepackage{amsfonts}
\usepackage{hyperref}
\usepackage{amsmath}
\usepackage{amssymb}
\usepackage{tcolorbox}
\usepackage{tikz}     
\newcommand{\qeq}{\mathrel{\stackrel{\makebox[0pt]{\mbox{\normalfont\tiny ?}}}{=}}}

\title{\boldmath\boldmath 3D Conformal Field Theory in Twistor Space}

\author{Aswini Bala, Sachin Jain, Dhruva K.S., Deep Mazumdar and Vibhor Singh}

\affiliation{Indian Institute of Science Education and Research,\\ Dr Homi Bhabha Road, Pashan, Pune, India}

\emailAdd{aswini.bala@students.iiserpune.ac.in}
\emailAdd{sachin.jain@iiserpune.ac.in}
\emailAdd{k.s.dhruva@students.iiserpune.ac.in}
\emailAdd{deepkamal.mazumdar@students.iiserpune.ac.in}
\emailAdd{singh.vibhorrajesh@students.iiserpune.ac.in}

\abstract{The aim of this paper is to study three dimensional Lorentzian conformal field theories in twistor space. We formulate the conformal Ward identities and solve for two and three point Lorentzian Wightman functions. We found that the Helicity operators apart from the conformal generators play an important role in fixing their functional form. The equations take the form of first order Euler equations which in addition to the usual solutions that are polynomials, also possess  weak solutions which are distributional in nature. All of these play an important role in our analysis. For instance, in the case of three point functions, the distributional solutions are indeed the ones realized by the CFT correlators. We also extend our analysis to parity odd Wightman functions which take an interesting form in twistor space. We verify our results by systematically analyzing the corresponding Wightman functions in momentum space and spinor helicity variables and matching with the twistor results via a half-Fourier transform.}

\begin{document}
\maketitle

\section{Introduction}
Conformal field theories are traditionally formulated and analyzed in position space \cite{Belavin:1984vu,Rattazzi:2008pe,Poland:2022qrs,Hartman:2022zik}. However, in the past decade or so, a momentum space analysis 
\cite{Maldacena:2011nz,McFadden:2011kk,Ghosh:2014kba,Coriano:2013jba,Bzowski:2013sza,Bzowski:2015pba,Bzowski:2017poo,Farrow:2018yni,Bzowski:2018fql,Bautista:2019qxj,Lipstein:2019mpu,Baumann:2020dch,Jain:2020rmw,Jain:2020puw,Jain:2021wyn,Baumann:2021fxj,Jain:2021qcl,Jain:2021vrv,Jain:2021gwa,Jain:2021whr,Isono:2019ihz,Gillioz:2019lgs,Baumann:2019oyu} has led to a variety of interesting results that are not obvious from the position space perspective. These include  double copy relations \cite{Lipstein:2019mpu,Jain:2021qcl}, the relation between parity-even and parity-odd correlators \cite{Caron-Huot:2021kjy,Jain:2021wyn,Jain:2021gwa,Jain:2021whr,Skvortsov:2018uru}. Further, in three dimensions, which is the case of interest to us in this paper, off-shell spinor helicity variables have been developed \cite{Maldacena:2011nz,Baumann:2020dch,Baumann:2021fxj,Jain:2021vrv} and they provide another viable vantage point to uncover hidden structures in CFT correlators and make precise their connection to one higher dimensional S matrices. Most of this work has been in Euclidean signature, $\mathbb{R}^3$. The analysis of $\mathbb{R}^{2,1}$ Lorentzian CFTs in momentum space has seen some progress \cite{Gillioz:2018mto,Gillioz:2019lgs,Gillioz:2019iye,Bautista:2019qxj,Gillioz:2020mdd,Gillioz:2020wgw,Karateev:2020axc,Gillioz:2021sce,Gillioz:2021kee,Nishikawa:2023zcv}, but not nearly as much as in the Euclidean case in part due to the additional technical difficulties one encounters. However, Lorentzian CFTs, rather than their Euclidean counterparts are more suitable to reformulate in twistor space as was shown by \cite{Baumann:2024ttn}. The advantage twistor space has over the usual momentum space and spinor helicity variables is that it carries a \textit{linear} action of the conformal group and thus simplifies the form of conformal correlators, in particular those involving conserved currents.

Twistor theory has a long and rich history, first having been introduced by Penrose \cite{Penrose:1967wn} and has since led to a novel formulation of four dimensional scattering amplitudes \cite{Nair:1988bq,Witten:2003nn,Mason:2009sa,Arkani-Hamed:2009hub}. A salient feature of twistor space is that it naturally carries an action of the (super-)conformal group as we discussed above. As such, it also holds a huge potential for the analysis of conformal field theories. In particular, our focus in this paper are three dimensional conformal field theories (CFT$_3$).  In an earlier work for theories with $\mathcal{N}=1,2$ supersymmetry, we found that the introduction of Grassmann twistor variables greatly simplifies the analysis of superspace correlation functions \cite{Jain:2023idr}. In \cite{Baumann:2024ttn}, the authors employed the embedding space formalism for $CFT_3$ to define the Penrose transformation to twistor space. They obtained two and three point correlators of conserved currents directly using just the projective rescaling invariance of the Penrose transform and the fact that that the integrand depends only on twistor dot products. Our approach to twistor space in this paper is complementary to it in the sense that we shall instead setup and solve the infinitesimal conformal Ward identities in twistor space. We shall also find that the helicity operators on top of conformal invariance, play an important role as they lead to first order Euler equations that fully fixes the form of two and three point correlators. Moreover, our analysis also accommodates parity odd correlators which are not obvious to obtain from the perspective of the Penrose transform. We check our results by systematically analyzing and bootstrapping two and three point Wightman functions in momentum space and comparing them with the half-Fourier transform of the twistor results. We find a perfect agreement between these two approaches.

Finally, the formalism presented in this paper will be combined with the Grassmann twistors of \cite{Jain:2023idr} to develop super-twistor space and shall appear soon \cite{Jain:2025new}.
The outline of the paper is as follows:

\subsection*{Outline}
In section \ref{sec:3dCFTreview}, we review some salient aspects of $CFT_3$ that will be essential for the reminder of the paper. In particular, we discuss the construction of spinor helicity variables and the general form of two and three point correlators of conserved currents. In section \ref{sec:Twistors}, we half-Fourier transform from spinor helicity variables and set up and solve the conformal Ward identities and helicity identity in twistor space for two and three point correlators of conserved currents. Following up, in section \ref{sec:Manifesttwistorsec}, we work in manifest twistor variables which makes the analysis of the previous section much more natural. These twistor results are complemented by the systematic analysis of Wightman functions in section \ref{sec:Wightman}, where we lay out the axioms and present the methodology to solve for two and three point Wightman functions in momentum space and spinor helicity variables. We also discuss in detail the analytic continuation from and to Euclidean space. The subject of section \ref{sec:parityodd} is correlators that are odd under parity. We present their analysis in momentum space, spinor helicity variables and finally present a solution of how they can arise from twistor space. Finally, in section \ref{sec:Discussion}, we summarize and discuss a number of interesting future directions.

The paper is supplemented with a large number of essential appendices. Appendix \ref{app:Notation} sets up the notation and conventions used in the paper. In appendix \ref{app:distrubution}, we discuss the possibility of first order differential equations having distributional solutions and derive said solutions in detail. Appendix \ref{app:3dTwistorsApproach} deals with the geometry of twistor space. This is followed by appendices \ref{subsec:spinorhelicity}, \ref{app:CPTinv} which deal with reality conditions in spinor helicity variables, the action of the CPT transformation on them with implications on correlators both in spinor helicity and twistor variabes. This is followed by appendix \ref{app:HalfFourierTransform} which deals with the details of the half-Fourier transform that connects twistor space to spinor helicity variables. Appendix \ref{app:generators} lists the conformal generators in twistor space.  The remaining appendices deal with Wightman functions in momentum space and spinor helicity variables. Then, in appendix \ref{app:4dto3d}, we discuss the dimensional reduction from four dimensions to three dimensions to obtain the induced off-shell spinor helicity variables. The role of appendix \ref{app:Derivation of current conservation Ward-Takahshi identity} is to show how the usual Ward-Takahashi identities of a time-ordered correlator follow from its definition in terms of Wightman functions. In appendix \ref{app:WightmanFunctionProperties} we discuss some important properties of Wightman functions such as their reality properties. The last two appendices \ref{app:WightmanTwoPoint} and \ref{app:WightmanExamples} illustrate the formalism of section \ref{sec:Wightman} through explicit examples. 

\subsection*{A word on the notation}
While in Minkowski space, we work with a mostly plus metric as it is best suited for analytic continuation to Euclidean signature. A Wightman function is written as,
\begin{align}
    \langle 0|\cdots |0\rangle,
\end{align}
that is, we make it explicit that it is a vacuum expectation value of a particular ordered string of operators. In contrast we denote the fully Bose symmetric Euclidean correlators as,
\begin{align}
    \langle \cdots \rangle.
\end{align}
The subscripts $F,\;B$ like in \eqref{CFT3Euclidthreepoint} stand for free fermion and free boson respectively, while $h,\;nh$ (for instance in \eqref{CFT3Euclidthreepointhandnh}) stand for homogeneous and non-homogeneous respectively. Further, all our momentum space and spinor helicity correlators are translation invariant and come with a factor of $(2\pi)^d \delta^d(p_1+\cdots+p_n)$ even when not specified explicitly.

\section{A Brief tour of three dimensional CFT}\label{sec:3dCFTreview}
In this section, we review the spinor-helicity formalism for $CFT_3$, discuss the form the generators in these variables and review the general structure of two and three point correlation functions of conserved currents.
\subsection{Spinor helicity variables}
In three dimensions, one can use a pair of spinors to represent a generic momentum as follows \cite{Maldacena:2011nz}:
\begin{align}
    p_{\mu}\to p_{a}^{b}=p_\mu(\sigma^{\mu})_{a}^{b}=\frac{\lambda_{a}\Bar{\lambda}^{b}+\Bar{\lambda}_a \lambda^b}{2}
\end{align}
where the $\sigma^\mu$ are the Pauli matrices\footnote{More precisely, in Euclidean signature, these are the usual $\sigma^x,\sigma^y,\sigma^z$ and in Lorentzian signature they are $\sigma^x,\sigma^t=-i \sigma^y,\sigma^z$.} that obey,
\begin{align}
    \{\sigma^\mu,\sigma^\nu\}=g^{\mu\nu}.
\end{align}
In Euclidean space we have $g^{\mu\nu}=\delta^{\mu\nu}$ whereas in Minkowski space, $g^{\mu\nu}=\eta^{\mu\nu}$ where we work with a mostly plus metric. 

Our objects of interest or traceless symmetric spin-$s$ conserved currents $J_s^{a_1\cdots a_{2s}}$. In three dimensions, these objects only have two independent components. In the language of spinor helicity variables they are given by,
\begin{align}\label{pmcurrentdef}
    &\frac{J_{s}^{\pm}(\lambda,\Bar{\lambda})}{p^{s-1}}=\frac{1}{p^{s-1}}\zeta_{a_1}^{\pm}\cdots \zeta_{a_{2s}}^{\pm}J_s^{a_1\cdots a_{2s}}(\lambda,\Bar{\lambda})~,~\zeta_{a}^{-}=\frac{\lambda_a}{\sqrt{p}},\zeta_{a}^{+}=\frac{\Bar{\lambda}_a}{\sqrt{p}},
\end{align}
which we refer to as the positive and negative helicity current respectively. The magnitude of the momentum $p$ is given by,
\begin{align}\label{mommag}
    p=-\frac{1}{2}\lambda_a\Bar{\lambda}^{a}.
\end{align}
Let us now study the action of the symmetry generators on these currents.
\subsection*{Symmetry generators in spinor helicity}
The form of the symmetry generators do not depend on the signature of spacetime. Before getting into more details, let us briefly discuss the philosophy of bootstrapping conserved current correlators in spinor helicity variables. Usually, one starts with a translation invariant ansatz that is also consistent with the helicities of the currents in the correlator. The most important role is then played by the generator of special conformal transformations which fully fixes the form of two and three point correlators up to constants.  The action of the translation and special conformal generators on rescaled currents $\frac{J_s^{\pm}(\lambda,\Bar{\lambda})}{p^{s-1}}$ in spinor helicity variables is as follows\footnote{The rescaling makes the action of special conformal transformations simple.}:
\begin{align}\label{spinorhelicitygen}
    &P_{ab}=\frac{1}{2}\lambda_{(a}\,\Bar{\lambda}_{b)},\qquad K_{ab}=2\frac{\partial^2}{\partial\lambda^{(a}\,\partial \Bar{\lambda}^{b)}}.
\end{align}
The actions of rotations/boosts and dilatations can be found in appendix \ref{app:generators}. The important point to note from \eqref{spinorhelicitygen} is that the translations act in a simple multiplicative way whereas SCTs are second order differential operators. This fact leads to complicated expressions for correlation functions and obscures conformal invariance. Twistor space, on the other hand, as we shall see, will treat all generators on an equal footing and make them all first order.
 In addition to these generators, we also define the helicity operator which will play an important role: Its action on a current in a correlator is\footnote{
Also, please note that the formula \eqref{helicityopSH} is agnostic to to the type of correlator and thus applies to both Euclidean correlators and Lorentzian correlators such as Wightman functions.},
\begin{align}\label{helicityopSH}
    h_j\langle\cdots \frac{J_{s_j}^{\pm}(\lambda_j,\Bar{\lambda}_j)}{p_j^{s_j-1}}\cdots\rangle=-\frac{1}{2}\bigg(\lambda_j^a\frac{\partial}{\partial\lambda_j^a}-\Bar{\lambda}_j^a\frac{\partial}{\partial\Bar{\lambda}_j^a}\bigg)\langle\cdots \frac{J_{s_j}^{\pm}(\lambda_j,\Bar{\lambda}_j)}{p_j^{s_j-1}}\cdots\rangle=\pm s_j \langle\cdots \frac{J_{s_j}^{\pm}(\lambda_j,\Bar{\lambda}_j)}{p_j^{s_j-1}}\cdots\rangle.
\end{align}
where the final equality is guaranteed by the definition of the currents \eqref{pmcurrentdef}.
Note that when working in spinor helicity variables, this operator is necessary to fix the form of correlators as it is what distinguishes one helicity configuration from another. Also, from the definition \eqref{helicityopSH}, we see that $\lambda$ has helicity $-\frac{1}{2}$ whereas $\Bar{\lambda}$ has helicity $+\frac{1}{2}$. As an illustration of its use, consider the two point functions $\langle J_s^{-}J_{s}^{-}\rangle$ and $\langle J_s^{+}J_{s}^{+}\rangle$. After using momentum conservation, \eqref{helicityopSH} immediately fixes $\langle J_s^{-}J_{s}^{-}\rangle\propto (\lambda_1\cdot \lambda_2)^{2s}=\langle 1 2\rangle^{2s}$ and similarly $\langle J_s^{+}J_s^{+}\rangle\propto \langle \Bar{1}\Bar{2}\rangle^{2s}$. Similarly, three and higher point functions are constrained by \eqref{helicityopSH} with of course, more structures being possible at higher points\footnote{Usually, when working in spinor helicity variables, one starts with an ansatz that automatically satisfies \eqref{helicityopSH} and thus the important role in constraining correlators is played by the symmetry generators.}.

\subsection{Euclidean correlators}
The study and classification of conformal correlation functions of conserved currents in three dimensions have been the subject of intense scrutiny. 
The most general form of the two point function consists of a term that is even under parity and another one special to three dimensions that is odd:
\begin{align}
    \langle J_sJ_s\rangle=c_{\text{even}}\langle J_s J_s\rangle_{\text{even}}+c_{\text{odd}}\langle J_s J_s\rangle_{\text{odd}}.
\end{align}
Their three point functions take the form (schematically) \cite{Giombi:2011rz},
\begin{align}\label{CFT3Euclidthreepoint}
    &\langle J_{s_1}J_{s_2}J_{s_3}\rangle=c_{s_1s_2s_3}^{(F)}\langle J_{s_1}J_{s_2}J_{s_3}\rangle_{F}+c_{s_1s_2s_3}^{(\text{odd})}\langle J_{s_1}J_{s_2}J_{s_3}\rangle_{\text{odd}}+c_{s_1s_2s_3}^{(B)}\langle J_{s_1}J_{s_2}J_{s_3}\rangle_{B}.
\end{align}
$F$ and $B$ stand for the corresponding correlators computed in the free fermionic and free bosonic theories respectively and the $c_{s_1s_2s_3}$ are OPE coefficients.  The odd structure, although cannot be directly computed like the free theory results, it can be obtained via an \textit{epsilon} transform as follows \cite{Jain:2021gwa} :
\begin{align}\label{cFT3EuclidEPT}
    \langle J_{s_1}J_{s_2}J_{s_3}\rangle_{\text{odd}}=\langle \epsilon\cdot J_{s_1}J_{s_2}J_{s_3}\rangle_{F}-\langle \epsilon\cdot J_{s_1}J_{s_2}J_{s_3}\rangle_{B},
\end{align}
where,
\begin{align}\label{EPTdef}
    \epsilon\cdot J_{s}^{\mu_1\cdots\mu_s}(p_1)=\frac{p_{1\beta}\epsilon^{\alpha \beta (\mu_1}}{p_1}J_s^{\mu_2\cdots \mu_s)}~_{\alpha}(p).
\end{align}
An important fact to note is that the odd structure is compatible with current conservation only when the spin-triangle inequality $s_1+s_2\ge s_3,s_2+s_3\ge s_1,s_3+s_1\ge s_2$ is satisfied. Using \eqref{cFT3EuclidEPT} in \eqref{CFT3Euclidthreepoint} and choosing a different basis we can re-write it as,
\begin{align}\label{CFT3Euclidthreepointhandnh}
    \langle J_{s_1}J_{s_2}J_{s_3}\rangle=c_{s_1s_2s_3}^{(h)}\langle J_{s_1}J_{s_2}J_{s_3}\rangle_{h}+c_{s_1s_2s_3}^{(odd)}\langle\epsilon\cdot J_{s_1}J_{s_2}J_{s_3}\rangle_{h}+c_{s_1s_2s_3}^{(nh)}\langle J_{s_1}J_{s_2}J_{s_3}\rangle_{nh},
\end{align}
where we have defined the homogeneous and non-homogeneous correlators \cite{Jain:2021gwa},
\begin{align}
    &\langle J_{s_1}J_{s_2}J_{s_3}\rangle_h=\langle J_{s_1}J_{s_2}J_{s_3}\rangle_{F}-\langle J_{s_1}J_{s_2}J_{s_3}\rangle_{B},\notag\\
    &\langle J_{s_1}J_{s_2}J_{s_3}\rangle_{nh}=\langle J_{s_1}J_{s_2}J_{s_3}\rangle_{F}+\langle J_{s_1}J_{s_2}J_{s_3}\rangle_{B},\notag\\&\text{where}\notag\\
    &c_{s_1s_2s_3}^{(h)}=\frac{1}{2}(c_{s_1s_2s_3}^{(F)}-c_{s_1s_2s_3}^{(B)}), c_{s_1s_2s_3}^{(nh)}=\frac{1}{2}(c_{s_1s_2s_3}^{(F)}+c_{s_1s_2s_3}^{(B)}).
\end{align}
It turns out that the representation \eqref{CFT3Euclidthreepointhandnh} in contrast to \eqref{CFT3Euclidthreepoint} is the most natural from multiple perspectives. One obvious advantage is that it makes manifest the form of the odd structure viz \eqref{cFT3EuclidEPT}. Further, the $h$ and $nh$ correlators can be distinguished by their Ward-Takahashi identities. They satisfy,
\begin{align}\label{hnhWTEuclid}
    &\langle p_1\cdot J_{s_1}J_{s_2}J_{s_3}\rangle_h=0\notag\\
    &\langle p_1\cdot J_{s_1}J_{s_2}J_{s_3}\rangle_{nh}=\text{Ward-Takahashi identity}.
\end{align}
In words, the homogeneous correlator has a trivial (zero) Ward-Takahshi identity whereas the non-homogeneous correlator saturates the Ward-Takahashi identity of the correlator. 
This statement can also be written in spinor helicity variables as,
\begin{align}\label{KactionSHEuclid}
    &K^\mu\langle J_{s_1}J_{s_2}J_{s_3}\rangle_h=0\notag\\
    &K^\mu\langle J_{s_1}J_{s_2}J_{s_3}\rangle_{nh}\propto \text{Ward-Takahashi identity}.
\end{align}

Another nice fact about these correlators is that from the AdS bulk perspective, the homogeneous correlators arise from higher order (non-minimal) interactions while the non-homogeneous correlators emerge from the leading order interactions.

The fact \eqref{hnhWTEuclid} in particular makes the homogeneous correlator relatively easy to bootstrap for all spins. In contrast, the non-homogeneous correlator is much more complicated and harder to solve for. Also, when the spin-triangle inequality is violated, we actually get no homogeneous solution but rather two non-homogeneous solutions which are quite complicated \cite{Jain:2021whr}.

\subsection{Wightman functions}
As we shall elucidate, if we work in Lorentzian, rather than Euclidean signature and focus in particular on Wightman functions, the Ward-Takahashi identity becomes trivial, \begin{align}
    \langle 0| p_1\cdot J_{s_1}J_{s_2}J_{s_3}|0\rangle=0.
\end{align} 
This is in contrast to time-ordered correlators, see appendix \ref{app:Derivation of current conservation Ward-Takahshi identity} for a discussion on this. In spinor helicity variables, the analog of the action of the SCT \eqref{KactionSHEuclid} becomes for Wightman functions,
\begin{align}\label{KactionSHWightman}
    &K^\mu\langle 0|J_{s_1}J_{s_2}J_{s_3}|0\rangle_h=0\notag\\
    &K^\mu\langle 0| J_{s_1}J_{s_2}J_{s_3}|0\rangle_{nh}=0,
\end{align}
where we stick to the labels $h$ and $nh$ as these Wightman functions are the analytic continuations of their Euclidean counterparts as we shall discuss in much more detail in section \ref{sec:Wightman}.
As a result many of the complications associated to the Euclidean  non-homogeneous correlators discussed above disappear.
These Wightman functions as mentioned, are conformally invariant and identically conserved and hence are the correlators obtained from twistor space as discussed by \cite{Baumann:2024ttn}. With this in mind, let us set up and solve the conformal Ward identities in twistor space.

\section{Conformal Wightman functions from twistor space}\label{sec:Twistors}
Our aim in this section is to formulate and solve for (parity even) two and three point correlation functions of conserved currents in twistor space. We shall adopt an approach that is in the spirit of Witten's twistor space construction for  four dimensional scattering amplitudes \cite{Witten:2003nn}. We first discuss the necessity of working in Lorentzian space-time to define real twistors. We then perform half-Fourier transforms to go to momentum twistor space where we shall solve the conformal Ward identities, accompanied by the all important helicity counting identities for two- and three-point functions of  conserved currents. 

\subsection{Towards twistor space: The signature of space-time}
In this subsection, we take first steps in defining twistor space. We shall see that in order to define \textit{real} twistor space, one really needs to work in Lorentzian, rather than Euclidean signature. 
\subsection*{Problem with defining twistor space correlators starting from Euclidean space}
In principle, one would think that one can obtain a twistor space correlator starting from a 3d Euclidean CFT correlator in spinor helicity variables and performing a half Fourier transform on the $(\lambda,\Bar{\lambda})$ variables. However, the associated Euclidean reality conditions as discussed below, do not allow this. Let us understand this obstacle  more in detail. The main difficulty with usual spinor helicity variables is that the translation generator is multiplicative whereas the SCT is a second order differential operator as in \eqref{spinorhelicitygen}. It is very tempting to perform a half Fourier transform which will make the action of $K$ first order (and $P$ as well) hence simplifying the analysis. However, our hope is short lived as $\lambda$ and $\bar \lambda$ are Hermitian conjugates:
\begin{align}\label{EuclideanReality1}
     \lambda_{a}^\dagger=\Bar{\lambda}^a~,~\Bar{\lambda}_{}^{a\dagger}=\lambda_{a}.
 \end{align}
See  Appendix \ref{subsec:spinorhelicity} for a more detailed discussion. This implies we cannot Fourier transform of $\lambda$ or $\bar \lambda$ without affecting the other. 

This discussion makes it clear that because of the reality conditions, it is not straightforward to directly obtain a Twistor correlator starting from a Euclidean correlator\footnote{It would be interesting to reconcile this with the approach of Woodhouse \cite{Woodhouse:1985id}.}. One way to circumvent this issue is to complexify the spacetime which entails treating $\lambda$ and $\Bar{\lambda}$ as complex and independent and then defining the twistor transformations. Another simpler way, is to work in Lorentzian, rather than Euclidean signature where the associated reality conditions allow for the construction of a real twistor space.
\subsection*{How Lorentzian signature admits the construction of real twistor space}
  
The analogue of the spinor helicity reality condition \eqref{EuclideanReality1}  in Lorentzian signature is given by,
  \begin{align}\label{MinkowskiReality1}
    \lambda_{a}^*=\lambda_{a}~,~\Bar{\lambda}^{a*}=\Bar{\lambda}^a,
 \end{align}
where we refer the reader to appendix \ref{subsec:spinorhelicity} for more details. 
This implies for real Minkowskian momenta, the spinors $\lambda$ and $\Bar{\lambda}$ are real and independent. This can be understood as a fact inherited via a dimensional reduction of a null Kleinian momentum as we discuss in appendix \ref{app:4dto3d}. Therefore, one can now perform a half-Fourier transform with respect to say $\Bar{\lambda}$ while keeping $\lambda$ fixed to go to twistor space.

\subsection*{Which Lorentzian correlator?}
Now that we have established that Lorentzian signature is most well suited to develop twistor space\footnote{As we shall discuss in appendix \ref{subsec:spinorhelicity}, it will become clear that three dimensional Lorentzian space is what lends itself to allow for the construction of  \textit{real twistor space}, $\mathbb{RP}^3$ where this twistor transform is a simple 
Fourier transform.  Alternatively, one can work with complex momenta from the get-go and construct twistors in complex, rather than real projective space, which is $\mathbb{CP}^3$ and then look for a subspace thereof to obtain the associated Euclidean/ Minkowskian  correlator. In this paper, we follow the first route, that is, we construct the real twistor space $\mathbb{RP}^3$, starting with the Lorentz signature momenta and then solve for the correlators.}, one can ask the question about what Lorentzian correlators are natural from this perspective. To elaborate, the fact is that in contrast to Euclidean signature, there exist many different types of Lorentzian correlators such as time-ordered correlators, anti-time-ordered correlators, retarded correlators and much more. The answer, it turns out is that the Wightman functions are most suitable. This is due to the following fact. The action of the special conformal generator $K^\mu$ on a conserved current in spinor helicity variables \eqref{spinorhelicitygen} inside a correlator leads to (schematically and assuming $s_1$ is the largest spin in the correlator),
\begin{align}
    &\langle [K_{ab},\frac{J_{s_1}^{h_1}(p_1)J_{s_2}^{h_2}(p_2)J_{s_3}^{h_3}(p_3)}{p_1^{s_1-1}p_2^{s_2-1}p_3^{s_3-1}}]\rangle = \langle [2\frac{\partial^{2}}{\partial \lambda^{a} \partial\bar{\lambda}^{b}},\frac{J_{s_1}^{h_1}(p_1)J_{s_2}^{h_2}(p_2)J_{s_3}^{h_3}(p_3)}{p_1^{s_1-1}p_2^{s_2-1}p_3^{s_3-1}}]\rangle \notag \\
    =&\frac{1}{p_1^{2}}\zeta_{a}^{h_1}\zeta_{b}^{h_1}\zeta_{a_1}^{h_1}\cdots \zeta^{h_1}_{a_{s_1-3}}\zeta_{a_{s_1-2}}^{h_1}p_{1a_{s_1-1} a_{s_1}}\frac{\langle J_{s_1}^{a_1\cdots a_{s_1}}(p_1)J_{s_2}^{h_2}(p_2)J_{s_3}^{h_3}(p_3)\rangle}{p_1^{s_1-1}p_2^{s_2-1}p_3^{s_3-1}}.
\end{align}
Note that the last line is the Ward-Takahashi identity. While the half-Fourier transform would make SCT generator a first order differential operator, the presence of $p_1=-\frac{\langle \lambda_1\Bar{\lambda}_1\rangle}{2}$ (see \eqref{mommag}) in the denominator would make it an integral operation in the last line. However, if the correlator has a zero Ward-Takahashi identity, the RHS is zero and then one need not worry about solving a complicated integral equation. The Lorentzian correlator that precisely has a zero Ward Takahashi identity is the Wightman function, see appendix \ref{app:Derivation of current conservation Ward-Takahshi identity} for example.

 In short, the reason is that our twistor space construction yields identically conserved correlators (no Ward-Takahashi identitites) is due to the fact that we are dealing with Wightman functions. From the perspective of the Penrose transformation taken by \cite{Baumann:2024ttn}, the twistor space correlators are conserved from the word-go and hence must correspond to Wightman functions rather than any other type of correlator. At this point one possibility to obtain the twistor space correlator is to directly perform a half Fourier transform of the Minkowskian 
 Wightman function in spinor helicity variables. However it turns out it is much more easier and convenient to obtain twistor space correlators by directly solving conformal ward identities written in half Fourier transformed variables.


\subsection{Twistor variables}

The idea of the momentum twistor transform (the \textit{Witten} transform \cite{Witten:2003nn}) is to take the pair $(\lambda,\Bar{\lambda})$ and perform a \textit{half-Fourier} transform, that is, a Fourier transform with respect to either $\lambda$ or $\Bar{\lambda}$ \cite{Witten:2003nn}. The two options are,
\begin{align}\label{twotwistoroptions}
  \text{Twistor}: \qquad &(\lambda,\Bar{\lambda})\to (\lambda,\Bar{\mu}),\notag\\
  \text{Dual-Twistor}: \qquad &(\lambda,\Bar{\lambda})\to (\mu,\Bar{\lambda}).
\end{align}
Note that for this to make sense, we must work in Lorentzian signature as only there are the two spinors real and independent as we discussed in \eqref{MinkowskiReality1}. The first of these transformations is called the twistor transform whereas the latter is called the dual twistor transform.

Consider the rescaled currents $\frac{J_s^{\pm}(\lambda,\Bar{\lambda})}{p^{s-1}}$. We have two options viz \eqref{twotwistoroptions} for each current in performing the twistor transform. They are,
\begin{align}\label{TwistorTrans1}
    &\bigg\{\hat{J}_s^{+}(\lambda,\Bar{\mu})=\int \frac{d^2\Bar{\lambda}}{(2\pi)^2}e^{i\Bar{\lambda}\cdot \Bar{\mu}}\frac{J_s^{+}(\lambda,\Bar{\lambda})}{p^{s-1}}~\textbf{or}~\tilde{J}_s^{+}(\mu,\Bar{\lambda})=\int \frac{d^2\lambda}{(2\pi)^2}e^{-i\lambda\cdot \mu}\frac{J_s^{+}(\lambda,\Bar{\lambda})}{p^{s-1}}\bigg\}\notag\\&\bigg\{\hat{J}_s^{-}(\lambda,\Bar{\mu})=\int\frac{d^2 \Bar{\lambda}}{(2\pi)^2}e^{i\Bar{\lambda}\cdot \Bar{\mu}}\frac{J_s^{-}(\lambda,\Bar{\lambda})}{p^{s-1}}~\textbf{or}~\tilde{J}_s^{-}(\mu,\Bar{\lambda})=\int\frac{d^2 \lambda}{(2\pi)^2}e^{-i\lambda\cdot \mu}\frac{J_s^{-}(\lambda,\Bar{\lambda})}{p^{s-1}}\bigg\}.
\end{align}

The interpretation of each of these transformations will become apparent when we discuss correlation functions of these currents.
\subsection*{The action of the symmetry generators in the twistor representation}
The analog of \eqref{spinorhelicitygen} for these twistor space currents can be obtained by appropriately half-Fourier transforming the equations. The action of the conformal generators on the $\Bar{\lambda}$ transformed currents $\hat{J}_s^{\pm}(\lambda,\Bar{\mu})$ is as follows\footnote{The other generators can be found in appendix \ref{app:generators}.}:
\begin{align}\label{action1twistor}
    &P_{ab}=i\lambda_{(a}\frac{\partial}{\partial\Bar{\mu}^{b)}},\qquad K_{ab}=i\Bar{\mu}_{(a}\frac{\partial}{\partial \lambda^{b)}},
\end{align}
while the helicity generator acts as follows on currents in correlators:
\begin{align}\label{helicity1twistor}
    h_j\langle\cdots \hat{J}_{s_j}^{\pm}(\lambda_j,\Bar{\mu}_j)\cdots\rangle=-\frac{1}{2}\bigg(\lambda_j^a\frac{\partial}{\partial\lambda_j^a}+\Bar{\mu}_j^a\frac{\partial}{\partial\Bar{\mu}_j^a}+2\bigg)\langle\cdots \hat{J}_{s_j}^{\pm}(\lambda_j,\Bar{\mu}_j)\cdots\rangle=\pm s_j \langle\cdots \hat{J}_{s_j}^{\pm}(\lambda_j,\Bar{\mu}_j)\cdots\rangle.
\end{align}
\subsection*{The action of the symmetry generators in the dual twistor representation}
On the $\lambda$ transformed currents $J^{\pm}(\mu,\Bar{\lambda})$ they are,
\begin{align}\label{action2twistor}
    &P_{ab}=-i\Bar{\lambda}_{(a}\frac{\partial}{\partial\mu^{b)}},\qquad K_{ab}=-i\mu_{(a}\frac{\partial}{\partial \Bar{\lambda}^{b)}},
\end{align}
and
\begin{align}\label{helicity2twistor}
    h_j\langle \cdots \tilde{J}_{s_j}^{\pm}(\mu_j,\Bar{\lambda}_j)\cdots\rangle=\frac{1}{2}\bigg(\Bar{\lambda_j}^a\frac{\partial}{\partial\Bar{\lambda}_j^a}+\mu_j^a\frac{\partial}{\partial\mu_j^a}+2\bigg)\langle\cdots\tilde{J}_{s_j}^{\pm}(\lambda_j,\Bar{\mu}_j)\cdots\rangle=\pm s_j \langle \cdots \tilde{J}_{s_j}^{\pm}(\mu_j,\Bar{\lambda}_j)\cdots\rangle.
\end{align}
The important point is that in both cases \eqref{action1twistor} and \eqref{action2twistor}, both $P$ and $K$ are first order differential operators and on equal footing. Very importantly, note that all the generators are invariant under the $r\in\mathbb{R}$ scaling $(\lambda,\Bar{\mu})\to (\frac{\lambda}{r},\frac{\Bar{\mu}}{r})$ for the twistor representation and $(\mu,\Bar{\lambda})\to (r\mu,r\Bar{\lambda})$ for the dual twistor representation. This fact is inherited from the little group invariance of the Lorentzian momenta \eqref{Lorentz3momenta}. This indicates that these coordinates are projective real coordinates and counting the components (two real components for each spinor), we see that they naturally act on the space $\mathbb{RP}^3$.

\subsubsection{Setting up conformal Ward identity}
Our aim now is to solve for two and three point functions of these currents. We are working of course, in Minkowski signature and wish to obtain Wightman functions. The conformal Ward identities for the Wightman functions read,
\begin{align}\label{twistorconformalWardId}
    \sum_{j=1}^{n}\langle 0| \cdots [\mathcal{L},J_{s_j}^{\pm}]\cdots|0\rangle=0.
\end{align}
where $\mathcal{L}\in\{P_{ab},K_{ab},M_{ab},D\}$. Depending on which representation of the currents we take among the ones in \eqref{TwistorTrans1}, the generators $\mathcal{L}$ act as in \eqref{action1twistor} or \eqref{action2twistor}. We also need to supplement \eqref{twistorconformalWardId} with the helicity counting identities \eqref{helicity1twistor}, \eqref{helicity2twistor}. We shall present a few examples where we solve for two and three point correlators in these variables. 
\subsubsection{Solving ward identities in twistor space}
Let us first discuss the form of the solutions to the conformal Ward identities \eqref{twistorconformalWardId} for general $n-$ point functions. The action of momentum ($P_{ab}$) and SCT ($K_{ab}$) are as follows:
\begin{align}\label{nptPandKtwistor}
    P_{ab}:\qquad &\bigg(\sum_{i=1}^{n}\lambda_{i(a}\frac{\partial}{\partial\Bar{\mu}_i^{b)}}\bigg)\langle 0|\hat{J}_{s_1}^{+}(\lambda_1,\Bar{\mu}_1)\cdots \hat{J}_{s_n}^{+}(\lambda_n,\Bar{\mu}_n)|0\rangle=0,\notag\\
    K_{ab}:\qquad &\bigg(\sum_{i=1}^{n}\Bar{\mu}_{i(a}\frac{\partial}{\partial\lambda_i^{b)}}\bigg)\langle 0|\hat{J}_{s_1}^{+}(\lambda_1,\Bar{\mu}_1)\cdots\hat{J}_{s_n}^{+}(\lambda_n,\Bar{\mu}_n)|0\rangle=0.
\end{align}
Lorentz invariance and scale invariance imply that the correlator is dimensionless and depends only on spinor dot products respectively. Putting these facts together with the \eqref{nptPandKtwistor} we see that,
\begin{align}\label{npointtwistorgensol}
    \langle 0|\hat{J}_{s_1}^{+}(\lambda_1,\Bar{\mu}_1)\cdots\hat{J}_{s_n}^{+}(\lambda_n,\Bar{\mu}_n)|0\rangle=F_n(\{\lambda_i\cdot \Bar{\mu}_j-\lambda_j\cdot\Bar{\mu}_i\})=F_n(x_1,x_2,\cdots, x_{\frac{n(n-1)}{2}}).
\end{align}
One observes that for $n-$ points function, there are $\frac{n(n-1)}{2}$ different dot products which we denote as the quantities $x_1,\cdots,x_{\frac{n(n-1)}{2}}$. However we have $n$ helicity identities and $\frac{n(n-1)}{2}$ variables which leaves behind a function of $\frac{n(n-3)}{2}$ unknowns which is precisely the number of conformal cross ratios at $n$ points. This suffices to fully fix two, three point functions but it is not the case for any higher points. The strategy is to translate these identities into equations for the $x_i$. We shall work this out in detail for two and three point functions.
In those cases, we find the helicity identities take the form of the following ODE,
\begin{align}\label{EulerEQODE}
    x_i\frac{df_i}{dx_i}+n_i f_i=0,~n_i\in \mathbb{Z}.
\end{align}
This equation is of Euler type and has two classes of solutions, see appendix \ref{app:distrubution} for more details. The standard solution that for instance, \textsc{\textsl{Mathematica}} outputs is,
\begin{align}\label{nptsol1}
    f_i(x)=\frac{1}{x_i^{n_i}}.
\end{align}
However there are solutions that are not functions, but are rather \textit{distributions}\cite{kanwal1998generalized}. For \eqref{EulerEQODE}, these come in two flavours. The first is when $n_i\ge 1$. Then $\exists$ a solution,
\begin{align}\label{nptsol2}
f_i(x)=\frac{d^{n_i-1}}{dx^{n_i-1}}\delta(x)=\delta^{[n_i-1]}(x),~n_i\ge 1.
\end{align}
For the cases $n_i\le 0$, there exists another type of solution called the weak solution viz,
\begin{align}\label{nptsol3}
    f_i(x)=\frac{\theta(x_i)}{x_i^{n_i}(-n_i)!},~n_i\le 0.
\end{align}
The Euler ODE \eqref{EulerEQODE} has a parity symmetry ($x_i\to -x_i$) so we can choose solutions that are of a definite parity. Thus we can add a multiple of the regular solution \eqref{nptsol1} to \eqref{nptsol3} to obtain,
\begin{align}\label{nptsol3a}
    f_i(x)=\frac{\text{sgn}(x)}{2x_i^{n_i}(-n_i)!}~,n_i\le 0.
\end{align}
The interesting point about this linear combination is that \eqref{nptsol3a} is the extension of \eqref{nptsol2} to negative $n_i$. Thus we define the generalized distribution,
\begin{align}\label{deltadistgen}
    &\delta^{[n]}(x)=\frac{d^n}{d x^n}\delta(x), n\ge 0,\notag\\
    &\delta^{[n]}(x)=\frac{\text{sgn}(x)}{2(-n-1)!x^{n+1}}, n<0
\end{align}

Therefore the general solution is a linear combination of these solutions
\begin{align}
    f_i(x)=\frac{a_{i}}{x_i^{n_i}}+b_i\delta^{[n_i-1]}(x).
\end{align}
We shall now solve \eqref{twistorconformalWardId} for two and three point functions in twistor space which will illustrate the general formalism outlined above in detail. Although we do not proceed to higher points in this paper, we hope to return to that problem in the near future.

\subsection*{Two point functions}
Consider the $(++)$ helicity two point function. We choose a twistor for the positive helicity current. Conformal invariance \eqref{npointtwistorgensol} implies that the correlator takes the form,
\begin{align}
    \langle 0|\hat{J}_{s_1}^{+}(\lambda_1,\Bar{\mu}_1)\hat{J}_{s_2}^{+}(\lambda_2,\Bar{\mu}_2)|0\rangle=F(\lambda_1\cdot \Bar{\mu}_2-\lambda_2\cdot\Bar{\mu}_1).
\end{align}
The helicity identity \eqref{helicity1twistor} then leads to the following two differential equations for the function $F$:
\begin{align}\label{e1q}
     x\frac{d F(x)}{d x}=-2(s_1+1) F(x)~,~x\frac{d F(x)}{d x}=-2(s_2+1) F(x),
\end{align}
where $x=\lambda_1\cdot \Bar{\mu}_2-\lambda_2\cdot \Bar{\mu}_1$. \eqref{e1q} immediately entails that $s_1=s_2=s$. Thus, we need to solve,
\begin{align}\label{e1cont}
    x \frac{d F(x)}{dx}=-2(s+1)F(x).
\end{align}
The standard solution to \eqref{e1cont} is of the form \eqref{nptsol1} and is given by,
\begin{align}\label{e1function}
    F(x)=\frac{c_s}{x^{2s+2}}.
\end{align}
However, there also exist solutions that are not functions but \textit{distributions} as we saw in \eqref{nptsol2}. In this case,
\begin{align}\label{e1distribution}
    F(x)=c_s'\delta^{[2s+1]}(x),
\end{align}
is also a solution to \eqref{e1cont} valid for $2s+1\ge 0\equiv s\ge -\frac{1}{2}$ which is anyway true since spin is a positive (half-) integer, see appendix \ref{app:distrubution} for more details. Thus the third solution \eqref{nptsol3a} does not occur for two point functions. We will see however, that it plays an important role for three point functions shortly. Therefore, the most general solution to \eqref{e1} is a sum of \eqref{e1function} and \eqref{e1distribution}:
\begin{align}\label{gensol2ptpp}
    F(\lambda_1,\Bar{\mu}_1;\lambda_2,\Bar{\mu}_2)=\frac{c_s}{(\lambda_1\cdot \Bar{\mu}_2-\lambda_2\cdot \Bar{\mu}_1)^{2s+2}}+c_s' \delta^{[2s+1]}(\lambda_1\cdot \Bar{\mu}_2-\lambda_2\cdot \Bar{\mu}_1).
\end{align}
\subsection*{Comparing with spinor helicity expressions}
Let us discuss the meaning of each of these solutions in detail. To this end, let us perform a inverse half-Fourier transform to \eqref{gensol2ptpp} in order to go back to spinor helicity variables. Based on the definition of the twistor transformation \eqref{TwistorTrans1}, we see that,
\begin{align}
    F(\lambda_1,\Bar{\lambda}_1;\lambda_2,\Bar{\lambda}_2)=\int d^2 \Bar{\mu}_1~d^2\Bar{\mu}_2~e^{-i\Bar{\lambda}_1\cdot \Bar{\mu}_1-i\Bar{\lambda}_2\cdot \Bar{\mu}_2}~F(\lambda_1,\Bar{\mu}_1;\lambda_2,\Bar{\mu}_2).
\end{align}
Plugging in \eqref{gensol2ptpp}, the result is (see appendix \ref{app:HalfFourierTransform} for the details),
\begin{align}\label{2ptpp2sols}
    F(\lambda_1,\Bar{\lambda}_1;\lambda_2,\Bar{\lambda}_2)=c_s\frac{\langle 1 2\rangle^{2s}}{p_1}+c_s'\text{sgn}(\langle 1 2\rangle)\frac{\langle 1 2\rangle^{2s}}{p_1}.
\end{align}
The second term is problematic as it switches sign depending on the external value of the momenta which is unlike the usual CFT correlators. Another way to understand it is that it is odd under $CPT$ \cite{Baumann:2024ttn} which is not characteristic of the kind of theories we are interested in, which is easy to see given the spinor helicity $CPT$ transformation given in appendix \ref{app:CPTinv}.
Therefore, the first solution is the consistent one which entails $c_s'=0$.

Thus, the two point function is given by,
\begin{align}\label{e1}
     \langle 0|\hat{J}_{s_1}^{+}(\lambda_1,\Bar{\mu}_1)\hat{J}_{s_2}^{+}(\lambda_2,\Bar{\mu}_2)|0\rangle=\frac{c_s}{(\lambda_1\cdot \Bar{\mu}_2-\lambda_2\cdot \Bar{\mu}_1)^{2s+2}}.
\end{align}
 A similar analysis can also be obtained for the $(--)$ helicity two point function which yields,
 \begin{align}
      \langle 0|\hat{J}_{s_1}^{-}(\lambda_1,\Bar{\mu}_1)\hat{J}_{s_2}^{-}(\lambda_2,\Bar{\mu}_2)|0\rangle=\frac{c_s}{(\lambda_1\cdot \Bar{\mu}_2-\lambda_2\cdot \Bar{\mu}_1)^{-2s+2}}.
 \end{align}
\subsection*{Three point functions}
Let us consider the $(+++)$ helicity three point function in the twistor representation. Conformal invariance \eqref{npointtwistorgensol} implies,
\begin{align}
    \langle 0|\hat{J}_{s_1}^+(\lambda_1,\Bar{\mu}_1)\hat{J}_{s_2}^+(\lambda_2,\Bar{\mu}_2)\hat{J}_{s_3}^+(\lambda_3,\Bar{\mu}_3)|0\rangle=F(\lambda_1\cdot \Bar{\mu}_2-\lambda_2\cdot\Bar{\mu}_1,\lambda_2\cdot \Bar{\mu}_3-\lambda_3\cdot\Bar{\mu}_2,\lambda_3\cdot \Bar{\mu}_1-\lambda_1\cdot\Bar{\mu}_3).
\end{align}
The helicity identities \eqref{helicity1twistor} can be recast into the following equations:
\begin{align}
    &x \frac{\partial F(x,y,z)}{\partial x}=-(s_1+s_2-s_3+1)F(x,y,z)~,~y \frac{\partial F(x,y,z)}{\partial y}=-(s_2+s_3-s_1+1)F(x,y,z),\notag\\
    &z \frac{\partial F(x,y,z)}{\partial z}=-(s_3+s_1-s_2+1)F(x,y,z),
\end{align}
where $x=\lambda_1\cdot \Bar{\mu}_2-\lambda_2\cdot \Bar{\mu}_1,y=\lambda_2\cdot \Bar{\mu}_3-\lambda_3\cdot\Bar{\mu}_2,z=\lambda_3\cdot \Bar{\mu}_1-\lambda_1\cdot \Bar{\mu}_3$. 
Let us solve the above equations using separation of variables,
\begin{align}
    F(x,y,z)=f_1(x)f_2(y)f_3(z).
\end{align}
We then obtain the following system of first order ODEs:
\begin{align}\label{3ptpppPDE1}
    x \frac{d f_1}{d x}=-(s_1+s_2-s_3+1)f_1, y\frac{d f_2}{d y}=-(s_2+s_3-s_1+1)f_1, z \frac{d f_3}{d z}=-(s_3+s_1-s_2+1)f_3.
\end{align}
Before proceeding any further, we must distinguish between two cases.
\begin{enumerate}
    \item Inside the triangle: $s_1+s_2>s_3,~ s_2+s_3>s_1,~ s_3+s_1>s_2$
    \item Outside the triangle: $s_1+s_2>s_3,~ \color{blue}{s_2+s_3<s_1}\textcolor{black}{,}~ \color{black}{s_3+s_1>s_2}$
\end{enumerate}
 Correlators which satisfy the first condition are said to be inside the triangle. In the second case, note the reversal in sign for the second inequality and in this case, the correlator is said to be outside the triangle\footnote{Without loss of generality, we choose $s_1$ to be the largest spin in the correlator.}. Let us focus on each of these separately.
\subsection*{Inside the triangle}
Coming back to the ODEs \eqref{3ptpppPDE1}, we focus on the first of these three. The standard solution is of the form \eqref{nptsol1},
\begin{align}\label{pppsol1}
    \frac{1}{x^{s_1+s_2-s_3+1}}.
\end{align}
However, just like the two point case, there exists a distrubutional solution like \eqref{nptsol2},
\begin{align}\label{pppsol2}
    \delta^{[s_1+s_2-s_3]}(x),
\end{align}
valid when $s_1+s_2-s_3\ge 0$
Thus the general solution takes the form,
\begin{align}\label{f1pppevensol}
    f_1(x)=\frac{a_1}{x^{s_1+s_2-s_3+1}}+b_1 \delta^{[s_1+s_2-s_3]}(x).
\end{align}
A similar analysis can be carried out for the functions $f_2(y),f_3(z)$ where we obtain analogous solutions with the conditions $s_2+s_3>s_1,s_3+s_1>s_2$ for the delta function solutions. The final result is thus given by,
\small
\begin{align}\label{pppFxyz}
    F(x,y,z)=\bigg(\frac{a_1}{x^{s_1+s_2-s_3+1}}+b_1 \delta^{[s_1+s_2-s_3]}(x)\bigg)\bigg(\frac{a_2}{y^{s_2+s_3-s_1+1}}+b_2 \delta^{[s_2+s_3-s_1]}(y)\bigg)\bigg(\frac{a_3}{x^{s_3+s_1-s_2+1}}+b_3 \delta^{[s_3+s_1-s_2]}(z)\bigg).
\end{align}
\normalsize
It is easy to see that the polynomial functions is not desirable as it leads to momentum dependent sign factors. Thus we set all the $a_i=0$. 
Thus, the correct solution is the product of three delta function solutions. Therefore, our result for the three point function inside the triangle reads,
\begin{align}\label{e2}
    &\langle 0|\hat{J}_{s_1}^+(\lambda_1,\Bar{\mu}_1)\hat{J}_{s_2}^+(\lambda_2,\Bar{\mu}_2)\hat{J}_{s_3}^+(\lambda_3,\Bar{\mu}_3)|0\rangle\notag\\&=\delta^{[s_1+s_2-s_3]}(\lambda_1\cdot \Bar{\mu}_2-\lambda_2\cdot\Bar{\mu}_1)\delta^{[-s_1+s_2+s_3]}(\lambda_2\cdot \Bar{\mu}_3-\lambda_3\cdot\Bar{\mu}_2)\delta^{[s_1-s_2+s_3]}(\lambda_3\cdot \Bar{\mu}_1-\lambda_1\cdot\Bar{\mu}_3).
\end{align}
\subsection*{Outside the triangle}
Returning to the ODEs \eqref{3ptpppPDE1}, we now consider the case when $s_1+s_2>s_3,s_1+s_3>s_2$ but $s_2+s_3<s_1$.  For $f_1(x)$ and $f_3(z)$, the solutions are the delta function ones since the equations are the same as the ones inside the triangle. However, the story is different for $f_2(y)$. We have,
\begin{align}\label{ODEforf2}
    y\frac{df_2}{dy}=-(s_2+s_3-s_1+1)f_2(y).
\end{align}
Although a polynomial solution like \eqref{pppsol1} goes through for $s_2+s_3<s_1$, the delta function solution \eqref{pppsol2} is not valid in this case since,
\begin{align}
    \delta^{[s_2+s_3-s_1]}(y),
\end{align}
would represent a negative number of derivatives acting on the delta function.
However, in contrast to the two point case where these exhausted the possible solutions, we have yet another solution to \eqref{ODEforf2} which is of the form we discussed earlier \eqref{nptsol3}:
\begin{align}
    \frac{\theta(y)}{y^{s_2+s_3-s_1+1}(-s_2-s_3+s_1-1)!}.
\end{align}
The most general solution thus is,
\begin{align}
    f_2(y)=\frac{a_2}{y^{s_2+s_3-s_1+1}}+\frac{b_2~\theta(y)}{y^{s_2+s_3-s_1+1}(-s_2-s_3+s_1-1)!}.
\end{align}
The following linear combination of these solutions has a definite parity as discussed in \eqref{nptsol3a}:
\begin{align}
    &\frac{1}{2(-s_2-s_3+s_1-1)!y^{s_2+s_3-s_1+1}}\big(2\theta(y)-1)\notag\\&=\frac{\text{sgn}(y)}{2(-s_2-s_3+s_1-1)!y^{s_2+s_3-s_1+1}}=\delta^{[s_2+s_3-s_1]}(y).
\end{align}
where we used \eqref{deltadistgen}. This quantity also gives rise to the correct spinor helicity correlator after a half-Fourier transform. Thus, outside the triangle with $s_1>s_2+s_3$, the result for the correlator is,
\begin{align}\label{e2out}
    &\langle 0|\hat{J}_{s_1}^+(\lambda_1,\Bar{\mu}_1)\hat{J}_{s_2}^+(\lambda_2,\Bar{\mu}_2)\hat{J}_{s_3}^+(\lambda_3,\Bar{\mu}_3)|0\rangle\notag\\&=\delta^{[s_1+s_2-s_3]}(\lambda_1\cdot \Bar{\mu}_2-\lambda_2\cdot\Bar{\mu}_1)\delta^{[-s_1+s_2+s_3]}(\lambda_2\cdot \Bar{\mu}_3-\lambda_3\cdot\Bar{\mu}_2)\delta^{[s_1-s_2+s_3]}(\lambda_3\cdot \Bar{\mu}_1-\lambda_1\cdot\Bar{\mu}_3),
\end{align}
with it being understood that in the middle delta function what we actually mean is,
\begin{align}\label{deltasignrelation}
    \delta^{[-s_1+s_2+s_3]}(\lambda_2\cdot \Bar{\mu}_3-\lambda_3\cdot\Bar{\mu}_2)=\frac{\text{sgn}(\lambda_2\cdot \Bar{\mu}_3-\lambda_3\cdot\Bar{\mu}_2)}{2(-s_2-s_3+s_1-1)!(\lambda_2\cdot \Bar{\mu}_3-\lambda_3\cdot\Bar{\mu}_2)^{s_2+s_3-s_1+1}}.
\end{align}
Henceforth, any correlator written in this notation will be understood with \eqref{deltasignrelation} implemented whenever the spin-triangle inequality is not obeyed.
\subsubsection{Solving ward identities in dual twistor space}
Performing a similar analysis in the dual twistor representation we get,\small
\begin{align}\label{e1a}
     &\langle 0|\tilde{J}_s^{+}(\mu_1,\Bar{\lambda}_1)\tilde{J}_s^{+}(\mu_2,\Bar{\lambda}_2)|0\rangle=\frac{c_s}{(\Bar{\lambda}_1\cdot \mu_2-\Bar{\lambda}_2\cdot \mu_1)^{-2s+2}},\langle 0|\tilde{J}_s^{-}(\mu_1,\Bar{\lambda}_1)\tilde{J}_s^{-}(\mu_2,\Bar{\lambda}_2)|0\rangle=\frac{c_s}{(\Bar{\lambda}_1\cdot \mu_2-\Bar{\lambda}_2\cdot \mu_1)^{2s+2}}.
\end{align}
\normalsize
For the $(+++)$ helicity three point function the result is,
\begin{align}\label{e2a}
    &\langle 0|\tilde{J}_{s_1}^+(\mu_1,\Bar{\lambda}_1)\tilde{J}_{s_2}^+(\mu_2,\Bar{\lambda}_2)\tilde{J}_{s_3}^+(\mu_3,\Bar{\lambda}_3)|0\rangle\notag\\&=\delta^{[-s_1-s_2+s_3]}(\Bar{\lambda}_1\cdot \mu_2-\Bar{\lambda}_2\cdot\mu_1)\delta^{[s_1-s_2-s_3]}(\Bar{\lambda}_2\cdot \mu_3-\Bar{\lambda}_3\cdot\mu_2)\delta^{[-s_1+s_2-s_3]}(\Bar{\lambda}_3\cdot \mu_1-\Bar{\lambda}_1\cdot\mu_3),
\end{align}
with it being understood in that a negative number of derivatives act on a delta function means a formula analogous to \eqref{deltadistgen}.
\subsubsection{Mixed correlators}
The $(++-)$ helicity configuration can be constructed by choosing a twistor representation for the first two operators and a dual twistor representation for the third. Performing analysis analogous to those above we get,
\begin{align}\label{mixedhelicitye1}
    &\langle 0|\hat{J}_{s_1}^+(\lambda_1,\Bar{\mu}_1)\hat{J}_{s_2}^+(\lambda_2,\Bar{\mu}_2)\tilde{J}_{s_3}^-(\mu_3,\Bar{\lambda}_3)|0\rangle\notag\\&=\delta^{[s_1+s_2-s_3]}(\lambda_1\cdot \Bar{\mu}_2-\lambda_2\cdot\Bar{\mu}_1)\delta^{[-s_1+s_2+s_3]}(\lambda_2\cdot \mu_3-\Bar{\mu}_2\cdot\Bar{\lambda}_3)\delta^{[s_1-s_2+s_3]}(\mu_3\cdot \lambda_1-\Bar{\lambda}_3\cdot \Bar{\mu}_1).
\end{align}
Choosing a dual twistor representation for the first two and a twistor representation for the third we get,
\begin{align}\label{mixedhelicitye2}
    &\langle 0|\tilde{J}_{s_1}^+(\mu_1,\Bar{\lambda}_1)\tilde{J}_{s_2}^+(\mu_2,\Bar{\lambda}_2)\hat{J}_{s_3}^+(\lambda_3,\Bar{\mu}_3)|0\rangle\notag\\&=\delta^{[-s_1-s_2+s_3]}(\Bar{\lambda}_1\cdot \mu_2-\Bar{\lambda}_2\cdot\mu_1)\delta^{[s_1-s_2-s_3]}(\mu_2\cdot \lambda_3-\Bar{\lambda}_2\cdot\Bar{\mu}_3)\delta^{[-s_1+s_2-s_3]}(\lambda_3\cdot \mu_1-\Bar{\mu}_3\cdot\Bar{\lambda}_1).
\end{align}

Continuing this way, one obtains similar results for the remaining helicity configurations. The important point to notice is that two and three point solutions obtained above can be expressed in terms of  particular combination of spinorial dot products which are given by,
\begin{align}\label{twistordotprods}
     &Z_i\cdot Z_j=\lambda_i\cdot \Bar{\mu}_j-\lambda_j\cdot\Bar{\mu}_i,\notag\\
    &Z_i\cdot W_j=\lambda_i\cdot \mu_j-\Bar{\mu}_i\cdot \Bar{\lambda}_j,\notag\\
    &W_i\cdot Z_j=\mu_i\cdot \lambda_j-\Bar{\lambda}_i\cdot\Bar{\mu}_j,\notag\\
    &W_i\cdot W_j=\Bar{\lambda}_i\cdot\mu_j-\Bar{\lambda}_j\cdot\mu_i.
\end{align}
We shall see now that these are the natural conformally invariant dot products that arise when contracting twistors $Z^A$ or dual-twistors $W_A$ that we shall define in the next section. For example, the $(+++)$ twistor representation correlator \eqref{e2} is given by,
\begin{align}
    \langle 0| \hat{J}_{s_1}^{+}(Z_1)\hat{J}_{s_2}^{+}(Z_2)\hat{J}_{s_3}^{+}(Z_3)|0\rangle=\delta^{[s_1+s_2-s_3]}(Z_1\cdot Z_2)\delta^{[s_2+s_3-s_1]}(Z_2\cdot Z_3)\delta^{[s_3+s_1-s_2]}(Z_3\cdot Z_1).
\end{align}
Motivated by this simplification, we now proceed to define variables that automatically enforce this.
\section{Manifest Twistor variables}\label{sec:Manifesttwistorsec}
Let us define the following four component spinors:
\begin{align}\label{twistors}
    Z_i^A=\lambda_i^a\oplus \Bar{\mu}_{ia'},\qquad W_{iA}=\mu_{ia}\oplus \Bar{\lambda}_i^a,
\end{align}
$Z_i$ are the twistor variables while $W_i$ are the dual twistor variables. Note that $a,a'$ are both fundamental indices of the \textit{same} $\mathfrak{sl}(2,\mathbb{R})$ rotation algebra.

The (dual-)twistor space currents \eqref{TwistorTrans1} can then be cast in their natural twistorial form:
\begin{align}\label{currentstwistorform}
    \hat{J}_s^{\pm}(\lambda,\Bar{\mu})=\hat{J}_s^{\pm}(Z)~,~\tilde{J}_s^{\pm}(\mu,\Bar{\lambda})=\tilde{J}_s^{\pm}(W).
\end{align}
The action of the conformal generators becomes extremely simple in this language. All ten conformal generators combine into to form a single generator $T_{AB}$: We have,
\begin{align}
    [T_{AB},\hat{J}_s^{\pm}(Z)]=Z_{(A}\frac{\partial}{\partial Z^{B)}}\hat{J}_s^{\pm}(Z),\qquad[T_{AB},\tilde{J}_s^{\pm}(W)]=W_{(A}\frac{\partial}{\partial W^{B)}}\tilde{J}_s^{\pm}(W).
\end{align}
Using these representations one can check that the algebra closes:
\begin{align}
    [T_{AB},T_{CD}]=\Omega_{AC}T_{BD}+\Omega_{AD}T_{BC}+\Omega_{BC}T_{AD}+\Omega_{BD}T_{AC}.
\end{align}
The individual conformal generators can be obtained from the various components of $T_{AB}$ as we show in appendix \ref{app:generators}.
The conformal Ward identity \eqref{twistorconformalWardId} simply becomes,
\begin{align}\label{manifesttwistorconformalWard}
    \langle 0|\cdots [T_{AB},J_s^{\pm}]\cdots|0\rangle=0.
\end{align}
The helicity counting identities \eqref{helicity1twistor}, \eqref{helicity2twistor} translate to,
\begin{align}\label{manifesttwistorhelicitycount}
    &h_j\langle\cdots \hat{J}_{s_j}^{\pm}(Z_j)\cdots\rangle=-\frac{1}{2}\big(Z_j^A\frac{\partial}{\partial Z_j^A}+2\big)\langle\cdots \hat{J}_{s_j}^{\pm}(Z_j)\cdots\rangle=\pm s_j\langle\cdots \hat{J}_{s_j}^{\pm}(Z_j)\cdots\rangle,\notag\\
    &h_j\langle\cdots \tilde{J}_{s_j}^{\pm}(W_j)\cdots\rangle=\frac{1}{2}\big(W_j^A\frac{\partial}{\partial W_j^A}+2\big)\langle\cdots \tilde{J}_{s_j}^{\pm}(W_j)\cdots\rangle=\pm s_j\langle\cdots \tilde{J}_{s_j}^{\pm}(W_j)\cdots\rangle.
\end{align}
The solutions to \eqref{manifesttwistorconformalWard} and \eqref{manifesttwistorhelicitycount} are simply invariants constructed out of the twistors $Z_i$ and $W_j$ defined in \eqref{twistors}. For two and three point functions, the invariants that we obtained earlier by working in the $(\lambda,\Bar{\mu})$ and $(\mu,\Bar{\lambda})$ variables (see \eqref{e1}, \eqref{e1a}, \eqref{e2},\eqref{e2a},\eqref{mixedhelicitye1}) are the ones in \eqref{twistordotprods}.
$Z_i^A$ and $W_{jA}$ can directly be contracted as the former has its index upstairs and the latter has its index downstairs. To contract $Z_i^A$ with $Z_j^B$ or $W_{iA}$ with $W_{jB}$, we see that it is the bi-twistor $\Omega_{AB}=\Omega^{AB}$ that does the job\footnote{Above, we have defined $Z_i\cdot Z_j=-Z_i^{A}\Omega_{AB}Z_j^{B}$ and $W_i\cdot W_j=W_{iA}\Omega^{AB}W_{jB}$.}. In matrix form it is given by,
\begin{align}\label{symplecticform}
    \Omega_{AB}=\begin{pmatrix}
        0&\delta_a^{b'}\\
        -\delta^b_{a'} & 0
    \end{pmatrix}.
\end{align}
In a more group theoretic parlance, $Z^A,W_A$ belong to the fundamental and anti-fundamental representations of the group $Sp(4)$ (which is isomorphic to the double cover of the three dimensional conformal group) and $\Omega_{AB}$ is the symplectic form.

Recall that the above construction is in the context of three dimensional Lorentzian space-time where $\lambda$ and $\Bar{\lambda}$ are real and independent \eqref{MinkowskiReality} and thus $Z$ and $W$ are real. However, one can take the point of view of working in a complexified setting where $\lambda$ and $\Bar{\lambda}$ are viewed as independent complex quantities with the twistor transform \eqref{TwistorTrans1} being modified to be over an appropriate contour in the complex plane. In this paper, for our purposes it suffices to work with the real twistor variables. In appendix \ref{app:3dTwistorsApproach}, we discuss some of these aspects of three dimensional twistor space in more detail.

With all of this in mind, let us now systematically reformulate the twistor space bootstrap in these manifest variables.

\subsection{Solving Ward identities in Twistor Space}
We first focus on the general solutions to the differential equations that arise due to conformal invariance in twistor space.
\subsubsection{Two point correlators}
Lets first consider the currents to have positive helicity. Working in the twistor representation, the conformal Ward identities \eqref{manifesttwistorconformalWard} read,
\begin{align}
    \big(Z_{1(A}\frac{\partial}{\partial Z_1^{B)}}+Z_{2(A}\frac{\partial}{\partial Z_2^{B)}}\big)\langle 0|\hat{J}_{s_1}^{+}(Z_1)\hat{J}_{s_2}^{+}(Z_2)|0\rangle=0.
\end{align}
The solution to the above differential equation is simply,
\begin{align}
    \langle 0|\hat{J}_{s_1}^{+}(Z_1)\hat{J}_{s_2}^{+}(Z_2)|0\rangle=F(Z_1\cdot Z_2).
\end{align}
The helicity identity \eqref{manifesttwistorhelicitycount} imposes on the function the following two equations:
\begin{align}
    Z_{1}^A\frac{\partial}{\partial Z_1^A}F(Z_1\cdot Z_2)=-2(s_1+1)F(Z_1\cdot Z_2)~,~ Z_{2}^A\frac{\partial}{\partial Z_2^A}F(Z_1\cdot Z_2)=-2(s_2+1)F(Z_1\cdot Z_2).
\end{align}
As we discussed when working component-wise \eqref{gensol2ptpp}, there exist two solutions (which both demand $s_1=s_2$), one polynomial and another of a distributional nature:
\begin{align}\label{twopointtwistor2sols}
    \langle 0|\hat{J}_s^{+}(Z_1)\hat{J}_s^{+}(Z_2)|0\rangle\supset \bigg\{\frac{1}{(Z_1\cdot Z_2)^{2(s+1)}},\delta^{[2s+1]}(Z_1\cdot Z_2)\bigg\}.
\end{align}
As we argued earlier, the former is the correct solution. Thus, the result is,
\begin{align}
     \langle 0|\hat{J}_s^{+}(Z_1)\hat{J}_s^{+}(Z_2)|0\rangle\propto \frac{1}{(Z_1\cdot Z_2)^{2(s+1)}}.
\end{align}
Similarly, the $(--)$ helicity two point function in twistor space is given by,
\begin{align}\label{ppZ1Z2}
    \langle 0|\hat{J}_s^{-}(Z_1)\hat{J}_s^{-}(Z_2)|0\rangle\propto \frac{1}{(Z_1\cdot Z_2)^{2(-s+1)}}
\end{align}

\subsubsection{Three point correlators}

Just as we did at two points, let us solve the conformal Ward identities \eqref{manifesttwistorconformalWard} and the helicity identities \eqref{manifesttwistorhelicitycount} to determine the functional form of three point functions. Let us focus on one helicity configuration in detail: $(+++)$. Conformal invariance for the $(+++)$ correlator with all currents either in the $Z$ representation demands,
\begin{align}
    &\big(Z_{1(A}\frac{\partial}{\partial Z_1^{B)}}+Z_{2(A}\frac{\partial}{\partial Z_2^{B)}}+Z_{3(A}\frac{\partial}{\partial Z_3^{B)}}\big)\langle 0|\hat{J}_{s_1}^{+}(Z_1)\hat{J}_{s_2}^{+}(Z_2)\hat{J}_{s_3}^{+}(Z_3)|0\rangle_h=0.
\end{align}
The reason for the subscript $h$ (homogeneous) will become clear after seeing what Wightman function these choices correspond to after a half-Fourier transform\footnote{This label is for correlators that are inside the triangle. When the spin-triangle inequality is not satisfied, $h$ should be replaced by $F-B$.}.
The solutions to these identities are,
\begin{align}
    &\langle 0| \hat{J}^{+}_{s_1}(Z_1)\hat{J}^{+}_{s_2}(Z_2)\hat{J}^{+}_{s_3}(Z_3)|0\rangle_{h}=F(Z_1\cdot Z_2,Z_2\cdot Z_3,Z_3\cdot Z_1).
\end{align}
The helicity identities \eqref{manifesttwistorhelicitycount},
\begin{align}
    Z_{1}^A\frac{\partial}{\partial Z_1^A}F=-2(s_1+1)F~,~ Z_{2}^A\frac{\partial}{\partial Z_2^A}F=-2(s_2+1)F~,~Z_{3}^A\frac{\partial}{\partial Z_3^A}F=-2(s_3+1)F,
\end{align}
yield two solutions: polynomials and delta functions \eqref{f1pppevensol} (but keeping in mind the analytic continuation of the delta functions to cases with a negative number of derivatives acting on it as is appropriate outside the spin-triangle). The former give rise to the correct spinor helicity correlator and hence are the correct solutions. Thus the solutions we obtain are,
\begin{align}\label{ppptwistor}
    &\langle 0|\hat{J}^{+}_{s_1}(Z_1)\hat{J}^{+}_{s_2}(Z_2)\hat{J}^{+}_{s_3}(Z_3)|0\rangle_{h}\propto \delta^{[s_1+s_2-s_3]}(Z_1\cdot Z_2)\delta^{[s_2+s_3-s_1]}(Z_2\cdot Z_3)\delta^{[s_3+s_1-s_2]}(Z_3\cdot Z_1),
\end{align}
for any (half) integer spins $s_1,s_2$ and $s_3$\footnote{With of course, the restriction that $s_1+s_2+s_3$ is integer.}. Again, the important point to note is that for $n\in \mathbb{Z}$ the formula \eqref{deltadistgen}.
\subsection*{Examples}
For instance consider the correlator $\langle 0| TJJ|0\rangle_h$, that is $s_1=2,s_2=s_3=1$ which is inside the triangle. Using  \eqref{ppptwistor} and focusing on the twistor representation for now, we get,
\begin{align}
    \langle 0|T^{+}(Z_1)J^{+}(Z_2)J^{+}(Z_3)|0\rangle_h=\delta^{[2]}(Z_1\cdot Z_2)\delta(Z_2\cdot Z_3)\delta^{[2]}(Z_3\cdot Z_1).
\end{align}
On the other hand consider $\langle 0|J_4^{+} J^{+}J^{+}|0\rangle_{F-B}$, that is, $s_1=4,s_2=s_3=1$ which is outside the triangle. We have,
\begin{align}
    \langle 0|J_4(Z_1)J(Z_2)J(Z_3)|0\rangle_{F-B}&=\delta^{[4]}(Z_1\cdot Z_2) \frac{\text{Sgn}(Z_2\cdot Z_3)}{(Z_2\cdot Z_3)^3}\delta^{[4]}(Z_3\cdot Z_1)\notag\\
    &\equiv \delta^{[4]}(Z_1\cdot Z_2)\delta^{[-2]}(Z_2\cdot Z_3)\delta^{[4]}(Z_3\cdot Z_1).
\end{align}
\subsection{Solving Ward identities in dual Twistor space}
Let us now perform the analogous analysis in dual twistor variables.
\subsubsection{Two point functions}
One can also solve for the two point function in the dual twistor representation. The result after solving Ward identities and the helicity identity is,
\begin{align}\label{twistorW2point}
     \langle 0|\tilde{J}_s^{+}(W_1)\tilde{J}_s^{+}(W_2)|0\rangle\propto\frac{1}{(W_1\cdot W_2)^{2(-s+1)}},
\end{align}
which of course, matches with the component-wise analysis \eqref{e1a}.
Similarly, one can compute the minus helicity two point function. The results in the dual twistor representations is given by,
\begin{align}\label{mmWW}
    &\langle 0|\tilde{J}_s^{-}(W_1)\tilde{J}_s^{-}(W_2)|0\rangle\propto\frac{1}{(W_1\cdot W_2)^{2(s+1)}}.
\end{align}
Although it seems like we have two answers for each helicity configuration (one twistor and one dual twistor), it turns out that they describe the same spinor helicity expression after a half-Fourier transform as we show in appendix \ref{app:HalfFourierTransform}\footnote{One can actually construct a mixed helicity object but after half-Fourier transform, it turns out that it has support only at zero momentum and hence we do not consider it. The details can be found in appendix \ref{app:HalfFourierTransform}.}.

\subsubsection{Three point functions}
The three point ward identities in dual-twistor space is given by,
\begin{align}
    \big(W_{1(A}\frac{\partial}{\partial W_1^{B)}}+W_{2(A}\frac{\partial}{\partial W_2^{B)}}+W_{3(A}\frac{\partial}{\partial W_3^{B)}}\big)\langle 0| \tilde{J}^{+}_{s_1}(W_1)\tilde{J}^{+}_{s_2}(W_2)\tilde{J}^{+}_{s_3}(W_3)|0\rangle_{nh}=0.
\end{align}
The solution is,
\begin{align}
    \langle 0| \tilde{J}^{+}_{s_1}(W_1)\tilde{J}^{+}_{s_2}(W_2)\tilde{J}^{+}_{s_3}(W_3)|0\rangle_{nh}=G(W_1\cdot W_2,W_2\cdot W_3,W_3\cdot W_1).
\end{align}
Finally, solving the helicity identities yields,
\begin{align}
    \langle 0|\tilde{J}^{+}_{s_1}(W_1)\tilde{J}^{+}_{s_2}(W_2)\tilde{J}^{+}_{s_3}(W_3)|0\rangle_{nh}\propto \delta^{[-s_1-s_2+s_3]}(W_1\cdot W_2)\delta^{[-s_2-s_3+s_1]}(W_2\cdot W_3)\delta^{[-s_3-s_1+s_2]}(W_3\cdot W_1),
\end{align}
which matches with the component-wise analysis \eqref{e2a}.
\subsection{Three point result in all helicities}
Proceeding to obtain the results in the other helicities where we employ twistors and dual-twistors, we obtain the following general result,
\begin{align}\label{twistorspace3points}
      &\langle 0| \hat{J}_{s_1}^{h_1}(T_1)\hat{J}_{s_2}^{h_2}(T_2)\hat{J}_{s_3}^{h_3}(T_3)|0\rangle_{h}\propto \delta^{[s_1+s_2-s_3]}(T_1\cdot T_2)\delta^{[s_2+s_3-s_1]}(T_2\cdot T_3)\delta^{[s_3+s_1-s_2]}(T_3\cdot T_1),\notag\\
      &\langle 0| \tilde{J}_{s_1}^{h_1}(U_1)\tilde{J}_{s_2}^{h_2}(U_2)\tilde{J}_{s_3}^{h_3}(U_3)|0\rangle_{nh}\propto \delta^{[-s_1-s_2+s_3]}(U_1\cdot U_2)\delta^{[-s_2-s_3+s_1]}(U_2\cdot U_3)\delta^{[-s_3-s_1+s_2]}(U_3\cdot U_1),
\end{align}
where,
\begin{align}\label{TUnotation}
    &T_i=Z_i~\text{if}~h_i=+1 ~\text{and}~ T_i=W_i ~\text{if}~ h_i=-1,\notag\\
    &U_i=W_i~\text{if}~h_i=+1 ~\text{and}~ U_i=Z_i ~\text{if}~ h_i=-1.
\end{align}
Let us now discuss the meaning of the proportionality constants in all these equations. As we shall see, they are responsible for differentiating between the even and odd correlators. The discussion of the parity odd correlators can be found in section \ref{sec:parityodd}. In this section, we focus on the even case.

\subsubsection{Parity even}
Here, we take the coefficient of the two point functions \eqref{ppZ1Z2} and \eqref{mmWW} to be real:
\begin{align}\label{even2ptallhel}
    &\langle 0|\hat{J}_s^{+}(Z_1)\hat{J}_s^{+}(Z_2)|0\rangle= \frac{c_{s,\text{even}}}{(Z_1\cdot Z_2)^{2(s+1)}}~,~\langle 0|\tilde{J}_s^{-}(W_1)\tilde{J}_s^{-}(W_2)|0\rangle=\frac{c_{s,\text{even}}}{(W_1\cdot W_2)^{2(s+1)}}.
\end{align}
This choice picks out the parity and time-reversal even two point function. This can be seen by performing the half-Fourier transform that yields,
\begin{align}
    \langle 0|J_s^{+}(p_1)J_s^{+}(p_2)|0\rangle=(2\pi)^3\delta^3(p_1+p_2)c_{s,\text{even}}\frac{\langle 1 2\rangle^{2s}}{p_1}~,~\langle 0|J_s^{-}(p_1)J_s^{-}(p_2)|0\rangle=(2\pi)^3\delta^3(p_1+p_2)c_{s,\text{even}}\frac{\langle \Bar{1} \Bar{2}\rangle^{2s}}{p_1},
\end{align}
which are indeed the two components of the momentum space parity and time-reversal even Wightman function \eqref{general2pointSH}.

At the level of three points, the coefficients for even, homogeneous part for all the helicity configurations are as follows:
\begin{align}\label{homWightmantwistor}
\langle 0| \hat{J}^{h_1}_{s_1}(T_1)\hat{J}^{h_2}_{s_2}(T_2)\hat{J}^{h_3}_{s_3}(T_3)|0\rangle&=c_{s_1s_2s_3}^{(h)}\;\delta^{[s_1+s_2-s_3]}(T_1\cdot T_2)\delta^{[s_2+s_3-s_1]}(T_2\cdot T_3)\delta^{[s_3+s_1-s_2]}(T_3\cdot T_1),
\end{align}
which holds for all helicity configurations $h_i,h_j,h_k$ with the $T_i$ given as in \eqref{TUnotation}.\\
The coefficients for even, non-homogenous contibution for all the helicity configurations are as follows:
\begin{align}\label{nonhomWightmantwistor}
\langle 0| \tilde{J}^{h_1}_{s_1}(U_1)\tilde{J}^{h_2}_{s_2}(U_2)\tilde{J}^{h_3}_{s_3}(U_3)|0\rangle_{nh}&=c_{s_1s_2s_3}^{(nh)}\;\delta^{[-s_1-s_2+s_3]}(U_1\cdot U_2)\delta^{[-s_2-s_3+s_1]}(U_2\cdot U_3)\delta^{[-s_3-s_1+-s_2]}(U_3\cdot U_1),
\end{align}
$\forall$ $h_i,h_j,h_k$ and the $U_i$ provided in \eqref{TUnotation}.
Both these results after half-Fourier transform reproduce the correct Wightman function results as can easily be checked.

\subsubsection{Summary}
To summarize, the parity even two point function is given by \eqref{even2ptallhel}. At the level of three points, the parity even homogeneous and non-homogeneous correlators are given by \eqref{homWightmantwistor} and \eqref{nonhomWightmantwistor} respectively. After a half-Fourier transform, one can see that they give rise to the correct homogeneous and non-homogeneous spinor-helicity correlators like for instance, \eqref{TJJhelicities}. So far, we have started from twistor space and bootstrapped correlators and then converted the results back to spinor helicity variables which we identified with the components of a Wightman function\footnote{Recall that the reality condition in Lorentzian signature \eqref{LorentzReality} that we have used restricts all the momenta to be space-like. Note that this requirement was essential to perform the half-Fourier transform from the twistor space to Lorentzian spinor-helicity variables. One might thus wonder whether the corresponding momentum-space Wightman functions then only hold for space-like momentum. However, recall that the Wightman functions came with a definite operator ordering  which enforce that certain momenta have to be time-like to respect the spectral conditions as we shall see in the next section. The point is that all possible Wightman functions of conserved currents (all operator orderings) yield the same results up to some heaviside theta functions that enforce the spectral conditions. Twistor space via a half-Fourier transform gives rise to this bare quantity which must then be supplemented by the appropriate spectral conditions.}. In the next section, we shall take a more first-principles approach to bootstrap Wightman functions which serves to complement this section.

\section{Wightman functions}\label{sec:Wightman}
Correlation functions of local operators are among the most fundamental and important observables in generic quantum field theories. In Lorentzian signature, the basic building blocks for all correlators are the Wightman functions. In this section, we begin with a study of some generalities about Wightman functions and discuss their relation to Euclidean correlators. We then specialize to a discussion on Wightman functions in conformal field theories.

\subsection{Definitions and generalities}\label{subsec:Definitions}

A Wightman function is the following vacuum expectation value:
\begin{align}\label{WightmanFunction}
    W_n(x_1,\cdots, x_n)=\langle 0|O_{1}(t_1,\Vec{x}_1)\cdots O_n(t_n,\Vec{x}_n)|0\rangle,
\end{align}
where $O_i$, $i=1,\cdots,n$ are generic (spinning) operators. For an $n$-point correlator, we can form $n!$ such objects as there are that many possible orderings of the operators. Note however, that not all of them might be independent. Further, if the separation between all operators are space-like, all Wightman functions are identical thanks to microcausality. Technically speaking, these ``functions" are actually  \textit{tempered} distributions and must be integrated against Schwartz functions to obtain finite results, although we shall use the terms Wightman function and Wightman distribution interchangeably \footnote{See for instance David Duffins' TASI lecture notes on Conformal Field Theory in Lorentzian Signature available on his \href{http://theory.caltech.edu/~dsd/.}{Caltech home page}.}.

In the Heisenberg picture we can write \eqref{WightmanFunction} as\footnote{We also used the fact that the vacuum is Poincare invariant viz $H|0\rangle=\Vec{P}|0\rangle=0$.},
\small
\begin{align}
    &W(x_1,\cdots,x_n)\notag\\&=\langle 0|O_1(0)e^{-iH(t_1-t_2)-H(\epsilon_1-\epsilon_2)+i \Vec{P}\cdot(\Vec{x}_1-\Vec{x}_2)}O_2(0)\cdots e^{-iH(t_{n-1}-t_n)-H(\epsilon_{n-1}-\epsilon_n)+i\Vec{P}\cdot(\Vec{x}_{n-1}-\Vec{x}_n)}O_n(0)|0\rangle,
\end{align}
\normalsize
where we have used the $i\epsilon$ prescription with the constraint $\epsilon_1>\epsilon_2>\cdots>\epsilon_n$ which defines this Wightman function has a particular operator ordering. The effect of this prescription is to exponentially suppress contributions from high energy states. The limit $\epsilon_i\to 0$ (keeping the ordering fixed which) is only to be taken after smearing with the Schwartz functions:
\begin{align}\label{Smearing}
    W(f_1,f_2,\cdots,f_n)=\int d^d x_1\cdots d^d x_n f_1(x_1)\cdots f_n(x_n)W(x_1,\cdots,x_n).
\end{align}
Once the Wightman functions have been obtained, one can form other correlation functions of interest. For instance, the $n$ point time ordered correlator is a sum of the $n!$ different Wightman functions multiplied by the appropriate Heaviside theta functions that enforce the time ordering:
\small
\begin{align}\label{TimeOrderedCorr}
    \langle 0|T\{O_1(t_1,\Vec{x}_1)O_2(t_2,\Vec{x}_2)\cdots O_n(t_n,\Vec{x}_n)\}|0\rangle=\theta(t_1>t_2>\cdots>t_n)\langle 0|O_1(t_1,\Vec{x}_1)O_2(t_2,\Vec{x}_2)\cdots O_n(t_n,\Vec{x}_n)|0\rangle+\cdots.
\end{align}
\normalsize

\subsection*{Wightman functions do not contain contact terms in Ward-Takahshi identities} In contrast to the time ordered correlators, Wightman functions do not contain any contact term contributions on the RHS of Ward-Takahashi identities. For example, if $J^\mu$ is a conserved $U(1)$ current and $\phi$ and $\chi$ are oppositely charged scalars we still have,
\begin{align}\label{WightmanZeroWT}
    \partial_{1\mu}\langle 0|J^\mu(x_1)\phi(x_2)\chi(x_3)|0\rangle=0.
\end{align}
Given \eqref{WightmanZeroWT}, we show in appendix \ref{app:Derivation of current conservation Ward-Takahshi identity} how the familiar current conservation Ward-Takahashi identity for a time ordered correlator \eqref{TimeOrderedCorr} arises:
\begin{align}\label{timeorderedcorrWT}
    \partial_{1\mu}\langle 0|T\{J^\mu(x_1)\phi(x_2)\chi(x_3)\}|0\rangle=(q_\phi \delta^d(x_1-x_2)+q_\chi \delta^d(x_1-x_3))\langle 0|T\{\phi(x_2)\chi(x_3)\}|0\rangle.
\end{align} 

\subsection{Euclidean$\to$Wightman correlator}\label{subsec:EuclidtoWightmangeneral}
It is useful to understand how to compute the Wightman correlator given a Euclidean space correlator. Given Euclidean correlator
\begin{align}
    \langle O_{1}(\tau_1,\Vec{x}_1)\cdots O_{n}(\tau_n,\Vec{x}_n)\rangle,
\end{align}
one can obtain the Wightman function \eqref{WightmanFunction} as follows. Let $\tau_i=it_i+\epsilon_i$. We now take $\epsilon_i\to 0$ keeping fixed the ordering $\epsilon_1>\epsilon_2>\cdots>\epsilon_n$ to maintain the Euclidean time ordering that damps contributions due to high energy states. This Wick rotates the Euclidean correlator to the Wightman function \eqref{WightmanFunction}. 
\begin{align}\label{PosSpaeEuclidToWightman}
    W(x_1,\cdots,x_2)= \langle O_{1}(t_1+i\epsilon_1,\Vec{x}_1)\cdots O_{n}(t_n+i\epsilon_n,\Vec{x}_n)\rangle~,\epsilon_1>\epsilon_2>\cdots>\epsilon_n.
\end{align}

Finally, different orderings of the $\epsilon_i$ give rise to the different Wightman functions.
Having discussed some generalities, we shall now move on to the specific cases where the theories of interest also possess conformal invariance, in addition to Poincare invariance. We shall also focus mainly on three dimensions, although some of our general statements carry over to arbitrary dimensions.

\subsection{Wightman functions in conformal field theory}\label{sec:WightmanCFT}
In addition to the general properties discussed above, conformal invariance places severe restrictions on the form of Wightman functions.  We present a few different ways to obtain Wightman functions in momentum space
\begin{itemize}
\item First, via analytic continuation from Euclidean space  in subsection \ref{subsec:EuclidtoWightmaninCFT}.
\item Second, by directly bootstrapping them using their characteristic properties in subsection \ref{subsec:BootstrapWightman}.
\end{itemize} 
We also discuss the inverse procedure to obtain the Euclidean correlator from the Wightman function in subsection \ref{subsec:WightmantoEuclidCFT}. We discuss two point Wightman functions in more detail in appendix \ref{app:WightmanTwoPoint}. In appendix \ref{app:WightmanExamples}, we present explicitly worked out prototypical three point Wightman functions involving conserved currents, scalars, spinors and even those that involve slightly broken higher spin currents.

\subsubsection{Euclid$\to$ Wightman correlator}\label{subsec:EuclidtoWightmaninCFT}
Consider an $n-$point momentum space Euclidean correlator,
\begin{align}\label{EuclidAnsatz}
    \langle J_{s_1}(z_1,p_1)\cdots J_{s_n}(z_n,p_n)\rangle=\sum_{\mathcal{I}}\mathcal{T}_{\mathcal{I}}(\{z_j\cdot z_k,z_j\cdot p_k,\cdots\})\mathcal{F}_{\mathcal{I}}(\{p_j\cdot p_k\}),
\end{align}
where the sum runs over linearly independent tensor structures $\mathcal{T}_{\mathcal{I}}$ and $\mathcal{F}_{\mathcal{I}}$ are the associated form factors that are a function of a linearly independent set of the Lorentz invariants constructed out of the momenta.
In position space, the analytic continuation to obtain the corresponding Wightman functions is \eqref{PosSpaeEuclidToWightman}. Given the knowledge about the analytic structure of the Euclidean CFT correlators, one can obtain the analogous relation in momentum space \cite{Bautista:2019qxj,Baumann:2024ttn}.
This is essentially a two step process:
\begin{itemize}
\item Analytically continue $p^0\to i p^0.$
Please note that it is understood that we always perform this analytic continuation to obtain the Lorentz signature momenta even when not stated explicitly.

\item The next step is to take the discontinuity of the Wick rotated Euclidean form factors with respect to the Lorentz invariant quantities constructed out of the momenta. This yields the corresponding Wightman function form factors\footnote{For instance, at two and three points, the Lorentz invariants are $p_1^2$ and $(p_1^2,p_2^2,p_3^2)$ respectively.}. 
Consider the Euclidean correlator form factors $\mathcal{F}_{\mathcal{I}}(\{p_j\cdot p_k\})$ \eqref{EuclidAnsatz}. Let us focus on the Lorentz invariant $p_1^2$. When Wick rotated, the form factor has a branch cut starting at $p_1^2=0$. Thus, we must take a discontinuity with respect to this variable \cite{Bautista:2019qxj}. Proceeding this way, one obtains discontinuities with respect to the other Lorentz invariants\footnote{Till three points, this procedure is clear. For four and higher points, a more careful analysis is required.}. Note that the tensor structures $\mathcal{T}_{\mathcal{I}}$ do not depend on these quantities and hence their analytic continuation is trivial.   

\end{itemize}

 We illustrate this process with examples of two- and three-point functions.
\subsubsection*{Two points}
At two points, the relation between a generic Euclidean conformal two point function and the corresponding Wightman function is the following \cite{Bautista:2019qxj}:
\begin{align}\label{TwopointsEuclidtoWightman}
    \langle 0|J_{s}(z_1,p_1)J_{s}(z_2,p_2)|0\rangle=\theta(-p_1^0-|\Vec{p}_1|)\text{Disc}_{p_1^2}\big(\langle J_{s}(z_1,p_1)J_{s}(z_2,p_2)\rangle\big),
\end{align}
where we take the discontinuity of the correlator about the branch cut starting at $p_1^2=0$:
\begin{align}
\text{Disc}_{x}f(x)=\lim_{\epsilon\to 0}f(x+i\epsilon)-f(x-i\epsilon).
\end{align} 
 The first step in \eqref{TwopointsEuclidtoWightman} is to take the Euclidean correlator $\langle J_{s}(z_1,p_1)J_{s}(z_2,p_2)\rangle$ and analytically continue $p_1^0\to i p_1^0$ and $p_2^0\to i p_2^0$. The branch points of this quantity are at $p_1^2=0\implies p_1^0=\pm |\Vec{p}_1|$. 
We make a choice by multiplying by the theta function that selects the branch cut starting at $p_1^0=-|\Vec{p}_1|$ rather than the other one at $p_1^0=+|\Vec{p}_1|$ which would result in the Wightman function with the opposite operator ordering. This Heaviside theta function is a consequence of the spectral condition (rightmost operator has positive energy and by energy conservation, the leftmost operator thus has negative energy). Please see appendix \ref{app:WightmanTwoPoint} for a detailed discussion with examples. Two point functions of operators with specific scaling dimensions require renormalization in the Euclidean context. We also discuss these cases in appendix \ref{app:WightmanTwoPoint}.
\subsubsection*{Three points: parity even}
A general parity preserving three point Euclidean correlator takes the form,
\begin{align}
    \langle J_{s_1}(z_1,p_1)J_{s_2}(z_2,p_2)J_{s_3}(z_3,p_3)\rangle=\sum_\mathcal{I} \mathcal{T}_{\mathcal{I}}(\{z_j\cdot z_k,z_j\cdot p_k,\cdots\}) \mathcal{F}_{\mathcal{I}}(p_1,p_2,p_3).
\end{align}
To obtain the corresponding Wightman function we take the discontinuity of each form factor with respect to $p_1^2,\;p_2^2$ and $p_3^2$.
\begin{align}\label{WightmanToEuclidSpinning}
    \langle 0| J_{s_1}(z_1,p_1)J_{s_2}(z_2,p_2)J_{s_3}(z_3,p_3)|0\rangle\propto\sum_{\mathcal{I}} \mathcal{T}_{\mathcal{I}}(\{z_j\cdot z_k,z_j\cdot p_k,\cdots\}) \prod_{i=1}^{3}\text{Disc}_{p_i^2}\big(\mathcal{F}_{\mathcal{I}}(p_1,p_2,p_3)\big),
\end{align}
where the discontinuity is given by\footnote{The equality in the final step follows from what we observed for a plethora of examples.},
\begin{align}\label{discff}
    \prod_{i=1}^{3}\text{Disc}_{p_i^2}f(p_1,p_2,p_3)&=\big(f(-p_1,-p_2,-p_3)-f(p_1,p_2,p_3)\big)+\big(f(-p_1,p_2,p_3)-f(p_1,-p_2-p_3)\big)\notag\\
    &+\big(f(p_1,-p_2,p_3)-f(-p_1,-p_2,p_3)\big)+\big(f(p_1,p_2,-p_3)-f(-p_1,-p_2,p_3)\big)\notag\\
    &=-2\bigg(f(p_1,p_2,p_3)-f(-p_1,p_2,p_3)-f(p_1,-p_2,p_3)-f(p_1,p_2,-p_3)\bigg).
\end{align}  
The proportionality sign, apart from numerical constants also contains information about
which Wightman function we desire.
For instance, with the operator ordering considered above we obtain\footnote{Note that our conventions for the Fourier transform differ from \cite{Bautista:2019qxj}: $O(p)_{\text{ours}}=\int d^d x~e^{+ip\cdot x}O(x)$ whereas $O(p)_{\text{theirs}}=\int d^d x~e^{-ip\cdot x}O(x)$.},
\begin{align}\label{thetafunctions}
    \theta(-p_1^0-|\Vec{p}_1|)\theta(-p_2^0+|\Vec{p}_2|)\theta(p_2^0+|\Vec{p}_2|)\theta(p_3^0-|\Vec{p}_3|).
\end{align}
which follows from the spectral condition\footnote{Physically, this follows from the fact that an operator acting on the vacuum from the right should create a positive energy excitation.}. In position space, the different Wightman functions differ by the $i\epsilon$ prescriptions which translates in momentum space to different theta functions like \eqref{thetafunctions} that accompany them to satisfy the spectral condition. The important point is that the remaining functional form (tensor structures and form-factors) are exactly identical for all the Wightman functions of conserved currents and thus presents a unified bootstrap paradigm for them all.

Thus, the general three point Wightman function, given the Euclidean form factors $\mathcal{F}_{\mathcal{I}}$ is as follows:
\small
\begin{align}\label{EvenWightmanFromEuclidean}
    &\langle 0| J_{s_1}(z_1,p_1)J_{s_2}(z_2,p_2)J_{s_3}(z_3,p_3)|0\rangle\notag\\&\propto\sum_{\mathcal{I}} \mathcal{T}_{\mathcal{I}}(\{z_j\cdot z_k,z_j\cdot p_k,,..\}) \big(\mathcal{F}_{\mathcal{I}}(p_1,p_2,p_3)-\mathcal{F}_{\mathcal{I}}(-p_1,p_2,p_3)-\mathcal{F}_{\mathcal{I}}(p_1,-p_2,p_3)-\mathcal{F}_{\mathcal{I}}(p_1,p_2,-p_3)\big).
\end{align}
\normalsize

\subsection*{Higher points}
We expect the same methodology to work for higher point functions although we shall not pursue the same in this paper. We now present a more intrinsic bootstrap method to obtain Wightman functions rather than resorting to any analytic continuation.

\subsubsection{Solving for Wightman functions}\label{subsec:BootstrapWightman}
In this section we bootstrap three point Wightman functions directly without refering to Euclidean space correlator.  
A general $n-$point Poincare invariant ansatz for a momentum space Wightman function takes the following form:
\begin{align}\label{WightmanAnsatz}
    \langle 0|J_{s_1}(z_1,p_1)\cdots J_{s_n}(z_n,p_n)|0\rangle=\sum_{\mathcal{I}}\mathcal{T}_{\mathcal{I}}(\{z_j\cdot z_k,z_j\cdot p_k,\cdots\})\mathcal{G}_{\mathcal{I}}(\{p_j\cdot p_k\}),
\end{align}
where the sum runs over linearly independent tensor structures $\mathcal{T}_{\mathcal{I}}$ and $\mathcal{G}_{\mathcal{I}}$ are the associated form factors that are a function of a linearly independent set of the Lorentz invariants constructed out of the momenta.

The ansatz, \eqref{WightmanAnsatz}, is constrained by the following facts
\begin{itemize}
\item Conformal invariance,
\item The fact that Wightman functions involving conserved currents have a trivial (zero) Ward-Takahashi identity like in \eqref{WightmanZeroWT},
\item That every independent form factor vanishes when any of the external momenta are zero,\footnote{See appendix \ref{app:WightmanFunctionProperties} for details of why this is the case.}
\item Finally, the output Wightman function must also satisfy the reality condition\footnote{See appendix \ref{app:WightmanFunctionProperties} for a derivation of this property.},
\begin{align}
    \langle 0|J_{s_1}(z_1,p_1)\cdots J_{s_n}(z_n,p_n)|0\rangle^*=\langle 0|J_{s_n}(z_n,-p_n)\cdots J_{s_1}(z_1,-p_1)|0\rangle.
\end{align}
\end{itemize}


It may seem that the first two points above are the most restrictive ones and the third may not play as important a role. However, we shall find (at three points, but expect it to carry over to higher points) that it is this 
 actually, especially for the Wightman version of the homogeneous correlators that plays the most crucial role.  Here, we summarize the results:
 \subsubsection{General structure Inside the triangle}
 Inside the triangle, the most general parity even Euclidean three point function of conserved currents is given by,
\begin{align}\label{CFT3Euclidthreepointhandnh1}
    \langle J_{s_1}J_{s_2}J_{s_3}\rangle=c_{s_1s_2s_3}^{(h)}\langle J_{s_1}J_{s_2}J_{s_3}\rangle_{h}+c_{s_1s_2s_3}^{(nh)}\langle J_{s_1}J_{s_2}J_{s_3}\rangle_{nh}.
\end{align}
The homogeneous structure is identically conserved whereas the non-homogeneous structure saturates the Ward-Takahashi identity of the correlator\footnote{One can convince themselves that only the largest spin in such a correlator will have a non-trivial Ward-Takahashi identity.}.
The corresponding Wightman function also comprises of two even structures:
\begin{align}\label{CFT3Wightmanthreepointhandnh1}
    \langle 0|J_{s_1}J_{s_2}J_{s_3}|0\rangle=c_{s_1s_2s_3}^{(h)}\langle 0|J_{s_1}J_{s_2}J_{s_3}|0\rangle_{h}+c_{s_1s_2s_3}^{(nh)}\langle 0|J_{s_1}J_{s_2}J_{s_3}|0\rangle_{nh}.
\end{align}
In contrast to \eqref{CFT3Euclidthreepointhandnh1}, both structures are identically conserved. However, after analytic continuation, we showed that the $nh$ structure appearing in \eqref{CFT3Wightmanthreepointhandnh1} indeed goes over to its counterpart in \eqref{CFT3Euclidthreepointhandnh1} that has a non-trivial Ward-Takahashi identity. 

\subsubsection{General structure outside the triangle}
Outside the triangle, we have in Euclidean space two parity even non-homogeneous structures and no homogeneous structure:
\begin{align}\label{CFT3Euclidthreepointoutside1}
    \langle J_{s_1}J_{s_2}J_{s_3}\rangle=c_{s_1s_2s_3}^{(nh_1)}\langle J_{s_1}J_{s_2}J_{s_3}\rangle_{F-B}+c_{s_1s_2s_3}^{(nh_2)}\langle J_{s_1}J_{s_2}J_{s_3}\rangle_{F+B}.
\end{align}
For Wightman functions we have a similar formula namely,
\begin{align}\label{CFT3Wightmanthreepointoutside}
    \langle 0|J_{s_1}J_{s_2}J_{s_3}|0\rangle=c_{s_1s_2s_3}^{(F-B)}\langle 0|J_{s_1}J_{s_2}J_{s_3}|0\rangle_{F-B}+c_{s_1s_2s_3}^{(F+B)}\langle 0|J_{s_1}J_{s_2}J_{s_3}|0\rangle_{F+B}.
\end{align}
In contrast to \eqref{CFT3Euclidthreepointoutside1} where both structures are non-homogeneous, the Wightman functions of \eqref{CFT3Wightmanthreepointoutside} are both identically conserved.

Having discussed several ways of obtaining Wightman functions, we now present the analytic continuation to get to a Euclidean correlator given a Wightman function.
\subsection{Wightman $\to$ Euclid correlator}\label{subsec:WightmantoEuclidCFT}
Wightman functions as we discussed are related to the corresponding Euclidean CFT correlators via taking a discontinuity as in \eqref{WightmanToEuclidSpinning} . 
Thus, a Euclidean correlator can be obtained by performing dispersive integrals (the ``inverse" of the discontinuity procedure) over their Wightman counterparts\footnote{The authors of \cite{Baumann:2024ttn} present a method to obtain the Euclidean correlator from the Wightman function directly in spinor helicity variables. This involves extracting a particular pre-factor from the spinor helicity
correlator and interpreting the remaining expression as the discontinuity of the Euclidean correlator. While their method is straightforward to follow for three point functions of identical spin currents, it may be more difficult for general correlators with non-identical spin as well as at higher points. We instead perform the analytic continuation in momentum space where the distinction between the tensor structures and form factors is clear. 
}. Given a generic Wightman function,
    \begin{align}
    \langle 0| J_{s_1}(z_1,p_1)J_{s_2}(z_2,p_2)J_{s_3}(z_3,p_3)|0\rangle=\sum_{\mathcal{I}} \mathcal{T}_{\mathcal{I}}(\{z_j\cdot z_k,z_j\cdot p_k,\cdots\}) \mathcal{G}_{\mathcal{I}}(p_1,p_2,p_3),
\end{align}
one can obtain the corresponding Euclidean correlator via,
\begin{align}\label{WightmantoEuclid}
    \langle  J_{s_1}(z_1,p_1)J_{s_2}(z_2,p_2)J_{s_3}(z_3,p_3)\rangle=\sum_{\mathcal{I}} \mathcal{T}_{\mathcal{I}}(\{z_j\cdot z_k,z_j\cdot p_k,\cdots\}) \prod_{i=1}^{3}\text{Disp}_{p_i^2}~\mathcal{G}_{\mathcal{I}}(p_1,p_2,p_3),
\end{align}
where we have defined the dispersive integrals as:
\small
\begin{align}
\prod_{i=1}^{3}\text{Disp}_{p_i^2}\mathcal{G}_{\mathcal{I}}(p_1,p_2,p_3)=\prod_{i=1}^{3}\bigg(\frac{1}{2\pi i}\oint_{C_i}\frac{d(\omega_i^2)}{\omega_i^2-p_i^2}\bigg)\mathcal{G}_{\mathcal{I}}(\omega_1,\omega_2,\omega_3)=\prod_{i=1}^{3}\bigg(\frac{1}{\pi i}\int_0^\infty\frac{\omega_id\omega_i}{\omega_i^2-p_i^2}\textrm{Disc}_{\omega_i^2}\bigg)\mathcal{G}_{\mathcal{I}}(\omega_1,\omega_2,\omega_3),
\end{align}
\normalsize
where the contours $C_i$ enclose the corresponding $p_i$ counterclockwise. The second equality is obtained by deforming the contour of integration to large arcs, whose contribution we assume vanishes. We have also taken into account the fact that the branch points of the Euclidean correlator are at $p_i^2=0$. This integral can be computed 
by closing the contour in lower half-plane \cite{Baumann:2024ttn}, which is tantamount to the sum of residues of poles i.e $\omega_i=\omega_{i,P}$ with negative imaginary part (each momentum magnitude is given a positive imaginary part). This process of dispersive integrals over Lorentzian form factors result in their Euclidean avatar.
\begin{align}\label{Dispersipn}
\prod_{i=1}^{3}\text{Disp}_{p_i^2}\mathcal{G}_{\mathcal{I}}(p_1,p_2,p_3)=\sum_{\omega_{3,P}} \underset{\omega_3=\omega_{3,P}}{\textrm{Res}}\frac{\omega_3}{\omega_3^2-p_3^2}\Big(\sum_{\omega_{2,P}} \underset{\omega_2=\omega_{2,P}}{\textrm{Res}}\frac{\omega_2}{\omega_2^2-p_2^2}\Big(\sum_{\omega_{1,P}} \underset{\omega_1=\omega_{1,P}}{\textrm{Res}}\frac{\omega_1}{\omega_1^2-p_1^2}\mathcal{G}_{\mathcal{I}}(\omega_1,\omega_2,\omega_3)\Big)\Big).
\end{align}
A key difference between \cite{Baumann:2024ttn} and our approach is that while they perform the dispersive integrals over the form factors in spinor-helicity expressions, we do the same over the momentum space form factors\footnote{It is also possible to obtain the \textit{alpha} vacua correlators from the Wightman function. Given a Bunch-Davies form factor $\mathcal{F}_{I}(p_1,p_2,p_3)$, the alpha vacua form factors are obtained by flipping the momenta magnitudes. In \eqref{Dispersipn}, this is achieved by changing the countour prescription. For instance if we want $\mathcal{F}_{\mathcal{I}}(p_1,p_2,-p_3)$ we enclose the pole at $\omega_3=-
(p_3+i\epsilon)$ rather than the pole at $p_3+i\epsilon$. Similarly, one can obtain form factors with the other two momenta magnitudes flipped and thus form the general alpha vacua expression.}.

\section{Parity odd Wightman functions}\label{sec:parityodd}
In this section, we discuss parity and time-reversal odd Wightman functions, both in momentum space, spinor helicity variables and twistor space.
\subsection{Momentum space}

In three dimensional CFTs, three point functions of conserved currents admit a parity odd structure \cite{Jain:2021wyn}. Given the parity even homogeneous Euclidean correlator, one can obtain its parity odd counterpart via an \textit{epsilon} transform as we discussed in \eqref{cFT3EuclidEPT}. The analogous statement for Wightman functions as we shall see now, is a little more involved.

 One might naturally be inclined to perform an epsilon transform to the parity even Wightman function \eqref{EvenWightmanFromEuclidean} to obtain its odd counterpart. This however, does not lead to the correct result and the actual relation is a little more involved. Start with the Euclidean parity odd correlator (which is say, obtained by epsilon transforming the Euclidean homogeneous correlator) which takes the form,
\begin{align}\label{gencorrEucOdd}
    \langle J_{s_1}J_{s_2}J_{s_3}\rangle_{\text{odd}}=\sum_{\mathcal{I}}\mathcal{T}_{\mathcal{I}}(\{\epsilon^{z_j z_k p_k},\epsilon^{z_j p_j p_k },\cdots\})\mathcal{F}_{\mathcal{I}}(p_1,p_2,p_3).
\end{align}
Now, rather than taking the discontinuity of each form factor $\mathcal{F}_{\mathcal{I}}(p_1,p_2,p_3)$ using \eqref{discff} (which results in zero in the first equality of \eqref{discff}), we directly use the last line of the formula \eqref{discff}. This results in,
\begin{align}\label{CorrectoddWightmanCorrelator}
    &\langle 0|J_{s_1}J_{s_2}J_{s_3}|0\rangle_{\text{odd}}\notag\\&=i\sum_{\mathcal{I}}\mathcal{T}_{\mathcal{I}}(\{\epsilon^{z_j z_k p_k},\epsilon^{z_j p_j p_k },\cdots\})\bigg(\mathcal{F}_{\mathcal{I}}(p_1,p_2,p_3)-\mathcal{F}_{\mathcal{I}}(-p_1,p_2,p_3)-\mathcal{F}_{\mathcal{I}}(p_1,-p_2,p_3)-\mathcal{F}_{\mathcal{I}}(p_1,p_2,-p_3)\bigg),
\end{align}
Note the explicit factor of $i$ in contrast to the corresponding Euclidean correlator \eqref{gencorrEucOdd}. This is actually essential to maintain the reality conditions for Wightman functions discussed in appendix \ref{app:WightmanFunctionProperties}.
 We can also obtain this quantity via a suitable modification of the epsilon transform of the even Wightman function \eqref{EvenWightmanFromEuclidean} as follows: First, we re-write \eqref{EvenWightmanFromEuclidean} as,
\begin{align}
    \langle 0|J_{s_1} J_{s_2} J_{s_3}|0\rangle=\langle 0|J_{s_1} J_{s_2} J_{s_3}|0\rangle_{0}-\langle 0|J_{s_1} J_{s_2} J_{s_3}|0\rangle_{1}-\langle 0|J_{s_1} J_{s_2} J_{s_3}|0\rangle_{2}-\langle 0|J_{s_1} J_{s_2} J_{s_3}|0\rangle_{3},
\end{align}
where the subscript $0,1,2,3$ indicate the sum of the terms in \eqref{EvenWightmanFromEuclidean} with the form factors $\mathcal{F}_{\mathcal{I}}(p_1,p_2,p_3)$, 
 $\mathcal{F}_{\mathcal{I}}(-p_1,p_2,p_3)$, $\mathcal{F}_{\mathcal{I}}(p_1,-p_2,p_3),\mathcal{F}_{\mathcal{I}}(p_1,p_2,-p_3)$ respectively. The parity odd Wightman function can then be obtained via,
\begin{align}\label{WightmanEPT}
    &\langle 0|J_{s_1}J_{s_2}J_{s_3}|0\rangle_{\text{odd}}=i\bigg(\langle 0|J_{s_1}\epsilon\cdot J_{s_2} J_{s_3}|0\rangle_{0}-\langle 0|J_{s_1}\epsilon\cdot J_{s_2} J_{s_3}|0\rangle_{1}+\langle 0|J_{s_1}\epsilon\cdot J_{s_2} J_{s_3}|0\rangle_{2}-\langle 0|J_{s_1}\epsilon\cdot J_{s_2} J_{s_3}|0\rangle_{3}\bigg).
\end{align}
Note that in \eqref{WightmanEPT}, we have performed an epsilon transform with respect to $J_{s_2}$ but one can also obtain the same answer via an appropriate epsilon transform with respect to the other two currents analogusly. Both these methods yield the same result. We also checked using the methods of subsection \ref{subsec:WightmantoEuclidCFT} that these odd Wightman functions when analytically continued back to Euclidean space yield the correct results. Explicitly worked out examples can be found in appendix \ref{app:WightmanExamples}.

\subsection{Spinor helicity variables}
Given a parity odd Wightman function obtained using \eqref{CorrectoddWightmanCorrelator} or \eqref{WightmanEPT}, we can convert it into spinor helicity variables. For example, consider the correlator between a stress tensor and two $U(1)$ currents, the momentum space expression for which is given in \eqref{TJJWightmanodd}. In spinor helicity variables we have,
\begin{align}\label{TJJhelicities1}
     &\langle 0|T^{-}J^{-}J^{-}|0\rangle_{\text{odd}}=-i c_{211}^{odd}\frac{\langle 1 2\rangle^2\langle 3 1\rangle^2 p_1}{E^4},\langle 0|T^{-}J^{-}J^{+}|0\rangle_{\text{odd}}=-i c_{211}^{odd}\frac{\langle 1 2\rangle^2\langle \Bar{3}1\rangle^2 p_1}{(E-2p_3)^4},\notag\\
     &\langle 0|T^{-}J^{+}J^{-}|0\rangle_{\text{odd}}=-i c_{211}^{odd}\frac{\langle 1 \Bar{2}\rangle^2\langle 3 1\rangle^2 p_1}{(E-2p_2)^4},\langle 0|T^{+}J^{-}J^{-}|0\rangle_{\text{odd}}=-i c_{211}^{odd}\frac{\langle \Bar{1} 2\rangle^2\langle 3 \Bar{1}\rangle^2 p_1}{(E-2p_1)^4},\notag\\
     &\langle 0|T^{+}J^{+}J^{+}|0\rangle_{\text{odd}}=+ic_{211}^{odd}\frac{\langle \Bar{1}\Bar{2}\rangle^2\langle\Bar{3}\Bar{1}\rangle^2 p_1}{E^4},\langle 0|T^{+}J^{+}J^{-}|0\rangle_{\text{odd}}=+ic_{211}^{odd}\frac{\langle \Bar{1}\Bar{2}\rangle^2\langle3\Bar{1}\rangle^2 p_1}{(E-2p_3)^4},\notag\\
     &\langle 0|T^{+}J^{-}J^{+}|0\rangle_{\text{odd}}=+ic_{211}^{odd}\frac{\langle \Bar{1}2\rangle^2\langle\Bar{3}\Bar{1}\rangle^2 p_1}{(E-2p_2)^4},\langle 0|T^{-}J^{+}J^{+}|0\rangle_{\text{odd}}=+ic_{211}^{odd}\frac{\langle 1\Bar{2}\rangle^2\langle\Bar{3}1\rangle^2 p_1}{(E-2p_1)^4}.
\end{align}
 The important point to note about \eqref{TJJhelicities1} in contrast to its parity even homogeneous counterpart\footnote{The expression for the even homogeneous term can be found in \eqref{TJJhelicities}.} is the relative sign between configurations with more negative helicity currents and those with more positive helicity currents. Parity odd correlators in Euclidean space are non-zero only in $(---)$ and $(+++)$ helicities and are actually the same as in \eqref{TJJhelicities1} with the mixed helicities being zero. This factor of $i$ and the relative signs between these pairs of helicities can also be understood as a consequence of the correlator being odd under parity, odd under time-reversal but even under their composition. For other correlators, the exact functional form of each helicity is different but the factors of $i$ and the relative signs is the same as this example.

\subsection{Wightman functions from twistor space}
Now that we are acquainted with momentum space and spinor helicity parity odd Wightman functions, we ask how twistor space encodes them.
\subsubsection{Two points}
The parity even Wightman function in twistor space was found in \eqref{even2ptallhel}. Given the fact that there were two solutions to the twistor space Ward-identities \eqref{twopointtwistor2sols} and the even correlator is given by the polynomial type solution, it is natural to ask if the parity odd correlator is given by the delta function type solution, perhaps multiplied by a factor of $i$ to ensure $CPT$ invariance in the language of \cite{Baumann:2024ttn}. Thus we ask,
\begin{align}
    \langle 0|\hat{J}_s^{+}(Z_1)\hat{J}_s^{+}(Z_2)|0\rangle_{\text{odd}}\qeq i \delta^{[2s+1]}(Z_1\cdot Z_2).
\end{align}
Our hopes are immediately squashed by looking at the corresponding spinor helicity results after a half-Fourier transform where we find a sign factor that depends on the momentum just like in \eqref{2ptpp2sols} which is uncharacteristic of the correct parity odd two point Wightman function that is provided in \eqref{general2pointSH}. The solution we propose is that the parity odd two point correlator is also given by the polynomial type solution except, the $(++)$ and $(--)$ helcities come with an opposite sign and a factor of $i$ to maintain consistency with the Wightman function result in \eqref{general2pointSH}. Thus, we have,
\begin{align}\label{odd2pttwistorallhel}
    &\langle 0|\hat{J}_s^{+}(Z_1)\hat{J}_s^{+}(Z_2)|0\rangle_{\text{odd}}=i \frac{c_{s,\text{odd}}}{(Z_1\cdot Z_2)^{2(s+1)}}~,~\langle 0|\tilde{J}_s^{-}(W_1)\tilde{J}_s^{-}(W_2)|0\rangle_{\text{odd}}=-i\frac{c_{s,\text{odd}}}{(W_1\cdot W_2)^{2(s+1)}}.
\end{align}
This choice picks out the parity and time-reversal odd two point function. This can be seen by performing the half-Fourier transform that yields,
\small
\begin{align}
    \langle 0|J_s^{+}(p_1)J_s^{+}(p_2)|0\rangle_{\text{odd}}=i(2\pi)^3\delta^3(p_1+p_2)c_{s,\text{odd}}\frac{\langle 1 2\rangle^{2s}}{p_1}~,~\langle 0|J_s^{-}(p_1)J_s^{-}(p_2)|0\rangle_{\text{odd}}=-i(2\pi)^3\delta^3(p_1+p_2)c_{s,\text{odd}}\frac{\langle \Bar{1} \Bar{2}\rangle^{2s}}{p_1},
\end{align}
\normalsize
which are the two components of the momentum space parity and time-reversal odd Wightman function found in \eqref{general2pointSH}.

\subsubsection{Three points}
The parity even three point function was found to take the form of a product of three delta functions \eqref{twistorspace3points}. For the parity odd case, do the polynomial solutions present in \eqref{f1pppevensol} accommodate them? The answer, it turns out is no. The reason is that after half-Fourier transform, they go over to Wightman functions which have momentum dependent signs that are inconsistent with the correct results such as  \eqref{TJJhelicities1}. The solution, just like at the level of two points is to realize that the odd twistor correlators take the same functional form as the even ones but the coefficients of the different helicities are different.


More precisely, the expressions for the odd Wightman function can be divided into two classes, depending on the number of negative and positive helicity currents in the correlator:
\begin{numcases}{\langle 0| \hat{J}_{s_1}^{h_1}(T_1)\hat{J}_{s_2}^{h_2}(T_2)\hat{J}_{s_3}^{h_3}(T_3)|0\rangle_{\text{odd}}=}
-i c_{s_1s_2s_3}^{(odd)}\delta^{[s_1+s_2-s_3]}(T_1\cdot T_2)\delta^{[s_2+s_3-s_1]}(T_2\cdot T_3)\delta^{[s_3+s_1-s_2]}(T_3\cdot T_1), \nonumber \\
\text{when sign}(h_1)+\text{sign}(h_2)+\text{sign}(h_3)<0, \nonumber\\
\nonumber\\
+i c_{s_1s_2s_3}^{(odd)}\delta^{[s_1+s_2-s_3]}(T_1\cdot T_2)\delta^{[s_2+s_3-s_1]}(T_2\cdot T_3)\delta^{[s_3+s_1-s_2]}(T_3\cdot T_1),\nonumber\\
\text{when sign}(h_1)+\text{sign}(h_2)+\text{sign}(h_3)>0.
\end{numcases}
These factors of $-i$ and $+i$ are necessary in order to maintain consistency with the spinor helicity results such as \eqref{TJJhelicities1} which also follows as a consequence of CPT invariance, see appendix \ref{app:CPTinv} for the details. The question on how to obtain this from a single Penrose transformation as in \cite{Baumann:2024ttn} remains unclear. However, the above is a consistent solution reminiscent of the relation between even and odd correlators in spinor helicity variables and gives rise to the correct spinor helicity correlators such as \eqref{TJJhelicities1}.

\section{Discussion and future directions}\label{sec:Discussion}
In this paper, we have set up and solved the twistor space conformal Ward identities for two and three point correlators of conserved currents. It is interesting that these identities accompanied by the helicity operator lead to differential equations of the Euler type which has as its solutions not only usual classical polynomial functions, but also weak solutions such as distributional solutions. We found that all these classes of solutions play an important role and occur in different twistor space correlators. We also analyzed correlators that are odd under parity and proposed one solution as to how they can be represented in twistor space. Further, we also took steps in the paper to systematically analyze and solve for Wightman functions in momentum space and spinor helicity variables and found the results to be in perfect accord with their twistor space counterparts. There are many interesting avenues to explore further of which we list a few below.

\subsection*{Conformal blocks and higher point functions}
We have explored the spinning three-point correlators in twistor space and seen its immense power in terms of the simplicity of correlators. It would be interesting to go beyond and explore spinning higher point functions using this approach. Although they are generally complicated, we believe twistor space might help in this pursuit. In this work, we have only performed the analysis for conserved currents. In order to explore higher point functions, one would have to do the same analysis for three-point functions with one non-conserved current. It would also be interesting to bootstrap conformal partial waves \cite{Simmons-Duffin:2012juh} and conformal blocks \cite{Costa:2011dw} in twistor space. One starting point may be bootstrap of the twist=1 conformal blocks \cite{Alday:2016njk,Jain:2023juk}, where all currents are conserved. Moreover, the helicity basis has also proved useful in $CFT_3$ \cite{Caron-Huot:2021kjy} and since twistor space is intertwined with helicity, it would be interesting to see if for instance, the Lorentzian inversion formula takes a simple form in twistor space.

\subsection*{Connecting Twistor space results to Euclidean results and vice versa} 
\begin{figure}
    \centering
    \begin{tikzpicture}[node distance=3.5cm, auto, >=latex]
    \small
    \node (CP3) {\(\mathbb{CP}^3\)};
    \node (RP3) [right of=CP3] {\(\mathbb{RP}^3\)};
    \node (Minkowski) [right of=RP3] {\(\mathbb{R}^{2,1}\)};
    \node (Euclidean) [below of=Minkowski] {\(\mathbb{R}^3\)};
    \node (EPT) [below of=CP3] {$\mathbb{EPT}$};
    \node (MomMinkowski) [right of=Minkowski] {\(\mathbb{R}^{2,1}_{\text{mom}}\)};
    \node (MomEuclidean) [right of=Euclidean] {\(\mathbb{R}^3_{\text{mom}}\)};

    \draw[->, thick] (CP3) -- (RP3) node[midway, above] {\(\mathsf{Real\ Slice}\)};
    \draw[->, thick] (RP3) -- (Minkowski) node[midway, above] {\(\mathsf{Penrose\ Transform}\)};
    \draw[<->, thick] (Minkowski) -- (Euclidean) node[midway, left] {\(\mathsf{Wick\ Rotation}\)};
    \draw[<->, thick] (Euclidean) -- (EPT) node[midway, below] {\(\mathsf{Penrose\text{-}Woodhouse}\)};
    \draw[<-, thick] (EPT) -- (CP3) node[midway, left] {\(\mathsf{Euclidean\ Slice}\)};
    
    \draw[<->, thick] (Minkowski) -- (MomMinkowski) node[midway, above] {\(\mathsf{Fourier}\)};
    \draw[<->, thick] (Euclidean) -- (MomEuclidean) node[midway, above] {\(\mathsf{Fourier}\)};

    \draw[->, thick, bend left=50] (RP3) to node[midway, above] {\(\mathsf{Witten\text{-}Transform}\)} (MomMinkowski);

    \draw[->, dashed, bend right=50] (EPT) to node[midway, above] {\(\mathsf{?}\)} (MomEuclidean);

    \draw[->, thick] (MomMinkowski) to[bend left=15] node[midway, right] {\(\mathsf{Dispersion}\)} (MomEuclidean);
    \draw[->, thick] (MomEuclidean) to[bend left=15] node[midway, left] {\(\mathsf{Discontinuity}\)} (MomMinkowski);

    \end{tikzpicture}
    \caption{The network of twistor space and real space}
    \label{fig:penrose-fourier}
\end{figure}
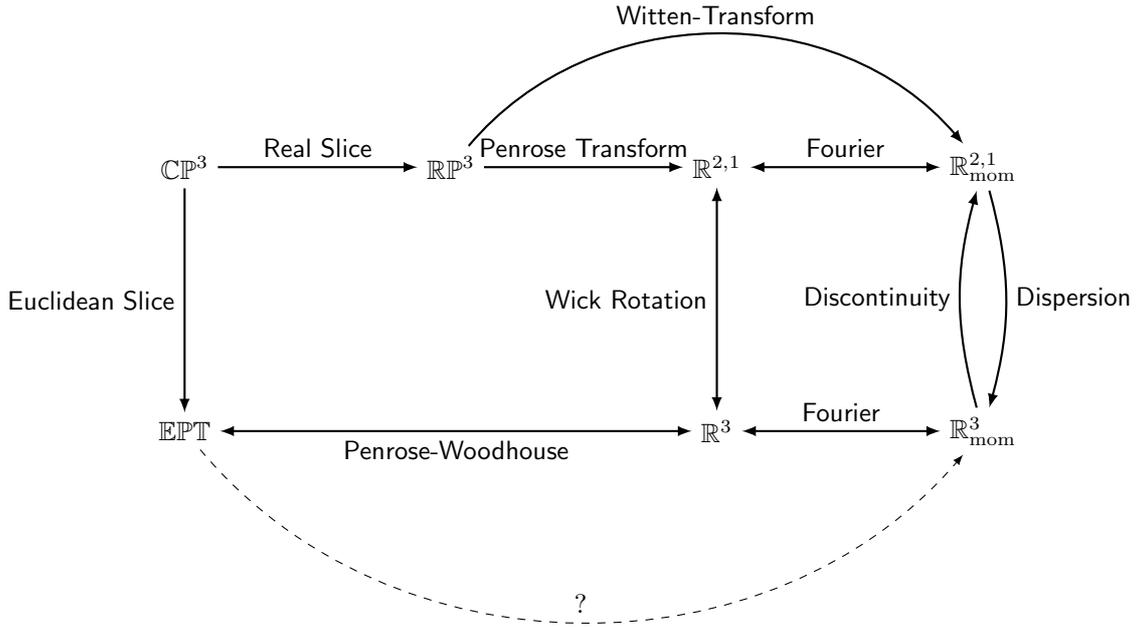

In the twistor space associated to four dimensional Euclidean space, there is a canonical way to go back and forth between twistor space and Euclidean space \cite{Woodhouse:1985id}. It would be interesting if we can implement a similar procedure to go back and forth between twistor space and 3d Euclidean correlators. In \cite{Baumann:2024ttn}, the Penrose transformation that relates a Wightman function to the twistor correlator is given by,
\begin{align}
    \langle 0|J_{s_1}\cdots J_{s_n}|0\rangle=\int DZ_1\cdots DZ_n~(Z_1\cdot \Upsilon_1^*)^{2s_1}F(Z_1,\cdots,Z_n),
\end{align}
where $Z_i^A=(\Lambda_i)^A_a \pi_i^a$ and $\Upsilon_i^*$ are the polarization spinors. $(\Lambda_i)^A_a$ are bi-spinors that are a function of the coordinates of $\mathbb{R}^{2,1}$. The LHS of the above equation can be Wick rotated to Euclidean space. It would be interesting to explore if the RHS then turns into a formula analogous to Woodhouse's result. On the other hand, as we argued from the perspective of this paper, one faces an obstruction in defining half-Fourier transform from spinor helicity variables to twistor space in Euclidean signature. One can then ask how this fact is reconciled with Woodhouse's result. We leave it to the future to further explore these issues and completely understand figure \ref{fig:penrose-fourier}.

\subsection*{$AdS$ amplitudes}
Over the last few years, and especially recently, there has been a lot of work to compute and bootstrap momentum space correlation functions in $AdS$. See \cite{DHoker:1999kzh,Raju:2010by,Albayrak:2020isk,Gadde:2022ghy,Albayrak:2023kfk,Albayrak:2020fyp,Armstrong:2022mfr,Chowdhury:2024dcy} and references for example. However, in the usual spinor helicity variables, even the four point MHV gluon correlators are quite complicated \cite{Armstrong:2020woi}. It would be interesting to see if these quantities become simpler when expressed in the language of twistors. Such an analysis could also potentially lead to developments of techniques analogous to BCFW recursion  relations \cite{Britto:2005fq} for AdS correlators as well, see \cite{Raju:2010by,Raju:2012zr} for previous work. It also opens the door to exploring the double copy on an AdS background. 

\subsection*{Cosmological correlators}
Real twistor space $\mathbb{RP}^{3}$ in three dimensions is naturally associated to Minkowski signature spacetime $\mathbb{R}^{2,1}$. Cosmological correlators on the other hand, are defined at a constant $\mathbb{R}^{3}$ Euclidean time-slice in four dimensional de-Sitter spacetime \cite{Maldacena:2011nz,McFadden:2011kk,Ghosh:2014kba}. 
If one's aim is to obtain cosmological correlators, then it is desirable to formulate a direct connection between Euclidean space and twistor space (which is an appropriate subset of $\mathbb{CP}^{3}$).

\subsection*{Chern-Simons matter theories and chiral higher spin theory in twistor space}
Chern-Simons matter theories are an important and interesting class of three dimensional conformal field theories that also exhibit (slightly broken) higher spin symmetry \cite{Giombi:2011kc,Aharony:2011jz,Maldacena:2011jn,Maldacena:2012sf}. Analysis performed in momentum space and spinor helicity variables uncovered their anyonic nature and also paved the way for the analysis of four and higher point functions \cite{Jain:2021gwa,Jain:2021whr,Jain:2021vrv,Jain:2021wyn,Jain:2021qcl,Gandhi:2021gwn,Jain:2022ajd,Kukolj:2024yyo}. It would be interesting to formulate these theories in twistor space. In \cite{Jain:2024bza,Aharony:2024nqs}, a subsector of these theories was found to be dual to chiral higher spin theory in $AdS_4$. We believe exploring such chiral subsector would be more natural from the Twistor space perspective. Also, given the fact that chiral higher spin theory already has a twistor space formulation \cite{Krasnov:2021nsq,Herfray:2022prf,Tran:2022tft}, it would be nice to reconcile the bulk and boundary notions of twistor space.


\subsection*{Dual conformal invariance, Yangian symmetry and higher spin symmetry}
In $\mathcal{N}=4$ SYM theory, the implementation of dual-super conformal symmetry and Yangian symmetry has been quite useful and led to a variety of interesting results such as \cite{Arkani-Hamed:2010zjl}, see \cite{Adamo:2011pv} for a nice review. Just like the scattering amplitudes, it would be interesting to see if we can formulate Yangian symmetry for conformal correlators. For this analysis, our differential equations viewpoint gives a nice way to formulate this problem. However, a more geometrical picture like in \cite{Arkani-Hamed:2012zlh,Arkani-Hamed:2013jha} and following the recent paper \cite{Baumann:2024ttn} would also be a nice approach.

Moreover, there are a class of CFTs such as free theories and Chern-Simons matter theories that have an infinite dimensional (slightly broken) higher spin symmetry. Thus it would be interesting to see whether there are any connections between higher spin symmetry and Yangian symmetry. 
On that note, it would also be interesting to formulate (slightly broken) higher spin symmetry \cite{Maldacena:2011jn,Maldacena:2012sf} in twistor space, which is another infinite dimensional symmetry algebra. 

\subsection*{Distributional solutions}
In twistor space, we see that distributional solutions play an important role. More generally, even in the usual position space study of CFT, distributional solutions, usually called contact terms are present, see \cite{Nakayama:2019mpz}. Many anomalies also present themselves through contact terms such as the trace anomaly, see \cite{Hartman:2024xkw} for a recent discussion. As an illustrative example, consider a two point function in one dimensional CFT. It admits in addition to the usual power-law type solution, a distributional solution in special cases\footnote{One way to see this is by solving the momentum space conformal Ward identities, see for instance equation (3.5) in \cite{S:2024zqp}.}. In particular,
\begin{align}
    \langle O_{\Delta_1}(t_1)O_{\Delta_2}(t_2)\rangle=\frac{c_{12}}{(t_1-t_2)^{2\Delta_1}}\delta_{\Delta_1,\Delta_2}+k_{12}\delta(t_1-t_2)\delta_{\Delta_1+\Delta_2,1}.
\end{align}
Sometimes, these contact terms can be removed via a redefinition of the correlator. However, there are cases such as in Chern-Simons matter theories \cite{Aharony:2012nh} where they cannot and thus represent actual physical quantities. It would thus be interesting to perform a complementary analysis to \cite{Nakayama:2019mpz} by determining contact term solutions via classifying the distributional solutions to the differential equations that arise due to conformal invariance.

\acknowledgments
We acknowledge our debt to the people of India for their steady support of research in basic sciences. We would like to thank M.Ali and N.Bhave for useful discussions. AB acknowledges
a UGC-NET fellowship.

\appendix
\section{Notation and conventions}\label{app:Notation}

While we employ spinor helicity variables, we frequently make use of various identities involving spinor brackets. In our conventions, they can be found in \cite{Jain:2023idr}. The contraction of spinor indices is performed using the two dimensional Levi-Civita symbol. Index raising and lowering is as follows:
\begin{align}
    A^a=\epsilon^{ab}A_b~,~A_b=\epsilon_{ab}A^a.
\end{align}
Contractions are performed as follows:
\begin{align}
    A\cdot B:=\langle A B\rangle=A_a B^a.
\end{align}
Finally, let us lay down our twistor space conventions. Given a twistor $Z^A$ ($W_A$) which are fundamental (anti-fundamental) representations of the conformal algebra of  $Sp(4)$\footnote{Recall the exceptional local isomorphism $SO(3,2)\cong Sp(4)$.}, we can lower (raise) its index using the associated symplectic form $\Omega_{AB}$:
\begin{align}
    Z_A=\Omega_{BA}Z^B~,~W^A=\Omega^{AB}W_B,
\end{align}
where \begin{align}\label{symplecticformapp}
    \Omega^{AB}=\Omega_{AB}=\begin{pmatrix}
        0&\delta_a^{b'}\\
        -\delta^b_{a'} & 0
    \end{pmatrix}.
\end{align}
The symplectic form satisfies,
\begin{align}
    \Omega_{AB}\Omega^{AC}=\delta_B^C,
\end{align}
where $\delta_B^C$ is the usual Kronecker delta function.

\section{Generalized solutions to differential equation}\label{app:distrubution}
In the framework of Generalized functions, Kanwal \cite{kanwal1998generalized} showed that in searching for solutions for a linear ordinary differential equation, we might have two classes of solutions:\\ 
\textbf{i}.)Classical Solutions, \\ \textbf{ii}.)Weak\,(distributional) solutions. \\ 
Let us obtain them one by one. From \eqref{e1cont} the form of the equation is,
\begin{align}\label{AppendixFeq1}
    & x \frac{d}{dx} F(x) + m F(x)=0 \,.
\end{align}
The first solution also called \textbf{Classical} solution of the above equation is,
\begin{align}\label{polynomialsolution2point}
   F_{1}(x) = \frac{C}{x^{m}} \, ,
\end{align}
where $C$ is the integration constant. There can be other solutions such as Weak functions (distribution type). A distribution $F(x)$ is a solution of \eqref{AppendixFeq1} if $\forall$ arbitrary smooth test function $\phi(x)$\,(which are well behaved at infinity), 
\begin{align}\label{AppendixFeq3}
    \langle\hat{\textbf{L}} F(x),\phi(x)\rangle = \int_{-\infty}^{+\infty} \phi(x) \hat{\textbf{L}}(F(x))= 0 ,
\end{align}
where $\hat{\textbf{L}} = x \frac{d}{dx} +m.$
The procedure we are going to follow to obtain such solutions is by writing an ansatz. For instance, consider, 
\begin{align}\label{distributionansatz}
    F_{2}(x) = \delta^{[n]}(x) ,
\end{align}
where, $\delta^{[n]}(x)$ is the $n^{th}$ derivative of $\delta(x)$ and n $\in$ $\mathbb{Z}^+$. Plugging in \eqref{distributionansatz} into \eqref{AppendixFeq3}, we obtain,
\begin{align}\label{AppendixFeq5}
    &\langle\hat{\textbf{L}} \delta^{[n]}(x),\phi(x)\rangle = 0 \notag \\
    \implies & \int_{-\infty}^{+\infty} dx \, \phi(x) \bigg(x\frac{d}{dx} + m \bigg)(\delta^{[n]}(x))  =0 \notag \\
    \implies & \int_{-\infty}^{+\infty} dx \, \bigg(\phi(x)\, x\frac{d}{dx}\delta^{[n]}(x)  + m \delta^{[n]}(x)\phi(x) \bigg)  =0 \notag \\ 
    \implies & \int_{-\infty}^{+\infty} dx \, \bigg(\frac{d}{dx}(x\,\delta^{[n]}(x) \phi(x) ) - \delta^{[n]}(x) \phi(x)- x\,\delta^{[n]}(x)\phi'(x)+ m \delta^{[n]}(x)\phi(x) \bigg) = 0 \notag \\
    \implies & \text{Boundary terms(B.T)} +\int_{-\infty}^{+\infty} dx\, (-(n+1) +m) \,\delta^{[n]}(x) \phi(x) =0 \, .
\end{align}
Thus, each boundary term (B.T) will vanish for arbitrary smooth functions $\phi(x)$ and the vanishing of the remaining integral also demands,
\begin{align}
    n = m-1 .
\end{align}
Hence, the second solution (also called \textbf{weak}) is given by ,
\begin{align}
F_2(x)=\delta^{[m-1]}(x) \forall  m\ge 1.
\end{align}
The analysis for the $m\le 0$ cases take a different form. The result there is given by,
\begin{align}\label{weaksol}
    &\delta^{[n]}(x) = \frac{x^{-n-1}}{(-n-1)!} \theta(x) ,\, n \leq -1.
\end{align}
In demanding a solution with a definite parity\footnote{We can observe that the helicity equation \eqref{AppendixFeq1} is invariant under parity hence we should look for solution with definite parity.}, we can take a linear combination of \eqref{weaksol} and \eqref{polynomialsolution2point} to get, 
\begin{align}\label{weaksolsign}
    F_2(x)= \frac{Sgn(x)}{2x^{n+1}(-n-1)!} ,\text{where} \, n \leq -1.
\end{align}
If we substitute \eqref{weaksolsign} into the ODE \eqref{AppendixFeq3} and following the similar steps as in \eqref{AppendixFeq5} will require that,
\begin{align}\label{nintermsofm}
         n = m-1
\end{align}
The second \textbf{weak} solution valid for $m\le 0$ is thus given by,
\begin{align}
   F_2(x)= \frac{Sgn(x)}{2x^{m}(-m)!} ,\, m \leq 0.
\end{align}
Hence, the second solution from the exercise above takes the following form:
   \begin{align}
   &F_2(x)= \begin{cases} 
\frac{d^n}{dx^n}\delta(x) & ,  n \geq 0 \\
\frac{Sgn(x)}{2x^{n+1}(-(n+1)!)} & , n \leq -1
\end{cases}
\end{align}
with $n$ given in \eqref{nintermsofm}.

\section{Twistor space}\label{app:3dTwistorsApproach}
In this appendix, we shall present the twistor space construction that we employ for the analysis of three dimensional conformal field theories. There are a few different ways to define twistor space \cite{Atiyah:2017erd}. The one that we shall choose is taking twistor space to be the space spanned by the projective spinors of the complexified conformal group. 

In three dimensions, the double cover of the conformal group $\text{Spin}(3,2)$ is isomorphic to $Sp(4,\mathbb{R})$. Its complexification is the group $Sp(4,\mathbb{C})$. The fundamental representation of $Sp(4,\mathbb{C})$ is a four component complex spinor representation. Our twistors are then,
\begin{align}
    Z^A=(Z^1,Z^2,Z^3,Z^4).
\end{align}
We then quotient by $\mathbb{C}$ to obtain the corresponding coordinates on $\mathbb{CP}^3$. Let us now express this twistor in terms of representations of the $2+1$ dimensional Lorentz group. The three dimensional complexified Lorentz group is $SL(2,\mathbb{C})$ which has two dimensional representations. Thus, we identify,
\begin{align}\label{Zinmulambda1}
    Z^A=\lambda^a\oplus \Bar{\mu}_{a'},
\end{align}
where $\lambda$ and $\Bar{\mu}$ are both in the fundamental representation of $SL(2,\mathbb{C})$. The incidence relation that connects twistor space to position space is given by,
\begin{align}\label{3dindidence}
    \Bar{\mu}_{a'}=x_{a a'}\lambda^a,
\end{align}
where $x_{aa'}$ is valued in $M_{\mathbb{C}}^3$. Taking the three dimensional Minkowski slice which corresponds to real $\lambda,\Bar{\mu}$ and $x=x^*$ (see the reality conditions \eqref{MinkowskiReality}), we obtain twistors that are valued in $\mathbb{RP}^3\subset \mathbb{CP}^3$ which are the twistors used in the main-text\footnote{In the main text, we understood this as resulting from a dimensional reduction from four dimensional Klein space. Here, we obtain a more intrinsic three dimensional picture of the twistor space construction.}.

Let us now compare and contrast this construction with the four-dimensional case. There, one can define twistor space as the space of all complex lines through the origin, see for instance \cite{Adamo:2017qyl}. This construction naturally leads to a twistor that is in the fundamental representation of the complexified conformal group $SL(4,\mathbb{C})$ in contrast to the $Sp(4,\mathbb{C})$ representation we obtained in three dimensions. The key difference is rather than \eqref{Zinmulambda1}, the four dimensional twistor is written as,
\begin{align}\label{4dtwistorinspinors}
    Z^A=\lambda^a\oplus \tilde{\mu}_{\Dot{a}},
\end{align}
 where $\lambda$ and $\tilde{\mu}$ are chiral and anti-chiral spinors of $\text{Spin}(4,\mathbb{C})$, which is the complexified rotation group in four dimensions.
 
In three dimensions, the analogous procedure does not lead to the kind of twistor we obtained above.
Instead, it leads to the so called mini-twistor space, see for instance \cite{CarrilloGonzalez:2022ggn}. We do not pursue that approach in this paper but rather work from the conformal group perspective as conformal correlators are the objects of interest here.

\section{Spinor helicity and reality conditions}\label{subsec:spinorhelicity}
Given a general three momentum, we can trade it for a pair of spinors $\lambda,\Bar{\lambda}$ which are defined implicitly via,
\begin{align}\label{SHvardef}
    p_{\mu}=\frac{1}{2}(\sigma_\mu)^a_b\lambda_{a}\Bar{\lambda}^b,
\end{align}
 and hence the momenta, $p_{\mu}$, are invariant under the little group transformations,
 \begin{align}\label{3deuclidlittlegroup}
     \lambda_{a}\to e^{-\frac{i\theta}{2}}\lambda_{a},\Bar{\lambda}_{a}\to e^{\frac{i\theta}{2}}\Bar{\lambda}_{a}.
 \end{align}
 At this point, it is important to make a distinction between three dimensional Euclidean and Minkowskian signature.
\subsection*{Reality conditions: Euclidean signature}
 The momenta in matrix form are given by,
 \begin{align}
     \slashed{p}=\begin{pmatrix}
         p_{z}& p_{x}-i p_{y}\\
         p_{x}+i p_{y}& -p_{z}.
     \end{pmatrix}
 \end{align}For Euclidean momenta which have $p_{x},p_{y},p_{z}\in \mathbb{R}$ we have the reality condition $\slashed{p}^\dagger=\slashed{p}$ which using \eqref{SHvardef} implies,
 \begin{align}\label{EuclideanReality}
     \lambda_{a}^\dagger=\Bar{\lambda}^a~,~\Bar{\lambda}_{}^{a\dagger}=\lambda_{a}.
 \end{align}
 Thus, for real Euclidean momenta, the spinors $\lambda$ and $\Bar{\lambda}$ are Hermitian conjugates of each other\footnote{This is reminiscent of the notion of conjugation in four dimensional Minkowski space, $\mathbb{R}^{3,1}$ as we discuss in appendix \ref{app:4dto3d}. }.

\section*{Reality conditions: Lorentzian signature}
 On the other hand, in Minkowski space we have $p_{x},p_{z}\in \mathbb{R}$ but $p_{y}=-i p_{t},p_{t}\in\mathbb{R}$ and thus the reality condition is instead $p^*=p$. This implies\footnote{By solving for $\lambda,\Bar{\lambda}$ in terms of the components of $p$, one sees that this is true only for spacelike momenta as also noted by \cite{Baumann:2024ttn}. For our purposes, we understand this reality conditions as arising from the dimensional reduction of a four dimensional Klein space momenta as detailed in appendix \ref{app:4dto3d}.},
  \begin{align}\label{MinkowskiReality}
     \lambda_{a}^*=\lambda_{a}~,~\Bar{\lambda}^{a*}=\Bar{\lambda}^a,
 \end{align}
thus showing that for real Minkowskian momenta, the spinors $\lambda$ and $\Bar{\lambda}$ are real and independent. This can be understood as a fact inherited via a dimensional reduction of a null Kleinian momentum as we discuss in appendix \ref{app:4dto3d}. Note that the Lorentzian momentum,
\begin{align}\label{Lorentz3momenta}
    p_{ab}=\frac{1}{2}(\lambda_{(a}\Bar{\lambda}_{b)}),
\end{align}
is invariant under $\lambda\to \frac{1}{r}\lambda,\Bar{\lambda}\to r \Bar{\lambda}$, $r\in \mathbb{R}$ which is a $\mathbb{R}$ transformation in contrast to the $U(1)$ little group redundancy in the Euclidean case \eqref{3deuclidlittlegroup}.

\section{CPT invariance in spinor helicity and twistor space}\label{app:CPTinv}
In this appendix, we discuss in detail the action of $CPT$ in spinor helicity variables and discuss its implication on the twistor and dual-twistor space correlation functions. For the correlators that we study, $CPT$ is simply $PT$ as all our correlators are even under $C$\footnote{For example a $U(1)$ current is odd under charge conjugation but the correlators that we consider for instance are of the form $\langle TJJ\rangle$ where $T$ is the stress tensor which is even under charge conjugation and the presence of two insertions of $J$ makes the entire correlator even under charge conjugation. If one were to study correlators like $\langle JJJ\rangle$, one would have to resort to making the currents non-Abelian. This is a simple generalization of our results so we do not present such an analysis here.}.
\subsection{Parity and time reversal in spinor helicity variables}
We work in $\mathbb{R}^{2,1}$ spanned by the coordinates $(t,x,z)$. A parity transformation corresponds to\footnote{This transformation is equivalent to flipping $x\to -x$ by a $SO(2)$ rotation.},
\begin{align}\label{parity1}
    P:(t,x,z)\to (t,x,-z).
\end{align}
Time-reversal naturally corresponds to,
\begin{align}\label{timereversal1}
    T:(t,x,z)\to (-t,x,z).
\end{align}
Our aim in this section is to derive the action of these transformations in momentum space and in particular, in spinor helicity variables. Recall that in three dimensional spinor helicity variables, 
\begin{align}
    \slashed{p}_a^b=\frac{1}{2}\big(\lambda_a\Bar{\lambda}^b+\Bar{\lambda}_a\lambda^b\big).
\end{align}
In matrix notation it reads,
\begin{align}\label{momentummatrix1}
    \slashed{p}=\begin{pmatrix}
        p_z&&p_x-p_t\\
        p_x+p_t&& -p_z
    \end{pmatrix}=\begin{pmatrix}
        \frac{\lambda_1\Bar{\lambda}_2+\Bar{\lambda}_1\lambda_2}{2}&&-\lambda_1\Bar{\lambda}_1\\
        \lambda_2\Bar{\lambda}_2&&-\frac{\lambda_2\Bar{\lambda}_1+\Bar{\lambda}_2\lambda_1}{2}
    \end{pmatrix}.
\end{align}
The subscripts $1,2$ here refer to the spinor components.
We shall now consider the action of parity and time-reversal on the momentum matrix and derive the corresponding actions on the spinors $\lambda$ and $\Bar{\lambda}$.
\subsubsection{Parity}
The parity transformation \eqref{parity1} acts on the momentum matrix by flipping $p_z\to -p_z$:
\begin{align}
    P:\begin{pmatrix}
        p_z&&p_x-p_t\\
        p_x+p_t&& -p_z
    \end{pmatrix}\to \begin{pmatrix}
        -p_z&&p_x-p_t\\
        p_x+p_t&& p_z
    \end{pmatrix}.
\end{align}
Using the expressions of each component of this matrix in terms of the spinors \eqref{momentummatrix1}, we can write this transformation as follows:
\begin{align}
    P:\big\{\lambda_1\Bar{\lambda}_2+\Bar{\lambda}_1\lambda_2\to-\lambda_1\Bar{\lambda}_2-\Bar{\lambda}_1\lambda_2,\lambda_1\Bar{\lambda}_1\to\lambda_1\Bar{\lambda}_1,\lambda_2\Bar{\lambda}_2\to \lambda_2\Bar{\lambda}_2,\lambda_2\Bar{\lambda}_1+\Bar{\lambda}_2\lambda_1\to -\lambda_2\Bar{\lambda}_1-\Bar{\lambda}_2\lambda_1\big\}.
\end{align}
This can be implemented by,
\begin{align}\label{parityonspinors}
    P:\begin{pmatrix}
        \lambda_1,&\lambda_2
    \end{pmatrix}\to\begin{pmatrix}
        -\Bar{\lambda}_1,&\Bar{\lambda}_2
    \end{pmatrix},\begin{pmatrix}
        \Bar{\lambda}_1,&\Bar{\lambda}_2
    \end{pmatrix}\to\begin{pmatrix}
        -\lambda_1,&\lambda_2
    \end{pmatrix}.
\end{align}
Let us see what this entails for spinor brackets: First, consider the contraction between two unbarred spinors $\lambda$ and $\chi$
\begin{align}
    \langle \lambda \chi\rangle=\lambda_1\chi_2-\lambda_2\chi_1.
\end{align}
Under parity \eqref{parityonspinors}, we see that,
\begin{align}
    P:\langle \lambda \chi\rangle\to \langle \Bar{\chi}\Bar{\lambda}\rangle~,~P:\langle \Bar{\lambda}\Bar{\chi}\rangle\to \langle \chi \lambda\rangle~,~P:p=-\frac{1}{2}\langle \lambda \Bar{\lambda}\rangle\to p.
\end{align}
\subsubsection{Time-reversal}
The time-reversal transformation \eqref{timereversal1} acts on the momentum matrix by flipping $p_z\to -p_z$ and $p_x\to -p_x$\footnote{Recall that time-reversal is anti-unitary and also does not flip the time component of the momentum.}:
\begin{align}
    T:\begin{pmatrix}
        p_z&&p_x-p_t\\
        p_x+p_t&& -p_z
    \end{pmatrix}\to \begin{pmatrix}
        -p_z&&-p_x-p_t\\
        -p_x+p_t&& p_z
    \end{pmatrix}.
\end{align}
Using the expressions of each component of this matrix in terms of the spinors \eqref{momentummatrix1}, we can write this transformation as follows:
\begin{align}
    T:\big\{\lambda_1\Bar{\lambda}_2+\Bar{\lambda}_1\lambda_2\to-\lambda_1\Bar{\lambda}_2-\Bar{\lambda}_1\lambda_2,\lambda_1\Bar{\lambda}_1\to\lambda_2\Bar{\lambda}_2,\lambda_2\Bar{\lambda}_2\to \lambda_1\Bar{\lambda}_1,\lambda_2\Bar{\lambda}_1+\Bar{\lambda}_2\lambda_1\to -\lambda_2\Bar{\lambda}_1-\Bar{\lambda}_2\lambda_1\big\},
\end{align}
which can be implemented via,
\begin{align}
     T:\begin{pmatrix}
        \lambda_1,&\lambda_2
    \end{pmatrix}\to\begin{pmatrix}
        \lambda_2,&-\lambda_1
    \end{pmatrix},\begin{pmatrix}
        \Bar{\lambda}_1,&\Bar{\lambda}_2
    \end{pmatrix}\to\begin{pmatrix}
        \Bar{\lambda}_2,&-\Bar{\lambda}_1
        \end{pmatrix}.
\end{align}
Spinor products are invariant under time reversal,
\begin{align}
    T:\langle \lambda \chi\rangle\to \langle \lambda \chi\rangle, \langle \Bar{\lambda}\Bar{\chi}\rangle\to \langle \Bar{\lambda}\Bar{\chi}\rangle,
\end{align}
and it preserves the momentum magnitude as well:
\begin{align}
    T: p=-\frac{1}{2}\langle \lambda \Bar{\lambda}\rangle \to p.
\end{align}
\subsubsection{PT}
To summarize, we have found that the $P$ and $T$ transformations act on the spinor brackets as follows:
\begin{align}\label{PandTspinorsfinal}&P\big(\langle \lambda \chi\rangle\big)=\langle \Bar{\chi}\Bar{\lambda}\rangle~,~P\big(\langle \Bar{\lambda}\Bar{\chi}\rangle\big)=\langle \chi \lambda\rangle~,~P\big(\langle \lambda \Bar{\lambda}\rangle\big)=\langle \lambda \Bar{\lambda}\rangle,\notag\\
        & T\big(\langle \lambda \chi\rangle\big)=\langle \lambda \chi\rangle~,~T\big(\langle \Bar{\lambda} \Bar{\chi}\rangle\big)=\langle \Bar{\lambda} \Bar{\chi}\rangle~,~T\big(\langle \lambda \Bar{\lambda}\rangle\big)=\langle \lambda \Bar{\lambda}\rangle.
    \end{align}
The combined $PT$ transformation thus acts as,
\begin{align}\label{PTspinors}
        PT\big(\langle \lambda \chi\rangle\big)=\langle \Bar{\chi}\Bar{\lambda}\rangle~,~PT\big(\langle\Bar{\lambda}\Bar{\chi}\rangle\big)=\langle \chi \lambda\rangle~,~ PT\big(\langle \lambda \Bar{\lambda}\rangle\big)=\langle \lambda \Bar{\lambda}\rangle.
    \end{align}
We shall investigate the behavior of two and three point current correlators in spinor helicity variables under these transformations now.
\subsection{Action on correlators}
One can classify correlation functions into two classes based on their behaviour under parity and time-reversal: Those that are odd under both parity and time-reversal and those that are even (the precise meaning of odd and even will be clear 
\subsection*{Parity even}
Let us first focus on two point functions. Consider the even spin $s$ conserved current two point function. In the two independent helicity configurations we have,
\begin{align}
    \langle\langle 0| J_s^{-}(p_1)J_s^{-}(p_2)|0\rangle\rangle_{\text{even}}=c_e \frac{\langle 1 2\rangle^{2s}}{p_1}~,~\langle\langle 0| J_s^{+}(p_1)J_s^{+}(p_2)|0\rangle\rangle_{\text{even}}=c_e \frac{\langle \Bar{1} \Bar{2}\rangle^{2s}}{p_1},
\end{align}
where $c_e\in \mathbb{R}$. Under a parity transformation \eqref{PandTspinorsfinal} we see that,
\begin{align}
    P:\langle\langle 0|J_s^{-}(p_1)J_s^{-}(p_2)|0\rangle\rangle_{\text{even}}=c_e \frac{\langle 1 2\rangle^{2s}}{p_1}\to (-1)^{2s}c_e \frac{\langle \Bar{1}\Bar{2}\rangle^{2s}}{p_1}=\langle\langle 0|J_s^{+}(p_1)J_s^{+}(p_2)|0\rangle\rangle_{\text{even}},
\end{align}
where we used the fact that $(-1)^{2s}=1$ since $s$ is a positive integer. Thus, the parity operation converts the even $(--)$ two point function into the even $(++)$ two point function. Similarly, under a time-reversal transformation \eqref{PandTspinorsfinal}, we see that the $(--)$ correlator is invariant:
\begin{align}
    T:\langle\langle 0|J_s^{-}(p_1)J_s^{-}(p_2)|0\rangle\rangle_{\text{even}}=c_e \frac{\langle 1 2\rangle^{2s}}{p_1}\to c_e \frac{\langle 1 2\rangle^{2s}}{p_1}=\langle\langle 0|J_s^{-}(p_1)J_s^{-}(p_2)|0\rangle\rangle_{\text{even}}.
\end{align}
Therefore, under time-reversal, the even $(--)$ two point function (and similarly the $(++)$ two point function) remains invariant.
Moving onto three point functions, let us focus on the $(---)$ and $(+++)$ helicities first. They take the form (both homogeneous and non-homogeneous),
\begin{align}
    &\langle\langle 0|J_{s_1}^{-}(p_1)J_{s_2}^{-}(p_2)J_{s_3}^{-}(p_3)|0\rangle\rangle=c_{s_1s_2s_3}\langle 1 2\rangle^{s_1+s_2-s_3}\langle 2 3\rangle^{s_2+s_3-s_1}\langle 3 1\rangle^{s_3+s_2-s_1}f(p_1,p_2,p_3),\notag\\
    &\langle\langle 0|J_{s_1}^{+}(p_1)J_{s_2}^{+}(p_2)J_{s_3}^{+}(p_3)|0\rangle\rangle=c_{s_1s_2s_3}\langle \Bar{1} \Bar{2}\rangle^{s_1+s_2-s_3}\langle \Bar{2} \Bar{3}\rangle^{s_2+s_3-s_1}\langle \Bar{3} \Bar{1}\rangle^{s_3+s_2-s_1}f(p_1,p_2,p_3)
\end{align}
where the function $f$ is different for the homogeneous and non-homogeneous correlators. Under time-reversal \eqref{PandTspinorsfinal}, it is clear that both these correlators are invariant. Under parity, they get interchanged. Thus, this correlator is even under $P$ and $T$ individually and thus even under $PT$ \eqref{PTspinors}.
\subsection*{Parity odd}
The parity odd two point function is given by,
\begin{align}
    \langle\langle 0| J_s^{-}(p_1)J_s^{-}(p_2)|0\rangle\rangle_{\text{odd}}=-ic_o \frac{\langle 1 2\rangle^{2s}}{p_1}~,~\langle\langle 0| J_s^{+}(p_1)J_s^{+}(p_2)|0\rangle\rangle_{\text{odd}}=+ic_o \frac{\langle \Bar{1} \Bar{2}\rangle^{2s}}{p_1},
\end{align}
where $c_o\in \mathbb{R}$. Under a parity transformation \eqref{PandTspinorsfinal} we see that,
\begin{align}
    P:\langle\langle 0|J_s^{-}(p_1)J_s^{-}(p_2)|0\rangle\rangle_{\text{odd}}=-ic_o \frac{\langle 1 2\rangle^{2s}}{p_1}\to -ic_o(-1)^{2s}c_o \frac{\langle \Bar{1}\Bar{2}\rangle^{2s}}{p_1}=-\langle\langle 0|J_s^{+}(p_1)J_s^{+}(p_2)|0\rangle\rangle_{\text{odd}},
\end{align}
where we used the fact that $(-1)^{2s}=1$ since $s$ is a positive integer. Thus, the parity operation converts the odd $(--)$ two point function into the negative of the odd $(++)$ two point function. Thus this correlator is odd under parity. Similarly, under a time-reversal transformation \eqref{PandTspinorsfinal}, we see that the $(--)$ and $(++)$ correlators flip sign due to the anti-linearity of the $T$ operation.
\begin{align}
    T:\langle\langle 0|J_s^{-}(p_1)J_s^{-}(p_2)|0\rangle\rangle_{\text{odd}}=-ic_o \frac{\langle 1 2\rangle^{2s}}{p_1}\to +i c_o\frac{\langle 1 2\rangle^{2s}}{p_1}=-\langle\langle 0|J_s^{-}(p_1)J_s^{-}(p_2)|0\rangle\rangle_{\text{odd}}.
\end{align}
Therefore, under time-reversal, the odd $(--)$ two point function (and similarly the $(++)$ two point function) flip a sign. However, we do see that the odd correlator is even under under the composition $PT$ \eqref{PTspinors}.

Moving onto three point functions, let us focus on the $(---)$ and $(+++)$ helicities first. They take the form,
\begin{align}
    &\langle\langle 0|J_{s_1}^{-}(p_1)J_{s_2}^{-}(p_2)J_{s_3}^{-}(p_3)|0\rangle\rangle=-ic_{s_1s_2s_3}\langle 1 2\rangle^{s_1+s_2-s_3}\langle 2 3\rangle^{s_2+s_3-s_1}\langle 3 1\rangle^{s_3+s_2-s_1}f(p_1,p_2,p_3),\notag\\
    &\langle\langle 0|J_{s_1}^{+}(p_1)J_{s_2}^{+}(p_2)J_{s_3}^{+}(p_3)|0\rangle\rangle=+ic_{s_1s_2s_3}\langle \Bar{1} \Bar{2}\rangle^{s_1+s_2-s_3}\langle \Bar{2} \Bar{3}\rangle^{s_2+s_3-s_1}\langle \Bar{3} \Bar{1}\rangle^{s_3+s_2-s_1}f(p_1,p_2,p_3)
\end{align}
Under time-reversal \eqref{PandTspinorsfinal}, it is clear that both these correlators flip sign due to the anti-linearity of $T$. Under parity, they get interchanged with an extra sign. Thus, this correlator is odd under $P$ and $T$ individually. However, it is then easy to check that it is even under the combination $PT$ \eqref{PTspinors}.

\subsection{$CPT$ in twistor space}
Lets now see how the $CPT$ invariance of spinor helicity correlators translates to twistor space. Lets start with two point functions. Let $p_{ab}=\frac{\lambda_{(a}\Bar{\lambda}_{b)}}{2}$ and $q_{ab}=\frac{\rho_{(a}\Bar{\rho}_{b)}}{2}$ be the two momenta. We have the half-Fourier transform formula for the $(--)$ helicity two point function.
\begin{align}
    \frac{\langle \lambda \rho\rangle^{2s}}{\langle \lambda \Bar{\lambda}\rangle^{2s-1}}\delta^3(p+q)=\int d^2\Bar{\mu}d^2\Bar{\nu}\frac{e^{-i\Bar{\lambda}\cdot\Bar{\mu}-i\Bar{\rho}\cdot \Bar{\nu}}}{(Z_1\cdot Z_2)^{-2s+2}}=\int d^2\Bar{\mu}d^2\Bar{\nu}\frac{e^{-i\Bar{\lambda}\cdot\Bar{\mu}-i\Bar{\rho}\cdot \Bar{\nu}}}{(\lambda\cdot\Bar{\nu}-\rho\cdot\Bar{\mu})^{-2s+2}}.
\end{align}
The $PT$ transformation \eqref{PTspinors} implies that the LHS transforms as follows,
\begin{align}
PT\bigg(\frac{\langle \lambda \rho\rangle^{2s}}{\langle \lambda \Bar{\lambda}\rangle^{2s-1}}\delta^3(p+q)\bigg)=\frac{\langle \Bar{\lambda} \Bar{\rho}\rangle^{2s}}{\langle \lambda \Bar{\lambda}\rangle^{2s-1}}\delta^3(p+q).
\end{align}
Let us see how the RHS changes under this transformation. First, we write the RHS component-wise for convenience as,
\begin{align}
    \int d\Bar{\mu}_1d\Bar{\mu}_2d\Bar{\nu}_1d\Bar{\nu}_2\frac{e^{-i\Bar{\lambda}_1\Bar{\mu}_2+i\Bar{\lambda}_2\Bar{\mu}_1-i\Bar{\rho}_1\Bar{\nu}_2+i\Bar{\rho}_2\Bar{\nu}_1}}{(\lambda_1\Bar{\nu}_2-\lambda_2\Bar{\nu}_1-\rho_1\Bar{\mu}_2+\rho_2\Bar{\mu}_1)^{-2s+2}}.
\end{align}
Under a $PT$ transformation \eqref{PTspinors} it turns into,
\begin{align}
    \int d\Bar{\mu}_1d\Bar{\mu}_2d\Bar{\nu}_1d\Bar{\nu}_2\frac{e^{i\lambda_2\Bar{\mu}_2-i\lambda_1\Bar{\mu}_1+i\rho_2\Bar{\nu}_2-i\rho_1\Bar{\nu}_1}}{(-\Bar{\lambda}_2\Bar{\nu}_2+\Bar{\lambda}_1\Bar{\nu}_1+\Bar{\rho}_2\Bar{\mu}_2-\Bar{\rho}_1\Bar{\mu}_1)^{-2s+2}}.
\end{align}
Since $\Bar{\mu},\Bar{\nu}$ are dummy variables we make the variable change,
\begin{align}
    (\Bar{\mu}_1,\Bar{\mu}_2)\to(-\mu_2,-\mu_1)~,~(\Bar{\nu}_1,\Bar{\nu}_2)\to(-\nu_2,-\nu_1).
\end{align}
We can then write the PT transformed RHS as,
\begin{align}
    &\int d\Bar{\mu}_1d\Bar{\mu}_2d\Bar{\nu}_1d\Bar{\nu}_2\frac{e^{-i(\lambda_1\mu_2-\lambda_2\mu_1)-i(\rho_1\nu_2-\rho_2\nu_1)}}{(\Bar{\lambda}_2\nu_1-\Bar{\lambda}_1\nu_2-\Bar{\rho}_2\mu_1+\Bar{\rho}_1\mu_2)^{-2s+2}}=\int d^2\mu d^2 \nu\frac{e^{i\lambda\cdot\mu+i\rho\cdot \nu}}{(\Bar{\lambda}\cdot \nu-\Bar{\rho}\cdot \mu)^{2s+2}}\notag\\
    &=\int d^2\mu d^2 \nu\frac{e^{i\lambda\cdot\mu+i\rho\cdot \nu}}{(W_1\cdot W_2)^{-2s+2}}.
\end{align}
To summarize we have obtained the following result:
\begin{align}
    &PT\bigg(\frac{\langle \lambda \rho\rangle^{2s}}{\langle \lambda \Bar{\lambda}\rangle^{2s-1}}\delta^3(p+q)=\int d^2\Bar{\mu}d^2\Bar{\nu}\frac{e^{-i\Bar{\lambda}\cdot\Bar{\mu}-i\Bar{\rho}\cdot \Bar{\nu}}}{(Z_1\cdot Z_2)^{-2s+2}}\bigg)\notag\\
    &\implies \frac{\langle \Bar{\lambda} \Bar{\rho}\rangle^{2s}}{\langle \lambda \Bar{\lambda}\rangle^{2s-1}}\delta^3(p+q)=\int d^2\mu d^2 \nu\frac{e^{i\lambda\cdot\mu+i\rho\cdot \nu}}{(W_1\cdot W_2)^{-2s+2}}.
\end{align}
In words, the fact that the $(--)$ helicity correlator can be written as a half-Fourier transform of the twistor correlator implies that the $(++)$ helicity correlator can be written as a half-Fourier transform of the dual-twistor correlator.

At the level of three points a similar exercise to the one above yields,
\begin{align}
    &PT\bigg(\langle 12\rangle^{s_1+s_2-s_3}\langle 2 3\rangle^{s_2+s_3-s_1}\langle 3 1\rangle^{s_3+s_1-s_2}f(p_1,p_2,p_3)=\int d^2 \Bar{\mu}_1 d^2 \Bar{\mu}_2 d^2 \Bar{\mu}_3 e^{-i\Bar{\lambda}_1\cdot\Bar{\mu}_1-i\Bar{\lambda}_2\cdot\Bar{\mu}_2-i\Bar{\lambda}_3\cdot\Bar{\mu}_3}F(Z_1,Z_2,Z_3)\bigg)\notag\\
    &\equiv \langle \Bar{1}\Bar{2}\rangle^{s_1+s_2-s_3}\langle \Bar{2} \Bar{3}\rangle^{s_2+s_3-s_1}\langle \Bar{3} \Bar{1}\rangle^{s_3+s_1-s_2}f(p_1,p_2,p_3)=\int d^2 \mu_1 d^2 \mu_2 d^2 \mu_3 e^{i\lambda_1\cdot\mu_1+i\lambda_2\cdot\mu_2+i\lambda_3\cdot\mu_3}F(W_1,W_2,W_3).
\end{align}
The function $f(p_1,p_2,p_3)$ can be the one associated to either the homogeneous or non-homogeneous correlator and $F(Z_1,Z_2,Z_3)$ is the associated twistor space function. The point is that for both cases, we see that the PT transformation of the $(---)$ helicity correlator written as a half-Fourier transform of a twistor integral is the $(+++)$ helicity correlator written as a half-Fourier transform of a dual-twistor integral. A similar exercise can be carried out in the remaining helicities.

\section{Details of half-Fourier transform from twistor space to spinor helicity variables}\label{app:HalfFourierTransform}
In this appendix, we shall present the explicit calculations of the half-Fourier transform required to obtain spinor helicity results from the twistor space expressions.
Let us begin with two-point functions.
\subsection{Two Point Function}\label{twopointdetail}
   Consider the two-point function in $(++)$ helicity for the $W$ representation, i.e. \eqref{twistorW2point}. We can convert it to spinor-helicity by performing the appropriate inverse half-Fourier transform of \eqref{TwistorTrans1}. Thus, we have:
       \begin{align}
         \langle0|J^{+}_{s}(\lambda_{1},\bar{\lambda}_{1})J^{+}_{s}(\lambda_{2},\bar{\lambda}_{2})|0\rangle =  \int d^{2}\mu_{1}d^{2}\mu_{2} e^{\textit{i}\lambda_{1}\cdot\mu_{1} +\textit{i}\lambda_{2}\cdot\mu_{2} } \langle0|\Tilde{J}^{+}_{s}(W_{1})\Tilde{J}^{+}_{s}(W_{2})|0\rangle,
       \end{align}
       This integral can be computed using the Schwinger parametrization given by -
    \begin{align}
        \frac{1}{x^{n}} = \frac{\pi}{\textit{i}^{n} (n-1)!} \int \frac{dk}{2\pi} k^{n-1} Sgn(k) e^{-\textit{i} k x}  , \,\,\, \text{for n}  >0   
    \end{align}
    
    as outlined in \cite{Baumann:2024ttn}. We have,
        \begin{align}\label{W2pt1}
            &\int d^{2}\mu_{1}d^{2}\mu_{2} \frac{1}{(W_{1}\cdot W_{2})^{2(-s+1)}} e^{\textit{i}\lambda_1\cdot\mu_1 + \textit{i}\lambda_2\cdot\mu_2} \\ \notag
         =& \frac{\pi}{\textit{i}^{-2s+2} (-2s +1)!} \int_{-\infty}^{+\infty} \frac{dc_{12}}{2\pi} \int d^{2}\mu_{1}d^{2}\mu_{2}\, \text{sign}(c_{12})c_{12}^{-2s+1} e^{\textit{i}c_{12} W_{1} \cdot W_{2}+\textit{i}\lambda_1\cdot\mu_1 + \textit{i}\lambda_2\cdot\mu_2}  \\ \notag
            =&\frac{\pi}{\textit{i}^{-2s+2} (-2s+1)!} \int \frac{dc_{12}}{2\pi} (2\pi)^{4} \text{sign}(c_{12})c_{12}^{-2s+1} \delta^{2}(-c_{12} \bar{\lambda}_{2}^{a}+ \lambda_{1}^{a})\delta^{2}(c_{12} \bar{\lambda}_{1}^{a}+\lambda_{2}^{a}),
            \end{align}
where we used \eqref{twistordotprods}. Working in the basis of $\bar{\lambda}_{1}$ and $\bar{\lambda}_{2}$ i.e.
            \begin{align}
\lambda_{2}^{a}= a_2 \bar{\lambda}_{1}^{a}+b_{2} \bar{\lambda}_{2}^{a} = \frac{\langle\bar{2}2\rangle}{\langle\bar{2}\bar{1}\rangle} \bar{\lambda}_{1}^{a} + \frac{\langle\bar{1}2\rangle}{\langle\bar{1}\bar{2}\rangle} \bar{\lambda}_{2}^{a},\\  \notag
             \lambda_{1}^{a}= a_1 \bar{\lambda}_{1}^{a}+b_{1} \bar{\lambda}_{2}^{a} = \frac{\langle\bar{2}1\rangle}{\langle\bar{2}\bar{1}\rangle} \bar{\lambda}_{1}^{a} + \frac{\langle\bar{1}1\rangle}{\langle\bar{1}\bar{2}\rangle} \bar{\lambda}_{2}^{a},
        \end{align}
and writing the delta function in \eqref{W2pt1} in terms of components in a particular basis, we obtain the result\footnote{We also use the formula,\begin{align}
&\delta^3(\vec{p_{1}}+\vec{p_{2}}) = \delta^3(\lambda_1^{(a}\bar{\lambda}_1^{b)}+\lambda_2^{(a}\bar{\lambda}_2^{b)} )  
        =\delta^3(c_1 \bar{\lambda}_1^{a}\bar{\lambda}_1^{b}+ c_2\bar{\lambda}_2^{a}\bar{\lambda}_2^{b}+ c_3\bar{\lambda}_1^{a}\bar{\lambda}_2^{b} )=\frac{1}{|\langle \bar{1}\bar{2}\rangle|^3} \delta(c_1)\delta(c_2)\delta(c_3),
         \end{align}
where, $c_1, c_2$ and $c_3$ are given by,
\begin{align} c_1=\frac{\langle\bar{2}1\rangle}{\langle\bar{2}\bar{1}\rangle},c_2=\frac{\langle\bar{1}2\rangle}{\langle\bar{1}\bar{2}\rangle},
c_3=\frac{\langle\bar{1}1\rangle}{\langle\bar{1}\bar{2}\rangle}+\frac{\langle\bar{2}2\rangle}{\langle\bar{2}\bar{1}\rangle}.
\end{align} },
        \begin{align}\label{WtwistorBacktoSH2pt}
             \notag \langle0|J^{+}_{s}(\lambda_{1},\bar{\lambda}_{1})J^{+}_{s}(\lambda_{2},\bar{\lambda}_{2})|0\rangle
             &=\frac{\pi}{\textit{i}^{-2s+2} (-2s+1)!}  (2\pi)^{3} \frac{\text{sign(2E)}}{(\text{sign}\langle\bar{1}\bar{2}\rangle)^{2}} \frac{\langle\bar{1}\bar{2}\rangle^{+2s}}{(2 p_{1})^{+2s-1}} \frac{\delta^{3}(\vec{p_{1}}+\vec{p_{2}})}{4} \\ 
             &=\frac{\pi}{\textit{i}^{-2s+2} (-2s+1)!}  (2\pi)^{3} \frac{\langle\bar{1}\bar{2}\rangle^{+2s}}{(2 p_{1})^{+2s-1}} \frac{\delta^{3}(\vec{p_{1}}+\vec{p_{2}})}{4} .
        \end{align}
Since we are in Minkowski space, the reality condition on the spinors $\lambda,\Bar{\lambda}$ (see \eqref{MinkowskiReality}) implies (sign$\langle\bar{1}\bar{2}\rangle)^2$= 1.

Similarly, the $(++)$ two point function in  the $Z$ representation can also be converted back into spinor helicity yielding,
       \begin{align}\label{ZtwistorBacktoSH2pt}
           \langle0|J^{+}_{s}(\lambda_{1},\bar{\lambda}_{1})J^{+}_{s}(\lambda_{2},\bar{\lambda}_{2})|0\rangle &=  \int d^{2}\bar{\mu}_{1}d^{2}\bar{\mu}_{2} e^{-\textit{i}\bar{\lambda}_{1}\cdot\bar{\mu}_{1} -\textit{i}\bar{\lambda}_{2}\cdot\bar{\mu}_{2} } \langle0|\hat{J}^{+}_{s}(Z_{1})\hat{J}^{+}_{s}(Z_{2})|0\rangle \notag \\
           &= (2\pi)^3 \frac{\delta^3(\vec{p_1}+\vec{p_2})}{4} \frac{\pi}{\textit{i}^{2s+2} (2s+1)!} \frac{\langle\bar{1}\bar{2}\rangle^{2s}}{(2p)^{2s-1}} .
       \end{align}
Both \eqref{WtwistorBacktoSH2pt} and \eqref{ZtwistorBacktoSH2pt} are equal up to an overall constant. 
We can verify that \eqref{WtwistorBacktoSH2pt} and \eqref{ZtwistorBacktoSH2pt} agree with the general structure of the two-point function \eqref{general2pointSH}.
Similarly, we can also obtain the spinor helicity expressions for (-~-) helicity.\\
Finally, we discuss the possibility of a mixed helicity two point function. Twistor space allows for such an expression but let us see what it corresponds to in spinor helicity variables. Performing the inverse half-Fourier transforms given in \eqref{TwistorTrans1} we obtain,
{\small
   \begin{align}
      &\langle0|J_{s}^{+}(\lambda_{1},\bar{\lambda}_{1})J_{s}^{-}(\lambda_{2},\bar{\lambda}_{2})|0\rangle \notag \\
       & = \int d^{2}\mu_{2}d^{2}\bar{\mu}_{1}\langle0|\hat{J}_{s}^{+}(Z_{1})\Tilde{J}_{s}^{-}(W_{2})|0\rangle  e^{-\textit{i} \lambda_2 \cdot \mu_2 - \textit{i}\bar{\lambda}_1\cdot\bar{\mu}_1}=\int d^{2}\mu_{2}d^{2}\bar{\mu}_{1}\frac{1}{(Z_{1}\cdot W_{2})^{2(s+1)}}  e^{\textit{i} \lambda_2 \cdot \mu_2 - \textit{i}\bar{\lambda}_1\cdot\bar{\mu}_1}\notag \\ 
       & \propto \frac{\langle1\bar{1}\rangle^{2s+1}}{\langle1\bar{2}\rangle^{2s+1}} |\langle1\bar{2}\rangle|^{2} \delta\left( \langle12\rangle \right)\delta\left(\langle\bar{1}\bar{2}\rangle\right) \delta\left( \langle2\bar{2}\rangle+ \langle1\bar{1}\rangle\right)\propto \frac{\langle1\bar{1}\rangle^{2s+1}}{\langle1\bar{2}\rangle^{2s+1}} |\langle1\bar{2}\rangle|^{2} \delta\left( \langle12\rangle \right)\delta\left(\langle\bar{1}\bar{2}\rangle\right) \delta\left(p_1+p_2\right)  \sim 0   \, .
       \end{align}}
\normalsize
The above expression is zero because it has support only for zero momentum. This is due to the fact that one of the delta function constraints sets $p_1+p_2=0$ which for space-like momenta implies that $p_1$ and $p_2$ are both zero. Thus, we do not consider such solutions in our analysis.
\subsection{Three points}\label{threepointdetail}
Let us now move on to the case of three point functions. As an illustrative example, we consider the case of a mixed helicity configuration. We have,
    \begin{align}\label{G.11}
 &\langle 0|J_{s_{1}}^{+}(\lambda_{1},\bar{\lambda}_{1})J_{s_{2}}^{+}(\lambda_{2},\bar{\lambda}_{2})J_{s_{1}}^{-}(\lambda_{3},\bar{\lambda}_{3})|0\rangle \\ \notag&=\int d^{2}\bar{\mu}_{1}d^{2}\bar{\mu}_{2}d^{2}\mu_{3} \langle0|\hat{J}_{s_{1}}^{+}(Z_{1})\hat{J}_{s_{2}}^{+}(Z_{2})\Tilde{J}_{s_{1}}^{-}(W_{3})|0\rangle e^{-\textit{i}\bar{\lambda}_1\cdot\bar{\mu}_1-\textit{i}\bar{\lambda}_2\cdot\bar{\mu}_2+\textit{i} \lambda_3\cdot \mu_3} \\ \notag&=\int d^{2}\bar{\mu}_{1}d^{2}\bar{\mu}_{2}d^{2}\mu_{3}  \delta^{[s_{1}+s_{2}-s_{3}]}(Z_{1}\cdot Z_{2})\delta^{[s_{2}+s_{3}-s_{1}]}(Z_{2}\cdot W_{3})\delta^{[s_{3}+s_{1}-s_{2}]}(W_{3}\cdot Z_{1}) e^{-\textit{i}\bar{\lambda}_1\cdot\bar{\mu}_1-\textit{i}\bar{\lambda}_2\cdot\bar{\mu}_2+\textit{i} \lambda_3\cdot \mu_3}\notag\\
 & =\frac{1}{4} \frac{\langle3\bar{1}\rangle^{s_3+s_1-s_2} \langle\bar{1}\bar{2}\rangle^{s_1+s_2-s_3} \langle\bar{2}3\rangle^{s_2+s_3-s_1}}{(E-2 p_3)^{s_1+s_2+s_3}}\delta^{3}(\vec{p_{1}}+\vec{p_{2}}+\vec{p_{3}}).       \end{align}
The above result was obtained using the integral representation of the $\delta^{[n]}(x)$, 
\begin{align}
   \delta^{[n]}(x) = \int dk\, k^{n} e^{\textit{i} k x }.   
\end{align}

Similar results can be obtained for the other helicity configurations.

\section{Generators of the conformal algebra in spinor helicity and twistor variables}\label{app:generators}
In the following appendix, we enlist the conformal generators in spinor helicity variables. The conformal algebra in three dimensions
consists of the following generators: The generator of translations $P_{ab}$, the generator of rotations $M_{ab}$, the generator of dilatations and the generator of special conformal transformations which are respectively denoted as $D$ and $K_{ab}$.
Their action on the spin-s conserved currents rescaled to scaling dimension 2 are given as follows:
\small
\begin{align}\label{spinorhelicitygena}
    P^\mu=\frac{1}{2}(\sigma^\mu)^a_b\lambda_a\,\Bar{\lambda}^b,\qquad&K^\mu,=2(\sigma^\mu)^{ab}\frac{\partial^2}{\partial\lambda^a\,\Bar{\partial \lambda}^b},\notag\\
    M_{\mu\nu}=\frac{1}{2}\epsilon_{\mu\nu\rho}(\sigma^\rho)^{a}_{b}\bigg(\,\Bar{\lambda}^b\frac{\partial}{\partial\,\Bar{\lambda}^a}+\lambda^b\frac{\partial}{\partial\lambda^a}\bigg),\qquad&D=\frac{i}{2}\bigg(\,\Bar{\lambda}^a\frac{\partial}{\partial\,\Bar{\lambda}^a}+\lambda^a\frac{\partial}{\partial\lambda^a}+2\bigg).
\end{align}
\normalsize
On the twistor transformed versions of corresponding currents $(\hat{J}_s/\Tilde{J}_s)$, the above generators are respectively given by:
\begin{align}\label{action1twistora}
    P_{ab}=i\lambda_{(a}\frac{\partial}{\partial\Bar{\mu}^{b)}},&\qquad K_{ab}=i\Bar{\mu}_{(a}\frac{\partial}{\partial \lambda^{b)}},\notag\\
\Tilde{M}_{ab}=i \bigg(\lambda_{(a}\frac{\partial}{\partial\lambda^{b)}}+\Bar{\mu}_{(a}\frac{\partial}{\partial\Bar{\mu}^{b)}}\bigg),&\qquad  D=\frac{i}{2}\bigg(\lambda^a\frac{\partial}{\partial\lambda^a}-\Bar{\mu}^a\frac{\partial}{\partial\Bar{\mu}^{a}}\bigg).
\end{align}
Similarly, their dual-twistor counterparts are as follows:
\begin{align}\label{action2twistora}
    P_{ab}=-i\Bar{\lambda}_{(a}\frac{\partial}{\partial\mu^{b)}},\qquad& K_{ab}=-i\mu_{(a}\frac{\partial}{\partial \Bar{\lambda}^{b)}}, \notag\\
\Tilde{M}_{ab}=i \bigg(\Bar{\lambda}_{(a}\frac{\partial}{\partial\Bar{\lambda}^{b)}}+\mu_{(a}\frac{\partial}{\partial\mu^{b)}}\bigg),\qquad& D=\frac{i}{2}\bigg(\Bar{\lambda}^a\frac{\partial}{\partial\Bar{\lambda}^a}-\mu^a\frac{\partial}{\partial\mu^{a}}\bigg).
\end{align}
 Finally, given the manifest twistor space generator $T_{AB} = Z_{(A} \frac{\partial}{\partial Z^{B)}}$, the individual conformal generators can be recovered from its components by writing it in block matrix form:
\begin{align}
    T_{AB} = \begin{pmatrix}
    -\bar{\mu}_{(a'} \frac{\partial}{\partial \lambda^{b)}} &  -\bar{\mu}_{a'} \frac{\partial}{\partial \bar{\mu}_{b}} + \lambda^{b'} \frac{\partial}{\partial \lambda^{a}}  \\[1em]
    -\bar{\mu}_{a'} \frac{\partial}{\partial \bar{\mu}_{b}}+ \lambda^{b} \frac{\partial}{\partial \lambda^{a'}} & +\lambda^{(a} \frac{\partial}{\partial\bar{\mu}_{b')}} 
    \end{pmatrix}
    =
    \begin{pmatrix}
    \textit{i} K_{a'b} &  -\textit{i} M^b_{a'} + \frac{2}{\textit{i}}\delta_{a'}^{b} D \\
    -\textit{i} M^b_{a'}+ \frac{2}{\textit{i}}\delta^{b}_{a'} D & -\textit{i} P^{ab'}
    \end{pmatrix}.
\end{align}
where $P,\; K,\; M,\; D$ are the conformal generators as given in \eqref{action1twistora}.  Note that we used the fact that $Z_A=\Omega_{BA}Z^B$, where $Z^A$ is given in \eqref{twistors} and the invariant symplectic form $\Omega_{AB}$ is given in \eqref{symplecticform}. 
Similarly, one can obtain the components of the dual twistor generator $W_{(A}\frac{\partial}{\partial W^{B)}}$ which takes a similar form but in terms of the dual-twistor component generators \eqref{action2twistora}.

\section{From four dimensional to three dimensional spinor helicity variables}\label{app:4dto3d}
In this appendix, we show how one can obtain a general three dimensional momenta starting from a null momentum in four dimensions.
\subsection{$\mathbb{R}^{3,1}\to \mathbb{R}^3$}
A four dimensional massless momenta can be written in an unconstrained form as a outer product of two spinors as follows: $p_a^{\Dot{a}}=\lambda_a\tilde{\lambda}^{\Dot{a}}$ with a little group redundancy $\lambda\to e^{-\frac{i\theta}{2}}\lambda,\tilde{\lambda}\to e^{+\frac{i\theta}{2}}\tilde{\lambda}$, $\theta\in\mathbb{R}$. In matrix form we have,
\begin{align}
    p_a^{\Dot{a}}\sim \begin{pmatrix}
        p_t+p_z & p_x-i p_y\\
        p_x+i p_y &p_t-p_z
    \end{pmatrix}.
\end{align}
The reality condition for the four momenta components is that $p$ is Hermitian and thus $(\lambda_a)^\dagger=\tilde{\lambda}^{\Dot{a}}$ and $(\tilde{\lambda}^{\Dot{a}})^\dagger=\lambda_a$. A three dimensional Euclidean momenta can be obtained from this null 4d Minkowskian momenta as follows: Introduce a special vector,
\begin{align}\label{4dto3depsilon}
    \epsilon_{a \Dot{a}}=\begin{pmatrix}
        0&1\\
-1&0
    \end{pmatrix}.
\end{align}
The three dimensional momentum is then given by,
\begin{align}\label{3dfrom4dmom}
    p_{ab}=-\frac{1}{2}\big(p_{(a}^{\Dot{a}}\epsilon_{b)\Dot{a}}\big).
\end{align}
In matrix form this procedure is as follows: First form,
\begin{align}
   p_a^{\Dot{a}}\epsilon_{b \Dot{a}}\sim \begin{pmatrix}
        p_t+p_z & p_x-i p_y\\
        p_x+i p_y &p_t-p_z
    \end{pmatrix}\begin{pmatrix}
        0&1\\
-1&0
    \end{pmatrix}=\begin{pmatrix}
        -p_x+i p_y&p_t+p_z\\
        -p_t+p_z&p_x+i p_y
    \end{pmatrix}.
\end{align}
Now symmetrize this matrix and multiply by a factor of $-\frac{1}{2}$. The result is,
\begin{align}
    p_{ab}=-\frac{1}{2}\begin{pmatrix}
        -p_x+i p_y&p_t+p_z\\
        -p_t+p_z&p_x+i p_y
    \end{pmatrix}-\frac{1}{2}\begin{pmatrix}
        -p_x+i p_y&-p_t+p_z\\
        p_t+p_z&p_x+i p_y
    \end{pmatrix}=\begin{pmatrix}
        -p_x+i p_y&p_z\\
        p_z&p_x+i p_y
    \end{pmatrix}.
\end{align}
Thus, we see that $p_t$ has completely dropped out in this procedure and left behind a three dimensional general momenta. In the langauge of spinors we have,
\begin{align}
    p_{ab}=-\frac{1}{2}\big(\lambda_{(a} \epsilon_{b)\Dot{a}}\tilde{\lambda}^{\Dot{a}}\big)=\frac{1}{2}\big(\lambda_{(a}\Bar{\lambda}_{b)}\big),
\end{align}
where we have defined,
\begin{align}
    \Bar{\lambda}_b=\epsilon_{\Dot{a}b}\tilde{\lambda}^{\Dot{a}}.
\end{align}
The induced reality condition on the three dimensional spinors is,
\begin{align}
    \lambda_a^\dagger=\Bar{\lambda}^a,(\Bar{\lambda}^a)^\dagger=\lambda_a.
\end{align}
\subsection{$\mathbb{R}^{2,2}\to \mathbb{R}^{2,1}$}
The procedure is very similar to the above case. A null momenta in the four dimensional Klein space can be written as an outerproduct of two spinors, $p_a^{\Dot{a}}=\lambda_a \tilde{\lambda}^{\Dot{a}}$ with a little group redundancy $\lambda\to \frac{1}{r}\lambda,\tilde{\lambda}\to r\tilde{\lambda}$, $r\in\mathbb{R}$. In matrix form we have,
\begin{align}
    p_a^{\Dot{a}}\sim \begin{pmatrix}
        p_t+p_z & p_x-p_w\\
        p_x+p_w &p_t-p_z
    \end{pmatrix}.
\end{align}
The $p_w$ component that appears here is related to the corresponding Euclidean $p_y$ via $i p_y=p_w$. The reality condition here is simply that the above matrix is real and hence gives $\lambda_a^*=\lambda_a$ and $(\tilde{\lambda}^{\Dot{a}})^*=\tilde{\lambda}^{\Dot{a}}$, that is, the spinors are real and independent.
One can now follow the same procedure as in the previous subsection, i.e., defining a special vector $\epsilon_{a \Dot{a}}$ \eqref{4dto3depsilon} and defining a 3d momentum via \eqref{3dfrom4dmom}. The difference now is that $p_w$ is a timelike coordinate unlike $p_y$ and thus what we end up with is a three dimensional Lorentzian momentum. Again, in the language of the spinors we have,
\begin{align}
    p_{ab}=-\frac{1}{2}\big(\lambda_{(a} \epsilon_{b)\Dot{a}}\tilde{\lambda}^{\Dot{a}}\big)=\frac{1}{2}\big(\lambda_{(a}\Bar{\lambda}_{b)}\big),
\end{align}
where we have defined,
\begin{align}
    \Bar{\lambda}_b=\epsilon_{\Dot{a}b}\tilde{\lambda}^{\Dot{a}}.
\end{align}
The reality condition induced on the three dimensional momenta is now,
\begin{align}\label{LorentzReality}
    \lambda_a^*=\lambda_a,(\Bar{\lambda}^a)^*=\Bar{\lambda}^a.
\end{align}

\subsection{Understanding discrete symmetries in Klein space}
We have seen in appendix \ref{app:4dto3d} that the natural dimensional reduction of Klein space leads to three dimensional Minkowski space. It is thus interesting to ask how the discrete symmetries in Klein space reduce to those in $\mathbb{R}^{2,1}$. The starting point is the position bi-spinor:
\begin{align}
    x_{a}^{\Dot{a}}=\begin{pmatrix}
        t+z&&x-w\\
        x+w&&t-z
    \end{pmatrix}.
\end{align}
Parity is the transformation $z\to -z$ whereas time-reversal is $t\to -t$. The alternate transformations $x\to -x$ and $w\to -w$ are connected to the above choices via $SO(2)$ spatial and temporal rotations respectively. Our aim is to translate these transformations to spinor helicity variables. A null Kleinian momenta is given by,
\begin{align}
    p_a^{\Dot{a}}=\begin{pmatrix}
        p_t+p_z&&p_x-p_w\\
        p_x+p_w&&p_t-p_z
    \end{pmatrix}
    =\begin{pmatrix}
        \lambda_1\tilde{\lambda}_2&&-\lambda_1\tilde{\lambda}_1\\
        \lambda_2\tilde{\lambda}_2&&-\lambda_2\tilde{\lambda}_1
    \end{pmatrix}.
\end{align}
Under parity, $p_z$ flips to $-p_z$. This leads to the following transformation for the spinors:
\begin{align}
   P:\begin{pmatrix}
       \lambda_1,&&\lambda_2
   \end{pmatrix}
   \to\begin{pmatrix}
       -\tilde{\lambda}_1,&&\tilde{\lambda}_2
   \end{pmatrix},
\end{align}
which is identical to the parity transformation on the spinors in $\mathbb{R}^{2,1}$ \eqref{parityonspinors} after a dimensional reduction.
What about under time-reversal? Lets say we flip $t\to -t$. Then should one flip $p_x,p_z$ and $p_w$ to $-p_x,-p_z$ and $-p_w$? In Minkowski space $\mathbb{R}^{3,1}$, time-reversal is anti-linear and thus under $t\to -t$, $(p_t,p_x,p_z,p_y)\to (p_t,-p_x,-p_z,-p_y)$. This also ensures that energy, that is $p_t$ is invariant under a time-reversal. Klein space is connected to Minkowski space by a Wick rotation $p_y=i p_w$. Thus, $p_y$ flipping under time-reversal implies that $p_w$ must not flip since $i\to -i$ takes care of the sign flip. Therefore, the Klein space time-reversal should not change $p_w\to -p_w$. Thus we have time-reveral taking $(p_t,p_w,p_x,p_z)\to (p_t,p_w,-p_x,-p_z)$. On the spinors it amounts to,
\begin{align}
    T:\begin{pmatrix}
        \lambda_1,&&\lambda_2
    \end{pmatrix}\to\begin{pmatrix}
        -\lambda_2,&&\lambda_1,
    \end{pmatrix}
\end{align}
and similarly for $\tilde{\lambda}$. This leads to the same time-reversal operation we defined in $\mathbb{R}^{2,1}$ in \eqref{timereversal1} after a dimensional reduction.

In short, what we have shown is that the discrete symmetries of four dimensional Klein space naturally give rise to the discrete symmetries of three dimensional Minkowski space.

\section{Derivation of Ward-Takahashi identity for a time ordered correlator}\label{app:Derivation of current conservation Ward-Takahshi identity}
In this appendix, we show how the current conservation Ward-Takahashi identity for a time ordered correlator arises in the canonical approach. 
A time ordered three point function of a $U(1)$ current $J^\mu$ and two charged scalars $\phi,\chi$ is given by,
\begin{align}\label{Jphiphistar}
    \langle 0|T\{J^\mu(x_1)\phi(x_2)\chi(x_3)\}|0\rangle&=\theta(t_{12})\theta(t_{23})\langle 0|J^\mu(x_1)\phi(x_2)\chi(x_3)|0\rangle+\theta(t_{13})\theta(t_{32})\langle 0|J^\mu(x_1)\chi(x_3)\phi(x_2)|0\rangle\notag\\
    &+\theta(t_{21})\theta(t_{13})\langle 0|\phi(x_2)J^\mu(x_1)\chi(x_3)|0\rangle+\theta(t_{23})\theta(t_{31})\langle 0|\phi(x_2)\chi(x_3)J^\mu(x_1)|0\rangle\notag\\
    &+\theta(t_{31})\theta(t_{12})\langle 0|\chi(x_3)J^\mu(x_1)\phi(x_2)|0\rangle+\theta(t_{32})\theta(t_{21})\langle 0|\chi(x_3)\phi(x_2)J^\mu(x_1)|0\rangle,
\end{align}
where we used the shorthand $t_{ij}=t_i-t_j$. Let us take the derivative of \eqref{Jphiphistar} with respect to $\partial_{1\mu}$. The only contributions are when it acts on the Heaviside theta functions as the six Wightman functions appearing on the RHS are identically conserved. The result is,
\begin{align}
    \partial_{1\mu}\langle 0|T\{J^\mu(x_1)\phi(x_2)\chi(x_3)\}|0\rangle&=\delta(t_{12})\theta(t_{23})\langle 0|J^0(x_1)\phi(x_2)\chi(x_3)|0\rangle+\delta(t_{13})\theta(t_{32})\langle 0|J^0(x_1)\chi(x_3)\phi(x_2)|0\rangle\notag\\
    &-\delta(t_{12})\theta(t_{13})\langle 0|\phi(x_2)J^0(x_1)\chi(x_3)|0\rangle+\theta(t_{21})\delta(t_{13})\langle 0|\phi(x_2)J^0(x_1)\chi(x_3)|0\rangle\notag\\
    &-\theta(t_{23})\delta(t_{31})\langle 0|\phi(x_2)\chi(x_3)J^0(x_1)|0\rangle-\delta(t_{31})\theta(t_{12})\langle 0|\chi(x_3)J^0(x_1)\phi(x_2)|0\rangle\notag\\
    &+\theta(t_{31})\delta(t_{12})\langle 0|\chi(x_3)J^0(x_1)\phi(x_2)|0\rangle-\theta(t_{32})\delta(t_{21})\langle 0|\chi(x_3)\phi(x_2)J^0(x_1)|0\rangle.
\end{align}
Grouping like terms together, this quantity can be written as,
\begin{align}\label{WTexample1a}
    \partial_{1\mu}\langle 0|T\{J^\mu(x_1)\phi(x_2)\chi(x_3)\}|0\rangle=\delta(t_{12})\langle 0|T\{[J^0(x_1),\phi(x_2)]\chi(x_3)\}|0\rangle+\delta(t_{13})\langle 0| T\{\phi(x_2)[J^0(x_1),\chi(x_3)]|0\rangle.
\end{align}
Recall that,
\begin{align}
    &[Q,\phi(t_2,\Vec{x}_2)]=\int_{\Sigma_{t_2}}d^{d-1}\Vec{x}[J^0(t_2,\Vec{x}),\phi(t_2,\Vec{x}_2)]=q_{\phi}~\phi(t_2,\Vec{x}_2),\notag\\
    &[Q,\chi(t_3,\Vec{x}_3)]=\int_{\Sigma_{t_3}}d^{d-1}\Vec{x}[J^0(t_3,\Vec{x}),\chi(t_3,\Vec{x}_3)]=q_{\chi}~\chi(t_3,\Vec{x}_3).
\end{align}
This implies,
\begin{align}
    [J^0(x_1),\phi(x_2)]=q_{\phi}~\delta^{d-1}(\Vec{x}_{12})\phi(x_2)~,~[J^0(x_1),\chi(x_3)]=q_{\chi}~\delta^{d-1}(\Vec{x}_{13})\chi(x_3).
\end{align}
Substituting these commutators back in \eqref{WTexample1a} we obtain the familiar charge conservation Ward-Takahashi identity:
\begin{align}\label{WTexample1b}
     \partial_{1\mu}\langle 0|T\{J^\mu(x_1)\phi(x_2)\chi(x_3)\}|0\rangle=\big(q_{\phi}\delta^d(x_{12})+q_{\chi}\delta^d(x_{13})\big)\langle 0|T\{\phi(x_2)\chi(x_3)\}|0\rangle.
\end{align}
Integrating \eqref{WTexample1b} with respect to $x_1$ over the entire manifold $\mathbb{R}^{1,d-1}$, we obtain (after using Gauss' theorem on the LHS),
\begin{align}
    0=(q_{\phi}+q_{\chi})\langle 0|T\{\phi(x_2)\chi(x_3)\}|0\rangle\implies q_\chi=-q_\phi,
\end{align}
thus imposing conservation of charge.

\section{Properties of conformal Wightman functions}\label{app:WightmanFunctionProperties}
In this Appendix we discuss several properties of Wightman functions which are useful for our discussion in the main text. 
\subsection{Vanishing of Wightman function  when any of the momenta vanish}

We give an argument following \cite{Bautista:2019qxj} on why three point Wightman functions vanish whenever any of the momenta vanish. The position space form factor of generic three point (scalar, spinor, spinning) Wightman functions take the form (with $\epsilon>0$),

{\tiny
\begin{align}
    \mathbf{ff}(x_1,x_2,x_3)=\frac{1}{\big(-(t_1-t_2)^2+(\Vec{x}_1-\Vec{x}_2)^2+i\epsilon(t_1-t_2)\big)^\alpha\big(-(t_2-t_3)^2+(\Vec{x}_2-\Vec{x}_3)^2+i\epsilon(t_2-t_3)\big)^\beta\big(-(t_1-t_3)^2+(\Vec{x}_1-\Vec{x}_3)^2+i\epsilon(t_1-t_3)\big)^\gamma},
\end{align}
}
\normalsize
where $\alpha,\beta,\gamma$ are functions of the spins and scaling dimensions of the external operators. Its Fourier transform is given by,
\begin{align}
    \int_{\mathbb{R}^d\times \mathbb{R}^d\times \mathbb{R}^d} d^d x_1 d^d x_2 d^d x_3 e^{ip_1\cdot x_1+ip_2\cdot x_2+p_3\cdot x_3}~\mathbf{ff}(x_1,x_2,x_3).
\end{align}
Let us now make the following change of variables:
\begin{align}
    x_{12}^\mu=x_1^\mu-x_2^\mu,x_{23}^\mu=x_2^\mu-x_3^\mu, X^\mu=\frac{x_1^\mu+x_2^\mu+x_3^\mu}{3}.
\end{align}
The Fourier transform is then given by,

\begin{align}
    &\int_{\mathbb{R}^d\times \mathbb{R}^d\times \mathbb{R}^d} \frac{d^d X d^d x_{12} d^d x_{23} e^{i(p_1+p_2+p_3)\cdot X+\frac{i}{3}(2 p_1-p_2-p_3)\cdot x_{12}+\frac{i}{3}( p_1+p_2-2 p_3)\cdot x_{23}}}{\big(-t_{12}^2+\Vec{x}_{12}^2+i\epsilon t_{12}\big)^\alpha\big(-t_{23}^2+\Vec{x}_{23}^2+i\epsilon t_{23}\big)^\beta\big(-(t_{12}+t_{23})^2+(\Vec{x}_{12}+\Vec{x}_{23})^2+i\epsilon(t_{12}+t_{23})\big)^\gamma}.
\end{align}
The integral over $X$ results in $(2\pi)^d\delta(p_1+p_2+p_3)$, that is, the momentum conserving delta function. Upon imposing it, we are left with the following integral:
\begin{align}
\int_{\mathbb{R}^d\times \mathbb{R}^d} \frac{ d^d x_{12} d^d x_{23} e^{i p_1\cdot x_{12}+i(p_1+p_2)\cdot x_{23}}}{\big(-t_{12}^2+\Vec{x}_{12}^2+i\epsilon t_{12}\big)^\alpha\big(-t_{23}^2+\Vec{x}_{23}^2+i\epsilon t_{23}\big)^\beta\big(-(t_{12}+t_{23})^2+(\Vec{x}_{12}+\Vec{x}_{23})^2+i\epsilon(t_{12}+t_{23})\big)^\gamma}.
\end{align}
We now wish to show that when say, $p_1=0$, the integral vanishes. To that end consider the $x_{12}$ integral decomposed into its time and space components:
\begin{align}
    \int_{\mathbb{R}\times \mathbb{R}^{d-1}} dt_{12}d^{d-1} \Vec{x}_{12} \frac{e^{-ip_1^0t_{12}+i\Vec{p}_1\cdot\Vec{x}_{12}}}{\big(-t_{12}^2+\Vec{x}_{12}^2+i\epsilon t_{12}\big)^\alpha\big(-(t_{12}+t_{23})^2+(\Vec{x}_{12}+\Vec{x}_{23})^2+i\epsilon(t_{12}+t_{23})\big)^\gamma}.
\end{align}
Our focus is now on the integral over $t_{12}$. Consider the case when $p_1=0\implies p_1^0=\Vec{p}_1=0$. The integral becomes,
\begin{align}
    \int_{\mathbb{R}\times \mathbb{R}^{d-1}} dt_{12}d^{d-1} \Vec{x}_{12} \frac{1}{\big(-t_{12}^2+\Vec{x}_{12}^2+i\epsilon t_{12}\big)^\alpha\big(-(t_{12}+t_{23})^2+(\Vec{x}_{12}+\Vec{x}_{23})^2+i\epsilon(t_{12}+t_{23})\big)^\gamma}.
\end{align}
Consider the case when $\alpha$ and $\gamma$ are integers. The integrand, viewed then has poles at the following locations:
\begin{align}
  \{\pm|\Vec{x}_{12}|+i\epsilon, -t_{23}\pm \sqrt{\Vec{x}_{12}^2+2 \Vec{x}_{12}\cdot \Vec{x}_{23}+\Vec{x}_{23}^2}+i\epsilon\}.
\end{align}
In particular, they are only in the upper half plane. If $\alpha,\gamma$ are fractions, then the situation is that the branch points are only in the upper half plane. Moreover, note that the spectral condition \eqref{thetafunctions} implies that $-p_1^0>|\Vec{p}_1|\implies p_1^0<0$. Either way, we can choose a semi-circular contour in the lower half plane to evaluate this integral. Since there are no poles/branch-points, the result is identically zero.

A similar argument shows that the correlator vanishes when the momentum $p_2$ or $p_3$ vanishes. 
\subsection{Wightman function conjugation property}
\subsection*{Two point correlators}
In position space, the two point Wightman functions of a scalar $O_{\Delta}$ are given by 
\begin{align}\label{CFTWightman1}
    &\langle 0|O_{\Delta}(t_1,\Vec{x}_1)O_{\Delta}(t_2,\Vec{x}_2)|0\rangle=\frac{1}{(-t_{12}^2+\Vec{x}_{12}^2+i\epsilon t_{12})^\Delta},\langle 0|O_{\Delta}(t_2,\Vec{x}_2)O_{\Delta}(t_1,\Vec{x}_1)|0\rangle=\frac{1}{(-t_{21}^2+\Vec{x}_{21}^2+i\epsilon t_{21})^\Delta},
\end{align}
where $\epsilon>0$. 
It is clear that the complex conjugation of the first Wightman function results in the second as $(i\epsilon t_{12})^*=-i\epsilon t_{12}=i\epsilon t_{21}$.
\begin{align}
    \langle 0|O_{\Delta}(t_1,\Vec{x}_1)O_{\Delta}(t_2,\Vec{x}_2)|0\rangle^*=\langle 0|O_{\Delta}(t_2,\Vec{x}_2)O_{\Delta}(t_1,\Vec{x}_1)|0\rangle.
\end{align}
Thus we see that complex conjugation reverses the operator ordering. Let us derive the analogous statement in Fourier space. We have by definition,
\begin{align}
    \langle 0|O_{\Delta}(p_1)O_{\Delta}(p_2)|0\rangle=\int~d^3 x_1\;d^3 x_2~e^{ip_1\cdot x_1+ip_2\cdot x_2} \langle 0|O_{\Delta}(t_1,\Vec{x}_1)O_{\Delta}(t_2,\Vec{x}_2)|0\rangle.
\end{align}
Taking complex conjugate on both sides we get,
\begin{align}
    \langle 0|O_{\Delta}(p_1)O_{\Delta}(p_2)|0\rangle^*&=\int~d^3 x_1~d^3 x_2~e^{-ip_1\cdot x_1-ip_2\cdot x_2} \langle 0|O_{\Delta}(t_1,\Vec{x}_1)O_{\Delta}(t_2,\Vec{x}_2)|0\rangle^*\notag\\
    &=\int d^3 x_1~d^3 x_2~e^{-ip_1\cdot x_1-ip_2\cdot \Vec{x}_2} \langle 0|O_{\Delta}(t_2,\Vec{x}_2)O_{\Delta}(t_1,\Vec{x}_1)|0\rangle\notag\\
    &=\langle 0|O_{\Delta}(-p_2)O_{\Delta}(-p_1)|0\rangle.
\end{align}
Thus, in Fourier space, we see that complex conjugation reverses the order of the operators and flips the sign of the momenta.
Let us illustrate this for the two point function in momentum space \eqref{CFT2pointWightmanmomspace}. We have ,
\begin{align}\label{CFT2pointWightmanmomspace1}
    &\langle 0|O_{\Delta}(p_1)O_{\Delta}(p_2)|0\rangle\propto(-p_1^2)^{\Delta-\frac{d}{2}}\theta(-p_1^0-|\Vec{p}_1|)(2\pi)^d\delta^d(p_1+p_2),\notag\\
    &\langle 0|O_{\Delta}(p_2)O_{\Delta}(p_1)|0\rangle\propto(-p_1^2)^{\Delta-\frac{d}{2}}\theta(p_1^0-|\vec{p}_1|)(2\pi)^d\delta^d(p_1+p_2).
\end{align}
It is easy to see that the above reality condition is indeed satisifed.
\subsection*{Three point correlators}
Similar to the reality condition for two points, we see at three points the following reality condition:
\begin{align}
    \langle 0|O_{\Delta}(t_1,\Vec{x}_1)O_{\Delta}(t_2,\Vec{x}_2)O_{\Delta}(t_3,\Vec{x}_3)|0\rangle^*=\langle 0|O_{\Delta}(t_3,\Vec{x}_3)O_{\Delta}(t_2,\Vec{x}_2)O_{\Delta}(t_1,\Vec{x}_1)|0\rangle.
\end{align}
Tracing through the same steps as for the two point function we obtain the analogous momentum space reality condition:
\begin{align}
    \langle 0|O_{\Delta}(p_1)O_{\Delta}(p_2)O_{\Delta}(p_3)|0\rangle^*=\langle 0|O_{\Delta}(-p_3)O_{\Delta}(-p_2)O_{\Delta}(-p_1)|0\rangle.
\end{align}
Please note that above, it is the vector that flip and not their magnitudes.
It is obvious that this property extends to arbitrary three point (even and odd) Wightman functions involving spinning or spinorial operators. Since the conjugation in Lorentzian signature imposes a reality condition \eqref{MinkowskiReality} on spinors $\lambda,\;\bar{\lambda}$; the imaginary parts of the Wightman functions are purely due to the $i\epsilon$ prescription which gets flipped by complex conjugation.

\section{Two point Wightman functions}\label{app:WightmanTwoPoint}
\subsection{Scalars}
Concentrating on scalars at the moment, we have at two points, two possible orderings and hence two distinct Wightman functions \eqref{CFTWightman1}. Note that at space-like separation, the $i\epsilon$ term does not play a role and the two correlators in \eqref{CFTWightman1} are identical, which is in accordance with our earlier discussion.\\The time ordered correlator, following from the definition \eqref{TimeOrderedCorr}, is given by,
\begin{align}\label{CFTtimeordered2point}
    \langle 0|T\{O_{\Delta}(x_1)O_{\Delta}(x_2)\}|0\rangle&=\frac{1}{(-t_{12}^2+\Vec{x}_{12}^2+i\epsilon t_{12})^\Delta}\theta(t_{12})+\frac{1}{(-t_{21}^2+\Vec{x}_{21}^2+i\epsilon t_{21})^\Delta}\theta(t_{21})\notag\\
    &=\frac{1}{(-t_{12}^2+\Vec{x}_{12}^2+i\epsilon)^\Delta}.
\end{align}
Unlike the time-ordered correlator \eqref{CFTtimeordered2point}, the Wightman function \eqref{CFTWightman1} are well defined for all values of $\Delta$ \cite{Bautista:2019qxj}. This is easiest to see by going to Fourier space. 
\subsection*{To Fourier space}
The two Wightman functions in Fourier space are given by,
\begin{align}\label{CFT2pointWightmanmomspace}
    &\langle 0|O_{\Delta}(p_1)O_{\Delta}(p_2)|0\rangle=\frac{\pi^{\frac{d}{2}}2^{d-2\Delta}}{\Gamma(\Delta)\Gamma(\Delta-\frac{d}{2}+1)}(-p_1^2)^{\Delta-\frac{d}{2}}\theta(-p_1^0-|\Vec{p}_1|)(2\pi)^d\delta^d(p_1+p_2),\notag\\
    &\langle 0|O_{\Delta}(p_2)O_{\Delta}(p_1)|0\rangle=\frac{\pi^{\frac{d}{2}}2^{d-2\Delta}}{\Gamma(\Delta)\Gamma(\Delta-\frac{d}{2}+1)}(-p_1^2)^{\Delta-\frac{d}{2}}\theta(p_1^0-|\vec{p}_1|)(2\pi)^d\delta^d(p_1+p_2).
\end{align}
 The $d$ vector $p=(p^0,\Vec{p})$ where $p^0$ and $\Vec{p}$ are conjugate to $t$ and $\Vec{x}$ respectively. We have also defined its magnitude squared,
\begin{align}
    p^2=-(p^0)^2+\Vec{p}\cdot\Vec{p}.
\end{align}
The time ordered correlator on the other hand takes the following form:
\begin{align}\label{CFT2pointTimeOrdermomspace}
    \langle 0|T\{O_{\Delta}(p_1)O_{\Delta}(p_2)\}|0\rangle=i \frac{\pi^{\frac{d}{2}}2^{d-2\Delta}}{\Gamma(\Delta)}\Gamma(\frac{d}{2}-\Delta)(p_1^2)^{\Delta-\frac{d}{2}}(2\pi)^d\delta^d(p_1+p_2).
\end{align}
\subsection*{Renormalization}
Unlike \eqref{CFT2pointTimeOrdermomspace} which diverges when $\Delta=\frac{d}{2}+k,k\in \mathbb{Z}_{\ge 0}$ and requires regularization and renormalization, the Wightman functions \eqref{CFT2pointWightmanmomspace} are finite for all $\Delta$. This is not surprising given the fact that the time ordered correlator is the natural analytic continuation of the Euclidean correlator, whose divergences have been thoroughly analyzed in the literature \cite{Bzowski:2015pba}. Its expression is,
\begin{align}\label{CFT2pointEuclidmomspace}
    \langle O_{\Delta}(p_1)O_{\Delta}(p_2)\rangle=\frac{\pi^{\frac{d}{2}}2^{d-2\Delta}}{\Gamma(\Delta)}\Gamma(\frac{d}{2}-\Delta)(p_1^2)^{\Delta-\frac{d}{2}}(2\pi)^d\delta^{d}(p_1+p_2).
\end{align}

The question then arises: Given the fact that the Euclidean correlators have UV infinities for certain values of $\Delta$, how is it that Wightman functions are divergence free? This is due to the momentum space the analytic continuation \eqref{TwopointsEuclidtoWightman} that relates the Wightman functions to the discontinuity of Euclidean correlators. With this, let us now address the divergence problem. The Euclidean correlator \eqref{CFT2pointEuclidmomspace} is divergent when $\Delta=\frac{d}{2}+k,k\in\mathbb{Z}_{\ge 0}$. In those cases, after renormalization, the correlator develops a logarithm:
\begin{align}\label{RenScalarTwopoint}
    \langle O_{\Delta}(p_1)O_{\Delta}(p_2)\rangle_{\Delta=\frac{d}{2}+k}\sim p^{2k}\log\big(\frac{p^2}{\mu}\big).
\end{align}
Performing the steps as outlined in \eqref{TwopointsEuclidtoWightman} we get using \eqref{RenScalarTwopoint},
\begin{align}
     \langle 0|O_{\Delta}(p_1)O_{\Delta}(p_2)|0\rangle_{\Delta=\frac{d}{2}+k}=p^{2k}(2\pi i),
\end{align}
which is finite and independent of the renormalization scale $\mu$. 
\subsection{Spinning even two point functions}

As for two point functions of integer spin conserved currents, an analogous analysis holds.  Their two point function in momentum space takes the following form in odd dimensional Euclidean space\footnote{Here, the $z_i$ are auxiliary null polarization vectors that have been contracted with the indices of the operators. For the conserved currents, we further impose the transversality condition $z_i\cdot p_i=0$.}:
\begin{align}
    \langle J_s(p_1,z_1)J_s(p_2,z_2)\rangle_{\text{even}}=c_{s}^{\text{(even)}}\frac{(z_1\cdot z_2)^s}{p_1^{4-2s-d}}.
\end{align}
The analytic continuation \eqref{TwopointsEuclidtoWightman} yields,
\begin{align}\label{twopointspinevenEuclidtoWightman}
    \langle 0| J_s(p_1,z_1)J_s(p_2,z_2)|0\rangle_{\text{even}}&=\text{Disc}_{p_1^2}\langle J_s(p_1,z_1)J_s(p_2,z_2)\rangle_{\text{even}}\theta(-p_1^0-|\Vec{p}_1|)\notag\\
    &=c_s^{(\text{even})}(z_1\cdot z_2)^s\text{Disc}_{p_1^2}\bigg(\frac{1}{p_1^{4-2s-d}}\bigg)\theta(-p_1^0-|\Vec{p}_1|).
\end{align}
In even dimensions, their Euclidean two point functions have UV divergences and must be renormalized \cite{Bzowski:2017poo}. The resulting momentum space correlator develops a log much like in \eqref{RenScalarTwopoint}. However, taking the discontinuity to obtain the corresponding Wightman function removes the logarithm and yields a renormalization scale independent result. 

\subsection{Parity and time-reversal odd two point functions}
In the special case of three dimensions, the two point function of conserved currents can acquire a parity odd contribution. In Euclidean momentum space it is the following:
\begin{align}\label{odd2ptin3d}
    \langle J_{s}(p_1,z_1)J_s(p_2,z_2)\rangle_{\text{odd}}=c_s^{(\text{odd})}\frac{\epsilon^{z_1 z_2 p_1}}{p_1}(z_1\cdot z_2)^{s-1}p_1^{2s-1}.
\end{align}
This quantity can be obtained by epsilon transforming the parity even correlator \cite{Jain:2021wyn}:
\begin{align}\label{2pointEPToddeven}
    \langle J_s^{\mu_1\cdots \mu_s}(p_1)J_s^{\nu_1\cdots\nu_s}(p_2)\rangle_{\text{odd}}\propto \frac{1}{p_1}{\epsilon^{ p_1 \alpha(\mu_1}}\langle J_s^{\alpha\cdots \mu_s)}(p_1)J_s^{\nu_1\cdots\nu_s}(p_2)\rangle_{\text{even}}.
\end{align}
The analytic continuation to obtain the Wightman function is identical to \eqref{twopointspinevenEuclidtoWightman}:
\begin{align}\label{twopointspinoddEuclidtoWightman}
   \langle 0| J_{s}(p_1,z_1)J_s(p_2,z_2)|0\rangle_{\text{odd}}=c_s^{(\text{odd})}\frac{\epsilon^{z_1 z_2 p_1}}{p_1}(z_1\cdot z_2)^{s-1}\text{Disc}_{p_1^2}\bigg(p_1^{2s-1}\bigg). 
\end{align}
Note however, that the $\frac{1}{p_1}$ factor is not included in the computation of the discontinuity. This procedure is also consistent with the Wightman function epsilon transform relation that turns out to be,
\begin{align}
    \langle 0| J_s^{\mu_1\cdots \mu_s}(p_1)J_s^{\nu_1\cdots\nu_s}(p_2)|0\rangle_{\text{odd}}\propto \frac{1}{p_1}{\epsilon^{ p_1 \alpha(\mu_1}}\langle 0|J_s^{\alpha\cdots \mu_s)}(p_1)J_s^{\nu_1\cdots\nu_s}(p_2)|0\rangle_{\text{even}}.
\end{align}

\subsection{General structure of Wightman function of currents in spinor helicity variable} 
  Converting the above momentum space expressions into spinor helicity variables, we see that the general structure for the two-point function of conserved currents in the two helicity configurations, i.e. (-~-) and (++) are respectively given as: 
  \begin{align}\label{general2pointSH}
     \langle 0|J_{s}^{-}(\lambda_1,\bar{\lambda}_1) J_{s}^{-}(\lambda_2,\bar{\lambda}_2)|0 \rangle &\sim (c_{s}^{(even)}-\textit{i} c_{s}^{(odd)}) \frac{\langle12\rangle^{2s}}{p_1},\notag\\
     \langle 0| J_{s}^{+}(\lambda_1,\bar{\lambda}_1) J_{s}^{+}(\lambda_2,\bar{\lambda}_2) |0\rangle &\sim (c_{s}^{(even)}+\textit{i} c_{s}^{(odd)}) \frac{\langle\bar{1}\bar{2}\rangle^{2s}}{p_1}.  
  \end{align}
\section{Examples of three point Wightman functions in momentum space and spinor helicity variables}\label{app:WightmanExamples}
In this appendix, we illustrate the two methods to obtain Wightman functions that we outlined in section \ref{sec:WightmanCFT}. We present one parity even and one parity odd example worked out in detail here as well as examples involving scalars and spinors.
\subsection{Inside the triangle: An even example}
The parity preserving Wightman function $\langle 0|TJJ|0\rangle$ is inside the triangle. We obtain it first via analytic continuation from Euclidean space and second, through the bootstrap.

\subsection*{From Euclid to Wightman}
The Euclidean parity-even correlators are given by,
\small
\begin{align}\label{TJJEuclidNHandH}
    \langle T(z_1,p_1)J(z_2,p_2)J(z_3,p_3)\rangle_{nh}&=c_{211}^{(nh)}\bigg(2(z_1\cdot p_2)\big((z_1\cdot z_2)(z_3\cdot p_1)+(z_2\cdot z_3)(z_1\cdot p_2)+(z_3\cdot z_1)(z_2\cdot p_3)\big)f_1(p_1,p_2,p_3)\notag\\&-(z_1\cdot z_2)(z_1\cdot z_3)(p_2+p_3+2f_4(p_1,p_2,p_3))\bigg),\notag\\
    \langle T(z_1,p_1)J(z_2,p_2)J(z_3,p_3)\rangle_{h}&=c_{211}^{(h)}\bigg((z_1\cdot p_2)^2(z_2\cdot z_3)h_1(p_1,p_2,p_3)+(z_1\cdot p_2)(z_1\cdot z_2)(z_3\cdot p_1)h_2(p_1,p_2,p_3)\notag\\
    &+(z_1\cdot p_2)(z_1\cdot z_3)(z_2\cdot p_3)h_3(p_1,p_2,p_3)+(z_1\cdot z_2)(z_1\cdot z_3)h_4(p_1,p_2,p_3)\bigg).
\end{align}
\normalsize
The constant $c_{211}^{(nh)}$ is related two point function coefficient of $\langle JJ \rangle$ which follows from the Ward-Takahashi identity. The other constant $c_{211}^{(h)}$ is an OPE coefficient which appears at the level of three point function. 
The explicit forms of the functions $h_1,h_2,h_3,h_4$ and $f_1,f_2,f_3,f_4$ are given by,
\begin{align}\label{TJJformfactors}
    &h_1=-\frac{p_1^3}{E^4},h_2=-\frac{p_1^2(p_1+p_2)}{E^4},h_3=-\frac{p_1^2(p_1+p_3)}{E^4},h_4=\frac{p_1^2(E-2p_3)(E-2p_2)}{2E^3},\notag\\
    &f_1=f_2=f_3=\frac{E+p_1}{E^2},f_4=-\frac{p_1^2}{E}.
\end{align}
 Following our procedure for analytic continuation \eqref{WightmanToEuclidSpinning} we have,
\begin{align}\label{TJJWightmanhandnh}
    &\langle 0| T(z_1,p_1)J(z_2,p_2)J(z_3,p_3)|0\rangle_{nh}\notag\\&=c_{211}^{(nh)}\bigg(2(z_1\cdot p_2)\big((z_1\cdot z_2)(z_3\cdot p_1)+(z_2\cdot z_3)(z_1\cdot p_2)+(z_3\cdot z_1)(z_2\cdot p_3)\big)\prod_{i=1}^{3}\text{Disc}_{p_i^2}f_1(p_1,p_2,p_3)\notag\\&-(z_1\cdot z_2)(z_1\cdot z_3)\prod_{i=1}^{3}\text{Disc}_{p_i^2}(p_2+p_3+2f_4(p_1,p_2,p_3))\bigg),\notag\\
    &\langle 0| T(z_1,p_1)J(z_2,p_2)J(z_3,p_3)|0\rangle_{h}\notag\\&=c_{211}^{(h)}\bigg((z_1\cdot p_2)^2(z_2\cdot z_3)\prod_{i=1}^{3}\text{Disc}_{p_i^2}h_1(p_1,p_2,p_3)+(z_1\cdot p_2)(z_1\cdot z_2)(z_3\cdot p_1)\prod_{i=1}^{3}\text{Disc}_{p_i^2}h_2(p_1,p_2,p_3)\notag\\
    &+(z_1\cdot p_2)(z_1\cdot z_3)(z_2\cdot p_3)\prod_{i=1}^{3}\text{Disc}_{p_i^2}h_3(p_1,p_2,p_3)+(z_1\cdot z_2)(z_1\cdot z_3)\prod_{i=1}^{3}\text{Disc}_{p_i^2}h_4(p_1,p_2,p_3)\bigg).
\end{align}
Also, using the three point discontinuity formula \eqref{discff}, we see that in contrast with the Euclidean correlator, the Wightman function involves form factors with flipped momenta magnitudes. In Euclidean signature, such solutions were ruled out by OPE consistency \cite{Maldacena:2011nz,Jain:2021whr} but for Wightman functions, they naturally appear.

Thus, we have shown that there exist two linearly independent parity and time reversal even Wightman functions which are the analytic continuations of the non-homogeneous and homogeneous Euclidean correlators. This is why we continue to use the labels $h$ and $nh$ for the Wightman functions even though they are all homogeneous in the sense they all have a zero Ward-Takahashi identity.

\subsection*{Solving for Wightman functions using consistency conditions}
Let us now solve for these Wightman functions using the principles outlined in subsection \ref{sec:WightmanCFT}. The most general parity and time-reversal invariant  Poincare invariant ansatz for the Wightman function is,
\small
\begin{align}
    \langle 0|T(z_1,p_1)J(z_2,p_2)J(z_3,p_3)|0\rangle_{\text{even}}&=(z_1\cdot p_2)\bigg((z_1\cdot p_2)(z_2\cdot z_3)A_1+(z_1\cdot z_2)(z_3\cdot p_1)A_2+(z_1\cdot z_3)(z_2\cdot p_3)A_3\bigg)\notag\\&+(z_1\cdot z_2)(z_1\cdot z_3)A_4.
\end{align}
\normalsize
Conformal invariance and the conservation equation \eqref{KactionSHWightman} yields the following form for the $A_i(p_1,p_2,p_3)$ \footnote{As we discussed earlier, in the Euclidean context, OPE consistency allows only for the functions with arguments $p_1,p_2$ and $p_3$ and not the flipped ones. For Wightman functions, however, we allow the more general ansatz and as we shall see shortly, this is indeed the correct procedure which will lead to the same result as taking the discontinuity \eqref{TJJWightmanhandnh}.}:
\begin{align}\label{TJJformfactors0}
    A_i&=a_0 h_i(p_1,p_2,p_3)+a_1 h_i(-p_1,p_2,p_3)+a_2 h_i(p_1,-p_2,p_3)+a_3 h_i(p_1,p_2,-p_3)\notag\\&+b_0 f_i(p_1,p_2,p_3)-b_0 f_i(-p_1,p_2,p_3)+b_2 f_i(p_1,-p_2,p_3)+b_2 f_i(p_1,p_2,-p_3),
\end{align}
where the functions $h_1,h_2,h_3,h_4$ and $f_1,f_2,f_3,f_4$ are given in \eqref{TJJformfactors}.
 Note that the coefficients of the $f_i$ (i.e., $b_0,b_2$) in \eqref{TJJformfactors0} appear in a particular fashion to ensure that \eqref{KactionSHWightman} is satisfied. This is because each individual term has a Ward-Takahashi identity which cancels out precisely in the way that they appear. In contrast, the coefficients of the $h_i$ in \eqref{TJJformfactors0} (namely $a_0,a_1,a_2$) are completely arbitrary at this stage since their coefficients individually satisfy \eqref{KactionSHWightman}\footnote{These coefficients as we shall see correspond to homogeneous correlators in Euclidean space and thus it makes sense that conservation yields no constraints on them.}.

This is where the third point, i.e., demanding that every form factor of the correlator vanishes when any of momenta are zero comes into play. It further enforces\footnote{Note that this also imposes $(2\leftrightarrow 3)$ permutation symmetry! (Recall that the different momentum space Wightman functions differ only by the arguments of the theta functions that accompany them  \eqref{thetafunctions}). This in hindsight is expected given the analytic continuation \eqref{WightmanToEuclidSpinning} that relates the Wightman function to its permutation symmetric Euclidean counterpart.},
\begin{align}\label{TJJconstraints}
    b_2=-b_0~,~a_1=a_2=a_3=-a_0.
\end{align}

Thus, we have obtained a two parameter family of solutions viz,
\small
\begin{align}\label{TJJevensols}
    \langle 0|T(z_1,p_1)J(z_2,p_2)J(z_3,p_3)|0\rangle&_{\text{even}}=b_0\langle 0|T(z_1,p_1)J(z_2,p_2)J(z_3,p_3)|0\rangle_{nh}+a_0\langle 0|T(z_1,p_1)J(z_2,p_2)J(z_3,p_3)|0\rangle_h\notag\\
    &=c_{211}^{(nh)}\langle 0|T(z_1,p_1)J(z_2,p_2)J(z_3,p_3)|0\rangle_{nh}+c_{211}^{(h)}\langle 0|T(z_1,p_1)J(z_2,p_2)J(z_3,p_3)|0\rangle_h,
\end{align}
\normalsize
where we have identified $a_0$ and $b_0$ with the OPE coefficients $c_{211}^{(h)}$ and $c_{211}^{(nh)}$ respectively. The reasoning for labeling these solutions as $nh$ and $h$ is due to their relation to the corresponding Euclidean non-homogeneous and homogeneous correlators. At this stage both $c_{211}^{(h)}$ and $c_{211}^{(nh)}$ are independent OPE data. However, by comparing with the results from analytic continuation \eqref{TJJWightmanhandnh}, we see that $c_{211}^{(nh)}$ is actually the coefficient of the $\langle JJ\rangle$ two point function through the Ward-Takahashi identity.
\subsection{Inside the triangle: An odd example}

Moving on, we can also construct a Wightman function that is odd under both parity and time-reversal separately \footnote{It is of course, even under the combination $PT$.}. After writing down an ansatz involving the three dimensional Levi-Civita symbol, we follow the same procedure as in the parity even case. The result is,
\begin{align}\label{TJJWightmanodd}
    &\langle 0|T(z_1,p_1)J(z_2,p_2)J(z_3,p_3)|0\rangle_{\text{odd}}=-i\bigg((z_1\cdot p_2)^2\epsilon^{z_2 p_2 z_3}\tilde{A}_1+(z_1\cdot p_2)\epsilon^{z_2 p_2 z_1}(z_3\cdot p_1)\tilde{A}_2\notag\\
    &+(z_1\cdot p_2)(z_1\cdot z_3)\epsilon^{z_2 p_2 p_3}\tilde{A}_3+(z_1\cdot z_3)\epsilon^{z_2 p_2 z_1}\tilde{A}_4\bigg).
\end{align}
with,
\begin{align}
    \tilde{A}_1=-\mathbf{\Delta}\bigg(\frac{p_1^3}{p_2 E^4}\bigg), \tilde{A}_2=-\mathbf{\Delta}\bigg(\frac{p_1^2(p_1+p_2)}{p_2 E^4}\bigg), \tilde{A}_3=\mathbf{\Delta}\bigg(\frac{p_1^2(p_1+p_3)}{p_2 E^4}\bigg), \tilde{A}_4=\mathbf{\Delta}\bigg(\frac{p_1^2(E-2p_3)(E-2p_2)}{2p_2 E^3}\bigg),
\end{align}
where the $\mathbf{\Delta}$ operation is as follows:
\begin{align}\label{mathbfDelta}
    \mathbf{\Delta}\bigg(f(p_1,p_2,p_3)\bigg)=f(p_1,p_2,p_3)-f(-p_1,p_2,p_3)-f(p_1,-p_2,p_3)-f(p_1,p_2,-p_3).
\end{align}
\eqref{TJJWightmanodd} is identically conserved and satisfies all the properties required of a Wightman function. One can also check that the Wightman epsilon transform relation \eqref{WightmanEPT} that relates \eqref{TJJWightmanodd} to the homogeneous Wightman function in  \eqref{TJJWightmanhandnh} is obeyed.

\subsection{Summary inside the triangle}
Thus, to summarize, we have three independent Wightman functions for $\langle 0| TJJ|0\rangle$: Two parity and time reversal even \eqref{TJJevensols} and one parity and time reversal odd \eqref{TJJWightmanodd}:

\small
\begin{align}\label{TJJallWightman}
    \langle 0|T(p_1,z_1)J(p_2,z_2)J(p_3,z_3)|0\rangle&=c_{211}^{(nh)}\langle 0|T(p_1,z_1)J(p_2,z_2)J(p_3,z_3)|0\rangle_{nh}+c_{211}^{(h)} \langle 0|T(p_1,z_1)J(p_2,z_2)J(p_3,z_3)|0\rangle_h\notag\\&+c_{211}^{(\text{odd})} \langle 0|T(p_1,z_1)J(p_2,z_2)J(p_3,z_3)|0\rangle_{\text{odd}}.
\end{align}
\normalsize

All of them are identically conserved which can be checked using \eqref{KactionSHWightman}. All of them vanish when any of the external momenta are zero. Before we move on, some comments about the pole structure of the correlators are in order.

\subsection*{Pole structure of Wightman function}
The Euclidean correlators only possess total energy singularities like $\frac{1}{E^{\#}}$ but not $\frac{1}{(E-2p_1)^{\#}}$,$\frac{1}{(E-2p_2)^{\#}}$,$\frac{1}{(E-2p_3)^{\#}}$. This is a consequence of consistency with the Euclidean operator product expansion. The Wightman functions on the other hand possess all four types of singularities. They arise from the Euclidean correlator from identities such as
\begin{align}
    \prod_{i=1}^{3}\text{Disc}_{p_i^2}\frac{1}{E}=-2\bigg(\frac{1}{E}-\frac{1}{E-2p_1}-\frac{1}{E-2p_2}-\frac{1}{E-2p_3}\bigg).
\end{align}

\subsection*{Helicity structure of Wightman function}

The non-homogeneous Euclidean correlator in \eqref{TJJEuclidNHandH} is non-zero only in the mixed helicity configurations, i.e, it is zero in the $(---)$ and $(+++)$ configurations. The homogeneous Euclidean correlator in \eqref{TJJEuclidNHandH} and the parity odd correlator which is its epsilon transform \eqref{TJJWightmanodd} on the other hand are non-zero only when all helicities coincide. In contrast, all their Wightman counterparts in \eqref{TJJallWightman} are non-zero in all eight helicity configurations.

\begin{align}\label{TJJhelicities}
     &\langle 0|T^{-}J^{-}J^{-}|0\rangle=(c_{211}^{(h)}-i c_{211}^{odd})\frac{\langle 1 2\rangle^2\langle 3 1\rangle^2 p_1}{E^4}+4c_{211}^{(nh)}\frac{\langle 12\rangle^2\langle 3 1\rangle^2 p_1}{(E-2p_3)^2(E-2p_2)^2},\notag\\&\langle 0|T^{-}J^{-}J^{+}|0\rangle=(c_{211}^{(h)}-i c_{211}^{odd})\frac{\langle 1 2\rangle^2\langle \Bar{3}1\rangle^2 p_1}{(E-2p_3)^4}+4c_{211}^{(nh)}\frac{\langle 12\rangle^2\langle \Bar{3} 1\rangle^2 p_1}{(E-2p_1)^2 E^2},\notag\\
     &\langle 0|T^{-}J^{+}J^{-}|0\rangle=(c_{211}^{(h)}-i c_{211}^{odd})\frac{\langle 1 \Bar{2}\rangle^2\langle 3 1\rangle^2 p_1}{(E-2p_2)^4}+4c_{211}^{(nh)}\frac{\langle 1\Bar{2}\rangle^2\langle 3 1\rangle^2 p_1}{(E-2p_1)^2 E^2},\notag\\&\langle 0|T^{+}J^{-}J^{-}|0\rangle=(c_{211}^{(h)}-i c_{211}^{odd})\frac{\langle \Bar{1} 2\rangle^2\langle 3 \Bar{1}\rangle^2 p_1}{(E-2p_1)^4}+4c_{211}^{(nh)}\frac{\langle \Bar{1}2\rangle^2\langle 3 \Bar{1}\rangle^2 p_1}{(E-2p_3)^2(E-2p_2)^2},\notag\\
     &\langle 0|T^{+}J^{+}J^{+}|0\rangle=(c_{211}^{(h)}+ic_{211}^{odd})\frac{\langle \Bar{1}\Bar{2}\rangle^2\langle\Bar{3}\Bar{1}\rangle^2 p_1}{E^4}+4c_{211}^{(nh)}\frac{\langle \Bar{1}\Bar{2}\rangle^2\langle \Bar{3} \Bar{1}\rangle^2 p_1}{(E-2p_3)^2(E-2p_2)^2},\notag\\&\langle 0|T^{+}J^{+}J^{-}|0\rangle=(c_{211}^{(h)}+ic_{211}^{odd})\frac{\langle \Bar{1}\Bar{2}\rangle^2\langle3\Bar{1}\rangle^2 p_1}{(E-2p_3)^4}+4c_{211}^{(nh)}\frac{\langle \Bar{1}\Bar{2}\rangle^2\langle 3 \Bar{1}\rangle^2 p_1}{(E-2p_1)^2 E^2},\notag\\
     &\langle 0|T^{+}J^{-}J^{+}|0\rangle=(c_{211}^{(h)}+ic_{211}^{odd})\frac{\langle \Bar{1}2\rangle^2\langle\Bar{3}\Bar{1}\rangle^2 p_1}{(E-2p_2)^4}+4c_{211}^{(nh)}\frac{\langle \Bar{1}2\rangle^2\langle \Bar{3} \Bar{1}\rangle^2 p_1}{(E-2p_1)^2 E^2},\notag\\&\langle 0|T^{-}J^{+}J^{+}|0\rangle=(c_{211}^{(h)}+ic_{211}^{odd})\frac{\langle 1\Bar{2}\rangle^2\langle\Bar{3}1\rangle^2 p_1}{(E-2p_1)^4}+4c_{211}^{(nh)}\frac{\langle 1\Bar{2}\rangle^2\langle \Bar{3} 1\rangle^2 p_1}{(E-2p_2)^2 (E-2p_3)^2}.
\end{align}
Notice in particular the change of sign of the parity odd contribution in the different helicity configurations. This is in accord with their Euclidean counterparts.

\subsection{Outside the triangle}
Let us now investigate the situation outside the triangle.  Here, we just present the final results and expressions in spinor helicity variables as the analysis is analogous to the case inside the triangle.
To summarize, in the conserved case, we obtained two independent Wightman functions for $\langle 0| J_4JJ|0\rangle$. Both of these are even under parity:
\small
\begin{align}\label{J4JJWightmansummary1}
    &\langle 0|J_4(p_1,z_1)J(p_2,z_2)J(p_3,z_3)\rangle|0\rangle\notag\\&=c_{211}^{(F+B)} \langle 0|J_4(p_1,z_1)J(p_2,z_2)J(p_3,z_3)\rangle|0\rangle_{F+B}+c_{211}^{(F-B)} \langle 0|J_4(p_1,z_1)J(p_2,z_2)J(p_3,z_3)\rangle|0\rangle_{F-B}.
\end{align}
\normalsize
 When current conservation is slightly broken, an odd structure becomes permissible and then the result is,
 \begin{align}\label{J4JJWightmansummary}
     &\langle 0|J_4(p_1,z_1)J(p_2,z_2)J(p_3,z_3)\rangle|0\rangle\notag\\&=c_{211}^{(F+B)} \langle 0|J_4(p_1,z_1)J(p_2,z_2)J(p_3,z_3)\rangle|0\rangle_{F+B}+c_{211}^{(F-B)} \langle 0|J_4(p_1,z_1)J(p_2,z_2)J(p_3,z_3)\rangle|0\rangle_{F-B}\notag\\&+c_{211}^{(\text{odd})} \langle 0|J_4(p_1,z_1)J(p_2,z_2)J(p_3,z_3)\rangle|0\rangle_{\text{odd}}.
 \end{align}
More discussion on this odd structure can be found in sub-appendix  \ref{sec:SBHS}. Here, we focus on the even solutions.
\subsection*{Pole structure}
Similar to the cases inside the triangle, the Wightman functions outside the triangle have in addition to total energy singularities, the other three types unlike their Euclidean counterpart.
\subsection*{Helicity structure}
The ($F-B$)  Euclidean correlator has support only in the $(---)$ and $(+--)$ helicities whereas the ($F+B$) Euclidean correlator is only non-zero in the remaining ones. In contrast, all Wightman functions are non-vanishing in all helicities. 
\begin{align}\label{J4JJhelicities}
     &\langle 0|J_4^{-}J^{-}J^{-}|0\rangle=c_{411}^{(F-B)}\frac{\langle 1 2\rangle^4\langle 3 1\rangle^4 p_1^3}{128 \langle 2 3\rangle^2E^6}+c_{411}^{(F+B)}\frac{\langle 1 2\rangle^4 \langle 3 1\rangle^4 p_1^3(E-2p_1)^6}{128 \langle 2 3\rangle^6(E-2p_3)^4(E-2p_2)^4}\notag\\&\langle 0|J_4^{-}J^{-}J^{+}|0\rangle=c_{411}^{(F-B)}\frac{\langle 1 2\rangle^4\langle \Bar{3} 1\rangle^4 p_1^3}{128 \langle 2 \Bar{3}\rangle^2(E-2p_3)^6}+c_{411}^{(F+B)}\frac{\langle 1 2\rangle^4 \langle \Bar{3} 1\rangle^4 p_1^3(E-2p_2)^6}{128 \langle 2 3\rangle^6E^4(E-2p_1)^4},\notag\\
     &\langle 0|J_4^{-}J^{+}J^{-}|0\rangle=c_{411}^{(F-B)}\frac{\langle 1 \Bar{2}\rangle^4\langle 3 1\rangle^4 p_1^3}{128 \langle \Bar{2} 3\rangle^2(E-2p_2)^6}+c_{411}^{(F+B)}\frac{\langle 1 \Bar{2}\rangle^4 \langle 3 1\rangle^4 p_1^3(E-2p_3)^6}{128 \langle \Bar{2} 3\rangle^6(E-2p_1)^4E^4},\notag\\&\langle 0|J_4^{+}J^{-}J^{-}|0\rangle=c_{411}^{(F-B)}\frac{\langle \Bar{1} 2\rangle^4\langle 3 \Bar{1}\rangle^4 p_1^3}{128 \langle 2 3\rangle^2(E-2p_1)^6}+c_{411}^{(F+B)}\frac{\langle \Bar{1} 2\rangle^4 \langle 3 \Bar{1}\rangle^4 p_1^3 E^6}{128 \langle 2 3\rangle^6(E-2p_2)^4(E-2p_3)^4},\notag\\
     &\langle 0|J_4^{+}J^{+}J^{+}|0\rangle=c_{411}^{(F-B)}\frac{\langle \Bar{1} \Bar{2}\rangle^4\langle \Bar{3} \Bar{1}\rangle^4 p_1^3}{128 \langle \Bar{2} \Bar{3}\rangle^2E^6}+c_{411}^{(F+B)}\frac{\langle \Bar{1} \Bar{2}\rangle^4 \langle \Bar{3} \Bar{1}\rangle^4 p_1^3 E^6}{128 \langle \Bar{2} \Bar{3}\rangle^6(E-2p_2)^4(E-2p_3)^4},\notag\\&\langle 0|J_4^{+}J^{+}J^{-}|0\rangle=c_{411}^{(F-B)}\frac{\langle \Bar{1} \Bar{2}\rangle^4\langle 3 \Bar{1}\rangle^4 p_1^3}{128 \langle \Bar{2} 3\rangle^2(E-2p_3)^6}+c_{411}^{(F+B)}\frac{\langle \Bar{1} \Bar{2}\rangle^4 \langle 3\Bar{1}\rangle^4 p_1^3 (E-2p_3)^6}{128 \langle \Bar{2} 3\rangle^6(E-2p_1)^4 E^4},\notag\\
     &\langle 0|J_4^{+}J^{-}J^{+}|0\rangle=c_{411}^{(F-B)}\frac{\langle \Bar{1} 2\rangle^4\langle \Bar{3} \Bar{1}\rangle^4 p_1^3}{128 \langle 2 3\rangle^2(E-2p_2)^6}+c_{411}^{(F+B)}\frac{\langle \Bar{1} 2\rangle^4 \langle \Bar{3} \Bar{1}\rangle^4 p_1^3 (E-2p_2)^6}{128 \langle 2 \Bar{3}\rangle^6 E^4(E-2p_1)^4},\notag\\&\langle 0|J_4^{-}J^{+}J^{+}|0\rangle=c_{411}^{(F-B)}\frac{\langle 1\Bar{2}\rangle^4\langle \Bar{3} 1\rangle^4 p_1^3}{128 \langle \Bar{2} \Bar{3}\rangle^2(E-2p_1)^6}+c_{411}^{(F+B)}\frac{\langle 1 \Bar{2}\rangle^4 \langle \Bar{3} 1\rangle^4 p_1^3 (E-2p_1)^6}{128 \langle \Bar{2} \Bar{3}\rangle^6(E-2p_3)^4(E-2p_2)^4}.
\end{align}

\subsection{Outside the triangle: Slightly broken conservation}\label{sec:SBHS}
If one allows the current conservation to be slightly broken\footnote{In a large $N$ theory, the non-conservation of such a current starts at $\order{\frac{1}{N}}$ \cite{Maldacena:2012sf}.} in the Euclidean case, a parity-odd structure is also allowed. The general form for such a correlator is then,
\begin{align}\label{CFT3Euclidthreepointoutside1SBHS}
    \langle J_{s_1}J_{s_2}J_{s_3}\rangle=c_{s_1s_2s_3}^{(nh_1)}\langle J_{s_1}J_{s_2}J_{s_3}\rangle_{F-B}+c_{s_1s_2s_3}^{(odd)}\langle J_{s_1}J_{s_2}J_{s_3}\rangle_{odd}+c_{s_1s_2s_3}^{(nh_2)}\langle J_{s_1}J_{s_2}J_{s_3}\rangle_{F+B}.
\end{align}
Interestingly, the odd structure is given by an epsilon transform of the $F-B$ structure\footnote{Note however, that this epsilon transform in contrast to the situation inside the triangle should only be performed with respect to the currents with the lowest two spins in the correlator \cite{Jain:2021gwa,Jain:2021whr}.},
\begin{align}\label{EPToutside}
    \langle J_{s_1}J_{s_2}J_{s_3}\rangle_{odd}=\langle J_{s_1}\epsilon\cdot J_{s_2}J_{s_3}\rangle_{F-B}.
\end{align}
For Wightman functions we find a similar formula namely,
\begin{align}\label{CFT3WightmanthreepointoutsideSBHS}
    \langle 0|J_{s_1}J_{s_2}J_{s_3}|0\rangle=c_{s_1s_2s_3}^{(F-B)}\langle 0|J_{s_1}J_{s_2}J_{s_3}|0\rangle_{F-B}+c_{s_1s_2s_3}^{(odd)}\langle 0|J_{s_1}J_{s_2}J_{s_3}|0\rangle_{odd}+c_{s_1s_2s_3}^{(F+B)}\langle 0|J_{s_1}J_{s_2}J_{s_3}|0\rangle_{F+B}.
\end{align}
The odd Wightman function, just like its Euclidean counterpart is also not conserved. In the Euclidean case, the epsilon transform \eqref{EPToutside} works because the Ward-Takahashi identity of the $F-B$ correlator after an epsilon transform becomes the non-conservation of the odd correlator. For the Wightman case on the other hand, the $F-B$ correlator has a zero Ward-Takahashi identity and thus this implies that there is no simple epsilon transform between these Wightman functions that relates one to the other. However, one can still bootstrap it using the methods outlined so far as we show explicitly with examples in appendix \ref{app:WightmanExamples}. 

\subsection{An example with a scalar operator}
Consider the correlator $\langle JJO_1\rangle$. In Euclidean space it is given by,
\begin{align}
\langle J(z_1,p_1)J(z_2,p_2)O_1(p_3)\rangle_{even}=&z_1\cdot z_2\Big(\frac{p_3^2-(p_1+p_2)^2}{2p_3E^2}\Big)-z_1\cdot p_2\;z_2\cdot z_1\Big(\frac{1}{p_3E^2}\Big).
\end{align}
Its Wightman counterpart (obtained via the bootstrap or analytic continuation via \eqref{WightmanToEuclidSpinning}) is given by,
\begin{align}
\notag\langle0| J(z_1,p_1)J(z_2,p_2)O_1(p_3)|0\rangle_{even}=&z_1\cdot z_2\Big(\frac{p_1p_2p_3}{E(E-2p_1)(E-2p_2)(E-2p_3)}\Big)\\+z_1\cdot p_2\;z_2\cdot z_1\Big(&\frac{p_1p_2(4(p_3^2-(p_1^2+p_2^2))+E(E-2p_1)(E-2p_2)(E-2p_3))}{p_3E^2(E-2p_1)^2(E-2p_2)^2(E-2p_3)^2}\Big).
\end{align}
A similar analysis can be performed for the parity odd case although we do not present the explicit details here.
\subsection{An example with spinors}
We present a prototypical example of our analysis for correlators with spin-half operators, focussing on the homogeneous correlator for simplicity. Consider $\langle O_\frac{1}{2}O_\frac{1}{2}J\rangle$ where the homogeneous part of the correlator is:
\begin{align}
\notag\langle O_\frac{1}{2}(\zeta_1,p_1)O_\frac{1}{2}(\zeta_2,p_2)J(z_3,p_3)\rangle_h=&\zeta_{1a}\zeta_{2b}\epsilon^{ab}\epsilon^{z_3p_1p_2}\frac{1}{p_1p_2E}-i\zeta_{1a}\zeta_{2b}\slashed{z_3}^{ab}\frac{(E-2p_1)(E-2p_2)}{2p_1p_2E}\\-&i\zeta_{1a}\zeta_{2b}\slashed{p_1}^{ab}z_3\cdot p_1\frac{(E-2p_2)}{p_1p_2E^2}-i\zeta_{1a}\zeta_{2b}\slashed{p_2}^{ab}z_3\cdot p_2\frac{(E-2p_1)}{p_1p_2E^2},
\end{align}
Its Wightman counterpart can be obtained using the methods outlined in section \ref{sec:Wightman} and is given by,
\begin{align}
\notag\langle0|O_\frac{1}{2}(\zeta_1,p_1)O_\frac{1}{2}(\zeta_2,p_2)J(z_3,p_3)|0\rangle_h=&\zeta_{1a}\zeta_{2b}\epsilon^{ab}\epsilon^{z_3p_1p_2}\frac{2p_3(p_1^2+p_2^2-p_3^2)}{p_1p_2E(E-2p_1)(E-2p_2)(E-2p_3)}\\\notag+i\zeta_{1a}\zeta_{2b}\slashed{z_3}^{ab}\Big(&\frac{p_3}{p_1p_2}-\frac{8p_1p_2p_3}{p_1p_2E(E-2p_1)(E-2p_2)(E-2p_3)}\Big)\\\notag+i\zeta_{1a}\zeta_{2b}\slashed{p_1}^{ab}z_3\cdot p_1&\frac{2p_3(3 p_{2}^{6} + p_{1}^{2} ( 9 p_{1}^{2} - 5 p_{3}^{2} ) - ( p_{1}^{2} - p_{3}^{2} )^3 + p_{2}^{2} ( -11 p_{1}^{4} + 10 p_{1}^{2} p_{3}^{2} + p_{3}^{4} )
)}{p_1p_2E^2(E-2p_1)^2(E-2p_2)^2(E-2p_3)^2}\\+i\zeta_{1a}\zeta_{2b}\slashed{p_2}^{ab}z_3\cdot p_2&\frac{2p_3(3 p_{1}^{6} + p_{1}^{4} ( 9 p_{2}^{2} - 5 p_{3}^{2} ) - ( p_{2}^{2} - p_{3}^{2} )^3 + p_{1}^{2} ( -11 p_{2}^{4} + 10 p_{2}^{2} p_{3}^{2} + p_{3}^{4} ))}{p_1p_2E^2(E-2p_1)^2(E-2p_2)^2(E-2p_3)^2},
\end{align}
A similar analysis can be carried out for the non-homogeneous as well as the parity-odd correlator, but we do not present it here. Analysis involving half-integer spin currents is also analogous.

\bibliographystyle{JHEP}
\bibliography{biblio}
\end{document}